\documentclass[imslayout,preprint]{imsart}
\def\journal@name{} 
\usepackage[margin=1.5in]{geometry}

\usepackage{mathtools}
\usepackage[colorlinks,citecolor=blue,urlcolor=black,linkcolor=blue,pdfborder={0 0 0}]{hyperref}
\usepackage{url}
\usepackage{comment}
\usepackage{graphicx}
\usepackage{enumitem}
\usepackage{framed}
\usepackage{dcolumn}
\usepackage{chngcntr}
\usepackage{subcaption}
\usepackage[sort]{natbib}
\usepackage[capitalize]{cleveref}  
\usepackage{datetime}
\usepackage{algorithm}
\usepackage{algpseudocode}
\usepackage[normalem]{ulem}
\usepackage{upgreek}
 \usepackage{multirow}
 \usepackage[table,xcdraw,usenames,dvipsnames]{xcolor}
 \usepackage{booktabs}

\crefname{nlem}{Lemma}{Lemmas}
\crefname{nprop}{Proposition}{Propositions}
\crefname{ncor}{Corollary}{Corollaries}
\crefname{nthm}{Theorem}{Theorems}
\crefname{nnon}{Conjecture}{Conjectures}
\crefname{assumption}{Assumption}{Assumptions}
\crefname{exa}{Example}{Examples}

\usepackage{jhh-misc2}
\usepackage{mathrsfs}
\usepackage{autonum}

\newcommand{\cf}[1]{\psi_{#1}} 
\newcommand{\ii}{i} 

\newcommand{\indicatorfn}{\mathds{1}}

\newcommand{\littleo}{o}

\newcommand{\empcov}{\hat\Sigma}

\newcommand{\todo}[1]{\textcolor{red}{#1}}

\newcommand{\optsym}{\circ}

\newcommand{\obsspace}{\mathbb X}
\newcommand{\numobs}{N}
\newcommand{\data}{x}
\newcommand{\datarv}{X}
\newcommand{\dataarg}[1]{x_{1:#1}}
\newcommand{\datarvarg}[1]{X_{1:#1}}

\newcommand{\alldatarv}{\datarvarg{\infty}}

\newcommand{\obs}[1]{x_{#1}}
\newcommand{\obsrv}[1]{X_{#1}}
\newcommand{\datameanarg}[1]{\bar\data_{#1}}
\newcommand{\datarvmeanarg}[1]{\bar\datarv_{#1}}

\newcommand{\postdist}[1]{\Pi_{#1}}
\newcommand{\postdistfull}[2]{\Pi(#1 \given #2)}
\newcommand{\ppostdist}[2]{\Pi_{#1}^{#2}}
\newcommand{\postdensity}[1]{\pi_{#1}}
\newcommand{\postdensityfull}[2]{\pi(#1 \given #2)}

\newcommand{\priordist}{\postdist{0}}
\newcommand{\priordensity}{\postdensity{0}}

\newcommand{\likdist}[1]{P_{#1}}
\newcommand{\likfun}[1]{p_{#1}}
\newcommand{\lik}[2]{\likfun{#2}(#1)}
\newcommand{\loglik}[2]{\ell_{#2}(#1)}
\newcommand{\loglikfun}[1]{\ell_{#1}}

\newcommand{\gradloglik}[2]{\dot\ell_{#2}(#1)}
\newcommand{\gradloglikfun}[1]{\dot\ell_{#1}}

\newcommand{\hessloglikfun}[1]{\ddot\ell_{#1}}

\newcommand{\marginallikfull}[1]{p(#1)}

\newcommand{\param}{\theta}
\newcommand{\paramspace}{\Theta}
\newcommand{\trueparam}{\param_{\dagger}}
\newcommand{\optparam}{\param_{\optsym}}
\newcommand{\paramsample}{\vartheta}

\newcommand{\bbparamsample}{\paramsample^{\bbsym}}
\newcommand{\bbparamsamplecopy}{\paramsample^{\bbsym\prime}}

\newcommand{\mle}[1]{\hat\param_{#1}}

\newcommand{\Ehessloglik}[1]{J_{#1}}
\newcommand{\Vargradloglik}[1]{I_{#1}}

\newcommand{\obsdist}{P_{\optsym}}
\newcommand{\empdist}{\Pr_{\numobs}}

\newcommand{\bbsym}{*} 
\newcommand{\bsnumobs}{M}

\newcommand{\bscount}[1]{K_{#1}}
\newcommand{\bscounts}{K_{1:\numobs}}
\newcommand{\bsobs}[1]{\obs{#1}^{\bbsym}}

\newcommand{\bsdata}{\data^{\bbsym}}
\newcommand{\bsdatarv}{\datarv^{\bbsym}}
\newcommand{\bsdatasample}[1]{\data^{\bbsym}_{(#1)}}

\newcommand{\bsdatarvarg}[1]{\bsdatarv_{1:#1}}

\newcommand{\bbpostdist}[1]{\postdist{#1}^{\bbsym}}
\newcommand{\bbpostdistfull}[2]{\Pi^{\bbsym}(#1 \given #2)}
\newcommand{\bbpostdensity}[1]{\postdensity{#1}^{\bbsym}}
\newcommand{\bbpostdensityfull}[2]{\postdensity{}^{\bbsym}(#1 \given #2)}

\newcommand{\bsdatarvmeanarg}[1]{\bar\datarv^{\bbsym}_{#1}}
\newcommand{\bbempdist}{\Pr_{\numobs}^{\bbsym}}

\newcommand{\bsscale}{c}

\newcommand{\concconst}{C_{\numobs}}

\def\norm#1{\left\|{#1}\right\|} 
\newcommand{\twonorm}[1]{\norm{#1}_2} 

\graphicspath{{figures/}}

\allowdisplaybreaks[2]

\begin{document}

\begin{frontmatter}

\title{Reproducible Parameter Inference \\ Using Bagged Posteriors}
\runtitle{~Reproducible Parameter Inference}
\runauthor{J.\ H.\ Huggins and J.\ W.\ Miller~}

\begin{aug}

\author[A]{\fnms{Jonathan H.} \snm{Huggins}\ead[label=e1]{huggins@bu.edu}} %
\and
\author[B]{\fnms{Jeffrey W.} \snm{Miller}\ead[label=e2]{jwmiller@hsph.harvard.edu}}
\address[A]{Department of Mathematics \& Statistics, Boston University, \printead{e1}}
\address[B]{Department of Biostatistics, Harvard University, \printead{e2}}

\end{aug}

\begin{abstract}
Under model misspecification, it is known that Bayesian posteriors often do not properly quantify uncertainty about true or pseudo-true parameters.
Even more fundamentally, misspecification leads to a lack of reproducibility in the sense that the same model will yield 
contradictory posteriors on independent data sets from the true distribution.
To define a criterion for reproducible uncertainty quantification under misspecification, we consider the probability that two confidence sets constructed from independent data sets have nonempty overlap, and we establish a lower bound on this overlap probability that holds for any valid confidence sets.
We prove that credible sets from the standard posterior can strongly violate this bound, particularly in high-dimensional settings
(i.e., with dimension increasing with sample size), indicating that it is not internally coherent under misspecification.
To improve reproducibility in an easy-to-use and widely applicable way, 
we propose to apply bagging to the Bayesian posterior (``BayesBag''); that is, 
to use the average of posterior distributions conditioned on bootstrapped datasets.
We motivate BayesBag from first principles based on Jeffrey conditionalization 
and show that the bagged posterior typically satisfies the overlap lower bound.
Further, we prove a Bernstein--Von Mises theorem for the bagged posterior, establishing its asymptotic normal distribution.
We demonstrate the benefits of BayesBag via simulation experiments and an application to crime rate prediction. %
\end{abstract}

\begin{keyword}
\kwd{Bagging}
\kwd{Bernstein--Von Mises theorem}
\kwd{Bootstrap}
\kwd{Model misspecification}
\kwd{Overlap probability}
\kwd{Uncertainty calibration}
\end{keyword}

\end{frontmatter}

\section{Introduction} \label{sec:introduction}

It is widely acknowledged that statistical models are usually not exactly correct in practice~\citep{Box:1979,Box:1980,Cox:1990,Lehmann:1990}.
This model misspecification is known to lead to unreliable inferences,
and in particular, Bayesian posteriors can be unstable and poorly calibrated under misspecification~\citep{Kleijn:2012,Greco:2008:robust,Jewson:2018:principles}.
Unfortunately, this leads to a lack of reproducibility, even when using the same method on a replicate data set from the same distribution
\citep{Yang:2018,Huggins:2019:BayesBagII}.
In this paper, we propose a criterion for reproducible uncertainty quantification and a general technique for achieving it.

Defining valid uncertainty quantification in misspecified models presents a conceptual problem since there is no ``''true parameter'' that 
indexes a model with distribution equal to the data-generating distribution. 
The usual solution is to focus on a pseudo-true parameter,
typically defined as the asymptotically optimal parameter in terms of Kullback--Leibler (KL) divergence \citep{Kleijn:2012,Grunwald:2012,Walker:2001,Muller:2013}. 
However, depending on the objectives of the analysis, it might not be desirable to concentrate at the KL-optimal parameter \citep{Miller:2018:coarsening,Bissiri:2016,Jewson:2018:principles}.
Thus, rather than adopting a particular definition of pseudo-truth, we introduce a truth-agnostic approach to assessing reproducibility. 
Specifically, we consider the probability that two credible sets constructed from independent data sets have nonempty overlap, 
and we establish a simple lower bound on this overlap probability that holds for any valid confidence sets. 
Under misspecification, we show that credible sets from the standard posterior can strongly violate this bound -- particularly when
the dimension grows with the number of observations -- indicating that it exhibits poor reproducibility.

To improve the reproducibility of Bayesian inference under misspecification, we propose to use \emph{BayesBag} \citep{Buhlmann:2014,Waddell:2002,Douady:2003}.
The idea of BayesBag is to apply bagging~\citep{Breiman:1996} to the Bayesian posterior.
More precisely, the \emph{bagged posterior} $\bbpostdensityfull{\param}{\data}$ is defined by taking bootstrapped copies $\bsdata \defined (\bsobs{1},\ldots,\bsobs{\bsnumobs})$ 
of the original dataset $\data \defined (\obs{1},\ldots,\obs{\numobs})$ and averaging over the posteriors obtained by treating each bootstrap dataset as the observed data:
\[
\bbpostdensityfull{\param}{\data} \defined \frac{1}{\numobs^{\bsnumobs}} \sum_{\bsdata} \postdensityfull{\param}{\bsdata}, \label{eq:bayesbag-definition}
\]
where $\postdensityfull{\param}{\bsdata} \propto \priordensity(\param)\prod_{m=1}^{\bsnumobs}\lik{\bsobs{m}}{\param}$ 
is the standard posterior density given data $\bsdata$ and the sum is over all $N^M$ possible bootstrap datasets of $\bsnumobs$ samples drawn 
with replacement from the original dataset.
In this work, we focus on parameter inference and prediction, complementing our work on BayesBag for model selection \citep{Huggins:2019:BayesBagII}. 
In theory and experiments, we consider both the case of fixed finite-dimensional parameters as well as high-dimensional cases where the dimension grows with the sample size.

We motivate the bagged posterior from first principles using Jeffrey conditionalization and show that bagged posterior credible sets typically 
satisfy our lower bound on the overlap probability, indicating that the bagged posterior quantifies uncertainty in a more reproducible way.
These results illustrate how the bagged posterior integrates the attractive features of Bayesian inference---such as flexible hierarchical modeling, automatic integration over nuisance parameters, and the use of prior information---with the distributional robustness of frequentist methods, nonparametrically accounting for sampling variability and model misspecification. 
Simulation experiments validate our theory and demonstrate the bagged posterior is particularly important for stability in high-dimensional settings. 
An application to crime rate prediction using a Poisson regression model with a horseshoe prior to induce approximate sparsity demonstrate that 
BayesBag-based analysis can also lead to different conclusions -- and better predictions -- than a standard Bayesian analysis.

In practice, we suggest approximating $\bbpostdensityfull{\param}{\data}$ by generating $B$ independent bootstrap datasets $\bsdatasample{1},\dots,\bsdatasample{B}$,
where $\bsdatasample{b}$ consists of $\bsnumobs$ samples drawn with replacement from $\data$, yielding the approximation
\[
\bbpostdensityfull{\param}{\data} \approx \frac{1}{B} \sum_{b=1}^{B} \postdensityfull{\param}{\bsdatasample{b}}.  \label{eq:bayesbag-approximation}
\]
Since the bagged posterior is just the average of standard Bayesian posteriors, one can use any algorithm for standard posterior inference to compute each of the $B$ posteriors, and then aggregate across them.  While this requires $B$ times as much computation as a single posterior, it is trivial to parallelize the computation of the $B$ posteriors.
Since \cref{eq:bayesbag-approximation} is a simple Monte Carlo approximation, the error of this approximation can easily be estimated in order to choose $B$ appropriately \citep{Huggins:2019:BayesBagII}.

Despite its many attractive features, there has been little practical or theoretical investigation of bagged posteriors prior to \citet{Huggins:2019:BayesBagII}.
In the only previous work of which we are aware, \citet{Buhlmann:2014} presented some simulation results for a simple 
Gaussian location model, while \citet{Waddell:2002} and \citet{Douady:2003} used bagged posteriors for phylogenetic tree inference in papers focused primarily on 
speeding up model selection and comparing Bayesian inference versus the bootstrap.

The article is organized as follows.
In \cref{sec:motivation}, we motivate the use of BayesBag for reproducible uncertainty quantification in terms of our overlap criterion as well as a Jeffrey conditionalization derivation.
In \cref{sec:theory-overlap}, we prove that the standard posterior often fails to satisfy the overlap criterion,
whereas the bagged posterior typically satisfies it, focusing on Gaussian location models, regular finite-dimensional models, and linear regression.
In \cref{sec:theory-bvm}, we prove a general Bernstein--Von Mises theorem establishing the asymptotic normal distribution of the bagged posterior, 
which is employed in the overlap theory of the preceding section.
\cref{sec:simulations} evaluates the performance of the bagged posterior in simulation studies,
and \cref{sec:application} illustrates with an application to crime rate prediction using Poisson regression.
We close with a discussion in \cref{sec:discussion}.

\section{Motivation} \label{sec:motivation}

\begin{figure}[tbp]
\begin{center}
\begin{subfigure}[b]{.49\textwidth}
\centering
\includegraphics[height=1.8in]{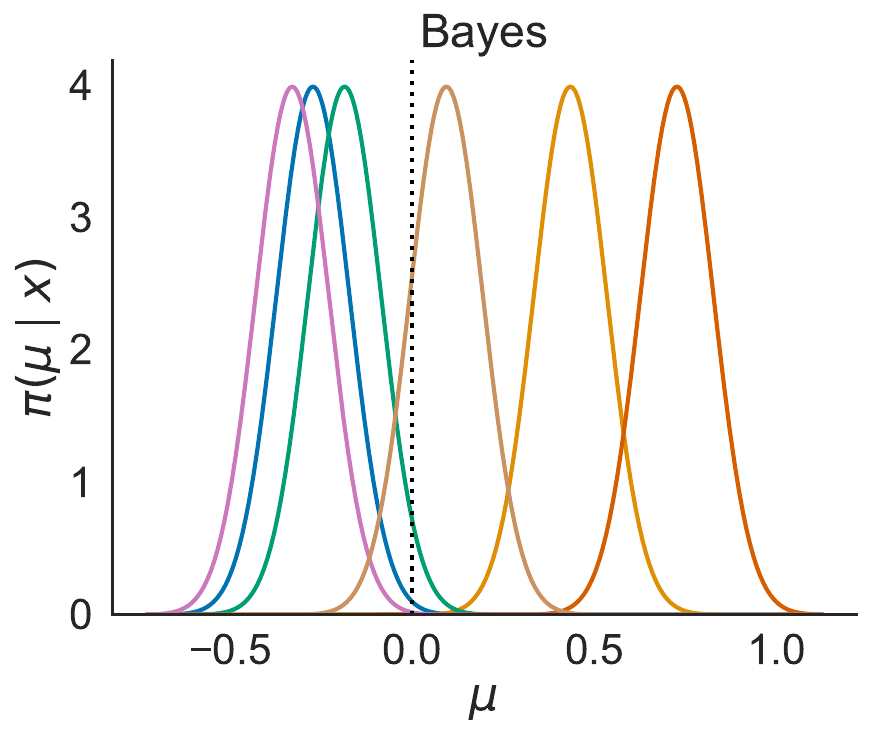}
\caption{}
\label{fig:location-model-overlap-Bayes}
\end{subfigure}
\begin{subfigure}[b]{.49\textwidth}
\centering
\includegraphics[height=1.8in]{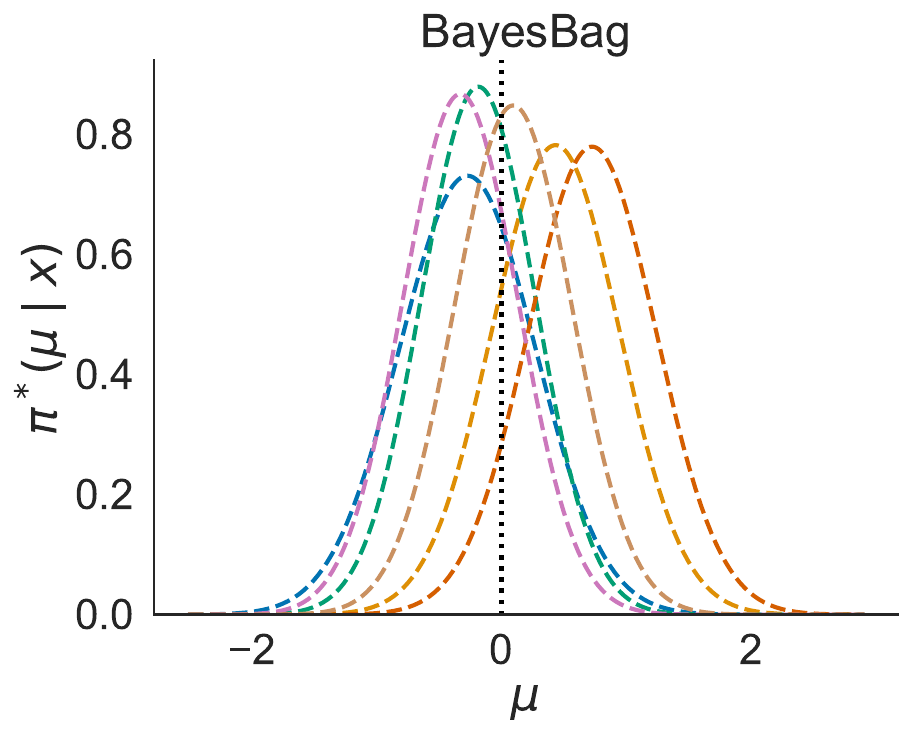}
\caption{}
\label{fig:location-model-overlap-BayesBag}
\end{subfigure}
\caption{Standard and bagged posterior distributions of the mean for a Gaussian location
model assuming the data are i.i.d.\ $\distNorm(\mu, 1)$, when the data are actually i.i.d.\ $\distNorm(0, 5^{2})$.
Posteriors for six randomly generated data sets of size $\numobs=100$ are shown.
\textbf{(a)} Many pairs of posterior distributions have essentially no overlap with each other, and 5 out 6 do not
contain the true mean in their 95\% central credible sets. 
\textbf{(b)} All pairs of bagged posterior distributions have significant overlap and 6 out of 6 contain the true mean 
in their 95\% central credible sets. 
}
\label{fig:location-model-overlap}
\end{center}
\end{figure}

When misspecified, a Bayesian model can be so unstable that it contradicts itself.
Specifically, given two independent data sets from the same distribution, 
the resulting two posteriors---for the same model---can place nearly all their
mass on disjoint sets.
\Cref{fig:location-model-overlap-Bayes} provides a simple illustration of the problem. 
Intuitively, it seems clear that this must violate some principle of coherent uncertainty quantification.
But if there is no true parameter for which the model is correct, then what is a posterior quantifying uncertainty about?
In most previous work, this question is dealt with by focusing on the pseudo-true parameter---that is, the model parameter value that is closest in Kullback--Leibler divergence to the true distribution
\citep{Kleijn:2012,Walker:2013,Hoff:2012,DeBlasi:2013}.
However, this choice---or any choice of pseudo-truth---is somewhat arbitrary and entails implicit assumptions about the goal of the analysis, such as minimizing a certain loss function.

In this section, we instead formulate a criterion for reproducible uncertainty quantification that does not require any assumptions of what is true in terms of models or parameters.
The basic idea is that two valid confidence sets constructed from independent data sets must intersect with a certain minimal probability.
We prove a simple lower bound on this overlap probability that holds
for any valid confidence sets, for any definition of pseudo-truth, and for any data distribution.
We then use this criterion to motivate the use of BayesBag via Jeffrey conditionalization.
\Cref{fig:location-model-overlap-BayesBag} illustrates how the bagged posterior does not suffer from the instability exhibited by  
the standard posterior. 

\subsection{Overlap criterion for reproducible uncertainty quantification}
\label{sec:overlap-criterion}

Suppose $x\mapsto A_x$ is a method of constructing confidence sets that takes data $x$ and produces a set $A_x$
that is intended to provide coverage of some unknown quantity of interest, $\eta$.
For any fixed value of $\eta$, let $X \given \eta$ be a random data set.
Here, $\eta$ does not have to be a model parameter.
Rather, it is simply some quantity that $X$ depends on.
\bnumdefn
We say that $x\mapsto A_x$ \textit{has coverage $1-\alpha$ with respect to $X \given  \eta$} if for all $\eta$, we have
$\mathbb{P}(\eta \in A_X \given  \eta) \geq 1-\alpha$.
\enumdefn
This definition is agnostic to making any assumptions of what is true in terms of models or parameters.

\bnprop \label{prop:overlap-bound}
Let $X$ and $Y$ be independent data sets, given $\eta$.
If $x\mapsto A_x$ and $y\mapsto B_y$ have coverage $1-\alpha$ and $1-\alpha'$ with respect to $X \given \eta$ and $Y \given \eta$, respectively, then
\[
\label{eq:overlap-criterion}
\mathbb{P}(A_X\cap B_Y \neq \varnothing \mid \eta) \geq (1-\alpha) (1-\alpha').
\]
\enprop
This provides a lower bound on the probability that two valid confidence sets intersect.
For example, if the coverage is $1-\alpha = 1-\alpha' = 0.95$, then the lower bound on the probability of intersection is $0.9025$.
We refer to $\mathbb{P}(A_X\cap B_Y \neq \varnothing \mid \eta)$ as the \emph{overlap probability}, 
and satisfying the bound is referred to as the \emph{overlap criterion}.
Failing to satisfy this criterion indicates a lack of stability and reproducibility across plausible datasets. 
While satisfying this bound is a necessary condition for coherent uncertainty quantification, it is 
not sufficient.
For example, choosing $A_{x} = A$, a constant, would satisfy the bound but would clearly be an ineffective method
for inference.

%

%
%

\subsection{Jeffrey conditionalization for reproducibility leads to BayesBag}
\label{sec:jeffrey-conditionalization}

For reproducibility, one needs to represent uncertainty across data sets from the true distribution.
A natural way to do this is via Jeffrey conditionalization, which turns out to lead to the bagged posterior. %
This interpretation elegantly unifies the Bayesian and frequentist elements of the bagged posterior 
that might otherwise seem challenging to interpret together in a principled way.
To explain, suppose we have a model $p(x,y)$ of two variables $x$ and $y$.
In the absence of any other data or knowledge,
we would quantify our uncertainty in $x$ and $y$ via
the marginal distributions $p(x) = \int p(x\mid y)p(y) \dee y$ 
and $p(y) = \int p(y\mid x)p(x) \dee x$, respectively.
Now, suppose we are informed that the true distribution of $x$ is $p_{\optsym}(x)$,
but we are not given any samples of $x$ or $y$.
We would then quantify our uncertainty in $x$ via $p_{\optsym}(x)$,
and a natural way to quantify our uncertainty in $y$ is 
via $q(y) \defined \int p(y\mid x) p_{\optsym}(x) \dee x$.
The idea is that $q(x,y) \defined p(y\mid x) p_{\optsym}(x)$ updates the model to have the correct distribution of $x$,
while remaining as close as possible to the original model $p(x,y)$.
This is referred to as Jeffrey conditionalization~\citep{Jeffrey:1968,Jeffrey:1990,Diaconis:1982}.

Suppose $\data = \dataarg{\numobs} \defined (\obs{1},\ldots,\obs{\numobs})$ is the data and $y = \param$ is a parameter, so that 
$p(x,y) = p(\dataarg{\numobs},\param)$ is the joint distribution of the data and the parameter.
If we are informed that the true distribution of the data is $p_{\optsym\numobs}(\dataarg{\numobs})$, then the 
Jeffrey conditionalization approach is to quantify our uncertainty in $\param$ by 
\[ \label{eq:jeffrey-conditionalization}
q(\param) = \int p(\param \given \dataarg{\numobs}) p_{\optsym\numobs}(\dataarg{\numobs}) \dee \dataarg{\numobs}.
\]

Now, suppose we are not informed of the true distribution exactly, but we are given data $\obsrv{1},\ldots,\obsrv{\numobs}\;\iid \dist p_{\optsym}$.
Since the empirical distribution $\empdist \defined \numobs^{-1}\sum_{n=1}^{\numobs}\delta_{\obsrv{n}}$ is a consistent estimator of $p_{\optsym}$,
and $p_{\optsym\numobs}(\dataarg{\numobs}) = \prod_{n=1}^{\numobs} p_{\optsym}(\obs{n})$,
it is natural to plug in $\prod_{n=1}^{\numobs}\empdist$ to approximate $p_{\optsym\numobs}$ in \cref{eq:jeffrey-conditionalization}.  Doing so, we arrive at the bagged posterior $\bbpostdensityfull{\param}{\data}$ from \cref{eq:bayesbag-definition}, in the case of $\bsnumobs = \numobs$:
\[
q(\param)\approx\int p(\param \given \dataarg{\numobs}) \prod_{n=1}^{\numobs} \empdist(\dee \obs{n}) =
\EE\big\{p(\param \given \datarvarg{\numobs}^{\bbsym}) \given \datarvarg{\numobs}\big\}
= \bbpostdensityfull{\param}{\data},
\]
where $\obsrv{1}^{\bbsym},\ldots,\obsrv{\numobs}^{\bbsym}\;\iid \dist \empdist$ given $\datarvarg{\numobs}$.
Thus, the bagged posterior represents uncertainty in $\theta$, integrating over data sets drawn from an approximation to the true distribution.
Hence, the bagged posterior naturally improves reproducibility across data sets.

\subsection{BayesBag combines Bayesian and frequentist uncertainty}
\label{sec:total-covariance}

In \cref{eq:jeffrey-conditionalization}, $p(\param \given \dataarg{\numobs})$ represents Bayesian model-based uncertainty 
and integrating with respect to $p_{\optsym\numobs}(\dataarg{\numobs})$ represents frequentist sampling uncertainty.
Remarkably, these two sources of uncertainty combine additively in the bagged posterior
whenever $\param \in \reals^{D}$.

To see this, let $\bsdatarv\given\data$ be a random bootstrap dataset given data $\data$, and let $\bbparamsample\given\bsdatarv\dist \postdensityfull{\param}{\bsdatarv}$ 
be distributed according to the standard posterior given data $\bsdatarv$.
Marginalizing out $\bsdatarv$, we have $\bbparamsample\given\data \dist \bbpostdensityfull{\param}{\data}$.
Let $\paramsample \given \data \dist \postdensityfull{\param}{\data}$ and define 
$\mu(\data) \defined \EE(\paramsample\given\data) = \int \param\,\postdensityfull{\param}{\data}\dee\param$
to be the standard posterior mean given $\data$.
By the law of total expectation, the mean of the bagged posterior is 
\[ \label{eq:bayesbag-posterior-mean}
\EE(\bbparamsample\given\data) = \EE\big\{\EE(\bbparamsample\given\bsdatarv)\given\data\big\} = \EE\{\mu(\bsdatarv)\mid\data\}
= \frac{1}{\numobs^{\bsnumobs}} \sum_{\bsdata} \mu(\bsdata).
\]
By the law of total covariance, the covariance matrix of the bagged posterior is
\[ \label{eq:bayesbag-posterior-covariance}
\begin{split}
\cov(\bbparamsample\given\data) &= \EE\big\{\cov(\bbparamsample\given\bsdatarv)\given\data\big\} + \cov\big\{\EE(\bbparamsample\given\bsdatarv)\given\data\big\} \\
 &= \EE\{\Sigma(\bsdatarv)\mid \data\} + \cov\{\mu(\bsdatarv)\mid \data\} ,
\end{split}
\]
where $\Sigma(\data) \defined \cov(\paramsample\given\data) = \int \{\param-\mu(\data)\}\{\param-\mu(\data)\}^{\top} \postdensityfull{\param}{\data}\dee\param$
is the standard posterior covariance.
In this decomposition of $\cov(\bbparamsample\given\data)$,
the first term approximates the mean of the posterior covariance matrix under the sampling distribution, and the
second term approximates the covariance of the posterior mean under the sampling distribution.
Thus, the first term reflects Bayesian model-based uncertainty averaged with respect to frequentist sampling variability,
and the second term reflects frequentist sampling-based uncertainty of a Bayesian model-based point estimate.

\section{Reproducibility using overlap probability} \label{sec:theory-overlap}

We now investigate if and when the standard and bagged posteriors satisfy the overlap criterion for reproducible uncertainty quantification. 
We focus on Gaussian location models, regular finite-dimensional models, and linear regression as representative cases, and 
consider settings where the dimension is fixed or growing with the sample size.
We show that under misspecification, the bagged posterior typically satisfies the overlap criterion %
whereas the standard posterior does not. 
But, for correctly specified models, both the standard and bagged posteriors usually satisfy the criterion.

First, however, as a check on the reasonableness of our criterion,
we establish that for any correctly specified Bayesian model, the overlap criterion is satisfied in expectation with respect to the prior.

\bnthm \label{thm:overlap-general}
Consider any model for data $X|\paramsample$ and any prior $\pi$ on $\paramsample$.
Suppose $x\mapsto A_x$ is a $100(1-\alpha)\%$ posterior credible set for $\paramsample$ under this model and prior, that is,
$\mathbb{P}(\paramsample\in A_x \mid x) \geq 1-\alpha$ for all $x$.
If $\paramsample\sim\pi$, and $X \mid \paramsample$, $\tilde{X}\mid \paramsample$ are independent data from the assumed model,
then $\EE\big\{\mathbb{P}(A_X\cap A_{\tilde{X}} \neq \varnothing \mid \paramsample)\big\} \geq (1-\alpha)^2$.
\enthm

\cref{thm:overlap-general} is a direct analogue of the classical result that posterior credible sets have correct frequentist coverage in expectation under the assumed prior.
All proofs are in \cref{app:proofs}.

\subsection{Gaussian location model}
\label{sec:multivariate-gaussian-location}

We first consider the simple Gaussian location model in which observations $\obs{n}$ are modeled as i.i.d.\ $\distNorm(\param,V)$ with fixed positive definite covariance matrix $V$, 
and assume a conjugate prior, $\param\dist\distNorm(0, V_{0})$.
Given data $\data=(\obs{1},\ldots,\obs{\numobs})$, the posterior is $\param\given\data \dist \distNorm(R_{\numobs}\datameanarg{\numobs}, V_{\numobs})$, where 
$\datameanarg{\numobs} \defined \numobs^{-1}\sum_{n=1}^{\numobs}\obs{n}$,
$R_{\numobs} \defined (V_{0}^{-1}V/\numobs + I)^{-1}$, and 
$V_{\numobs} \defined (V_{0}^{-1} + \numobs V^{-1})^{-1}$.
For intuition, one can think of $R_{\numobs} \approx I$ since $\|R_{\numobs} - I\| = O(\numobs^{-1})$.
The bagged posterior mean and covariance are 
\[
\EE(\bbparamsample\given\data)
&= \EE(R_{\bsnumobs}\bsdatarvmeanarg{\bsnumobs}\given\data) 
= R_{\bsnumobs}\datameanarg{\numobs} \label{eq:bayesbag-posterior-mean-normal-location} \\
\cov(\bbparamsample\given\data)
&= \EE(V_{\bsnumobs}\given\data) + \cov(R_{\bsnumobs}\bsdatarvmeanarg{\bsnumobs}\given\data) 
= V_{\bsnumobs} + \bsnumobs^{-1}R_{\bsnumobs}\empcov_{\numobs}R_{\bsnumobs}, \label{eq:bayesbag-posterior-covariance-normal-location}
\]
where $\empcov_{\numobs} \defined \numobs^{-1}\sum_{n=1}^{\numobs} (\obs{n}-\datameanarg{\numobs})(\obs{n}-\datameanarg{\numobs})^{\top}$
is the sample covariance.
In particular, when $\bsnumobs = \numobs$, these expressions simplify to $\EE(\bbparamsample\given\data)  = \EE(\paramsample\given\data)$ and 
$\cov(\bbparamsample\given\data)  = \cov(\paramsample\given\data) +  \numobs^{-1}R_{\numobs}\empcov_{\numobs}R_{\numobs}$.
Unlike the standard posterior, which simply assumes the data have covariance $V$, the bagged posterior accounts for the true covariance of the data through the inclusion 
of the term involving $\empcov_{\numobs}$. 

\subsubsection{Overlap probability for Gaussian location model with fixed dimension}

Consider the Gaussian location model above.
Fix $\alpha\in(0,1)$ and $u\in\reals^{D}\setminus\{0\}$, and let $A_{x_{1:\numobs}}$ be a $100(1-\alpha)\%$ central credible interval for $u^{\top} \param$
given data $x_{1:\numobs}$.
For BayesBag, let $A^*_{x_{1:\numobs}}$ denote the $100(1-\alpha)\%$ central interval for the normal distribution matching the mean and variance of the 
bagged posterior distribution of $u^{\top} \param$ given $x_{1:\numobs}$.
For readability, we abbreviate $p(\mathrm{overlap}) \defined \Pr(A_{X_{1:\numobs}} \cap A_{Y_{1:\numobs}} \neq \varnothing)$
and $p^*(\mathrm{overlap}) \defined \Pr(A^*_{X_{1:\numobs}} \cap A^*_{Y_{1:\numobs}} \neq \varnothing)$.

\bnthm \label{thm:overlap-gaussian}
Suppose the true data distribution $\obsdist$ has positive definite covariance $\Sigma_{\optsym}$. 
Let $X_1,X_2,\dots\;\iid\dist\obsdist$ and $Y_1,Y_2,\dots\;\iid\dist\obsdist$ independently.
Define $W\sim\distNorm(0,1)$ and $z_{\alpha/2}\in\reals$ such that $\Pr(|W| > z_{\alpha/2}) = \alpha$.
Then as $\numobs \to \infty$, for the standard posterior,
\[
p(\mathrm{overlap}) \longrightarrow \Pr\bigg(|W| \leq z_{\alpha/2} \sqrt{2} \Big(\frac{u^\top V u}{u^\top \Sigma_{\optsym} u}\Big)^{1/2}\bigg);
\]
and, assuming $\bsnumobs = \bsnumobs(\numobs)$ satisfies $\lim_{\numobs \to \infty}\bsnumobs/\numobs = c > 0$,
for the bagged posterior,
\[
p^*(\mathrm{overlap}) \longrightarrow \Pr\bigg(|W| \leq z_{\alpha/2} \sqrt{2} \Big(\frac{u^\top ((V + \Sigma_{\optsym})/c) u}{u^\top \Sigma_{\optsym} u}\Big)^{1/2}\bigg),
\]

\enthm

If the model is correct then $V = \Sigma_{\optsym}$, so the standard and bagged posteriors have the same asymptotic behavior when $\bsnumobs = 2 \numobs$, specifically,
the overlap probability converges to $\Pr(|W| \leq z_{\alpha/2} \sqrt{2})$.
However, in misspecified cases where $u^\top V u < u^\top \Sigma_{\optsym} u$, the overlap probability for the standard posterior can be arbitrarily small.
On the other hand, the bagged posterior satisfies
$\lim_{\numobs\to\infty} p^*(\mathrm{overlap}) \geq \Pr(|W| \leq z_{\alpha/2}) = 1-\alpha$
when $0 < c \leq 2$
since $u^\top V u \geq 0$.  Thus, BayesBag is guaranteed to satisfy the overlap criterion necessary for reproducible uncertainty quantification (\cref{eq:overlap-criterion}) when $0 < c \leq 2$, while standard Bayes is not.

\subsubsection{Overlap probability for Gaussian location model with growing dimension}

To study the case of growing dimension $D$ in the Gaussian location model, we establish finite sample expressions for the overlap probability in the special case of $V = I$ and a flat prior ($V_0^{-1} = 0$), assuming Gaussian data.

\bnthm \label{thm:overlap-gaussian-growing-dimension}
Consider the same setup as in \cref{thm:overlap-gaussian}.
Suppose $\obsdist = \distNorm(0,\Sigma_{\optsym})$, $V = I$, $V_0^{-1} = 0$, and $\|u\|=1$.
Then for the standard posterior,
\[\label{eq:overlap-gaussian-standard}
p(\mathrm{overlap}) = \Pr\bigg(|W| \leq \frac{z_{\alpha/2} \sqrt{2}}{(u^\top \Sigma_{\optsym} u)^{1/2}}\bigg)
\]
where $W\sim\distNorm(0,1)$, and for the BayesBag posterior, when $\numobs \geq 2$,
\[\label{eq:overlap-gaussian-bayesbag}
p^*(\mathrm{overlap}) \geq  \Pr\big(|T_{2 N - 2}| \leq z_{\alpha/2} \sqrt{(N-1)/M} \big)
\]
where $T_{2 N - 2}$ is $t$-distributed with $2 N - 2$ degrees of freedom.
\enthm
Note that the right-hand side of \cref{eq:overlap-gaussian-standard} does not depend on $N$, and the right-hand side of \cref{eq:overlap-gaussian-bayesbag} does not depend on $D$.
\cref{eq:overlap-gaussian-standard} can be arbitrarily small
as $D$ grows, since $u^\top \Sigma_{\optsym} u$ can be arbitrarily large.
For instance, this will often be the case when $\Sigma_{\optsym}$ has order $D^2$ nonnegligible entries.
Thus, as the dimension $D$ grows, the standard posterior can severely violate the overlap criterion.
Meanwhile, if $M/N \to 1$ as $N\to\infty$, then the lower bound in \cref{eq:overlap-gaussian-bayesbag} converges to $\Pr(|W|\leq z_{\alpha/2}) = 1-\alpha > (1-\alpha)^2$ since $T_{2 N - 2} \convD\distNorm(0,1)$ as $N\to\infty$.
Therefore, for all $N$ sufficiently large, for all $D$, BayesBag satisfies the overlap criterion.

\subsection{Regular finite-dimensional models}

Asymptotically, sufficiently regular finite-dimensional models behave like the Gaussian location model.
We have
$\numobs^{1/2}(\paramsample - \mle{\numobs}) \given \datarvarg{\numobs} \convD \distNorm(0, \Ehessloglik{\optparam}^{-1})$
by the Bernstein--Von Mises theorem, and 
$\numobs^{1/2}(\mle{\numobs} - \optparam) \convD \distNorm(0, \Ehessloglik{\optparam}^{-1}\Vargradloglik{\optparam}\Ehessloglik{\optparam}^{-1})$
by classical theory, 
where $\mle{\numobs}$ is the maximum likelihood estimator, $\optparam$ is the Kullback--Leibler optimal parameter, and $\Ehessloglik{\optparam}$, $\Vargradloglik{\optparam}$ are information matrices; see \cref{sec:theory-bvm} for details.
In \cref{sec:theory-bvm}, we prove that for the bagged posterior, 
$\numobs^{1/2}(\bbparamsample -  \mle{\numobs}) \given \datarvarg{\numobs} \convD \distNorm(0, \,\Ehessloglik{\optparam}^{-1}/\bsscale + \Ehessloglik{\optparam}^{-1}\Vargradloglik{\optparam}\Ehessloglik{\optparam}^{-1}/\bsscale)$
where $\bsscale \defined \lim_{\numobs \to \infty} \bsnumobs/\numobs$.

Fix $u\in\reals^D\setminus\{0\}$ and $\alpha\in(0,1)$.
Let $p_\infty(\mathrm{overlap})$ and $p^*_\infty(\mathrm{overlap})$ denote the asymptotic overlap probabilities 
of $100(1-\alpha)\%$ central credible intervals for $u^\top \param$
under these asymptotic normal distributions for the standard and bagged posteriors, respectively,
assuming $\Ehessloglik{\optparam}$ and $\Vargradloglik{\optparam}$ are positive definite.

\bnthm \label{thm:overlap-bvm}
Let $W\sim\distNorm(0,1)$.  For the standard posterior,
\[
p_\infty(\mathrm{overlap}) = \Pr\bigg(|W| \leq z_{\alpha/2} \sqrt{2} \bigg(\frac{u^\top \Ehessloglik{\optparam}^{-1} u}{u^\top \Ehessloglik{\optparam}^{-1}\Vargradloglik{\optparam}\Ehessloglik{\optparam}^{-1} u}\bigg)^{1/2}\bigg),
\]
and for the bagged posterior,
\[
p^*_\infty(\mathrm{overlap}) &= \Pr\bigg(|W| \leq z_{\alpha/2} \sqrt{2} \bigg(\frac{u^\top (\Ehessloglik{\optparam}^{-1}/\bsscale + \Ehessloglik{\optparam}^{-1}\Vargradloglik{\optparam}\Ehessloglik{\optparam}^{-1}/\bsscale) u}{u^\top \Ehessloglik{\optparam}^{-1}\Vargradloglik{\optparam}\Ehessloglik{\optparam}^{-1} u}\bigg)^{1/2}\bigg) \\
&\geq \Pr(|W| \leq z_{\alpha/2} \sqrt{2/c})
.
\]
\enthm

In general, the ratio $(u^\top \Ehessloglik{\optparam}^{-1} u) / (u^\top \Ehessloglik{\optparam}^{-1}\Vargradloglik{\optparam}\Ehessloglik{\optparam}^{-1} u)$ can be arbitrarily large or small. In particular, $p_\infty(\mathrm{overlap})$ can be arbitrarily small, implying that the asymptotic standard posterior can strongly violate the overlap criterion in \cref{eq:overlap-criterion}.
On the other hand, as long as $c\leq 2$, we have $p^*_\infty(\mathrm{overlap}) \geq 1-\alpha$, implying that the asymptotic bagged posterior satisfies the overlap criterion.

\subsection{Linear regression model}
\label{sec:linear-regression}

Consider data consisting of regressors $Z_{n} \in \reals^{D}$ and outcomes $Y_{n} \in \reals~(n=1,\dots,\numobs)$,
and let $Z\in\reals^{\numobs\times D}$ denote the complete design matrix, and $Y\in\reals^\numobs$ the vector of outcomes.
We analyze the linear regression model
\[
Y\mid Z,\beta \sim \distNorm(Z \beta, \sigma^2 I),
\]
where $\beta\in\reals^D$ is the vector of coefficients, $\sigma^2>0$ is the outcome variance, 
To simplify the analysis, assume $Z^\top Z$ is invertible, $\sigma^2$ is fixed but possibly unknown, and $\beta$ is given a flat prior.
For any $u\in\reals^D\setminus\{0\}$, the resulting posterior on $u^\top \beta$ is
\[
u^\top \beta \mid  Z,Y \sim \distNorm(v^\top Y, \, \sigma^2 \|v\|^2)
\]
where $v := Z(Z^\top Z)^{-1} u$.
Now, suppose the true distribution is $Y\mid Z\sim \distNorm(\mu_{\dagger},\Sigma_{\dagger})$ where $\mu_{\dagger}$ and $\Sigma_{\dagger}$ are functions of $Z$, say, $\mu_{\dagger} = m(Z)$ and $\Sigma_{\dagger} = K(Z)$.
Note that the model is correctly specified when $m(Z) = Z\beta$ and $K(Z) = \sigma^2 I$.

Consider two replicate experiments with data $Y\mid Z\sim \distNorm(m(Z),K(Z))$ and $\tilde{Y}\mid \tilde{Z}\sim \distNorm(m(\tilde{Z}),K(\tilde{Z}))$, respectively, where $Z^\top Z$ and $\tilde{Z}^\top \tilde{Z}$ are invertible
and the model variances are $\sigma^2$ and $\tilde{\sigma}^2$.
Letting $A = v^\top Y \pm z_{\alpha/2} \sigma \|v\|$ and $\tilde{A} = \tilde{v}^\top \tilde{Y} \pm z_{\alpha/2} \tilde{\sigma} \|\tilde{v}\|$ be the corresponding $100(1-\alpha)\%$ central credible sets for $u^\top \beta$, the overlap probability is 
$p(\mathrm{overlap} \mid Z,\tilde{Z}) = \Pr(A\cap\tilde{A}\neq\varnothing \mid Z,\tilde{Z})$.

\bnthm \label{thm:overlap-linreg}
Consider the linear regression model above and let $W\sim\distNorm(0,1)$.
\benum
\item If $m(Z) = Z \beta_{\dagger}$ and $K(Z) = \sigma_{\dagger}^2 I$, then
\[ \label{eq:thm-overlap-linreg-1}
p(\mathrm{overlap} \mid Z,\tilde{Z}) 
&= \Pr\bigg(|W| \leq \frac{z_{\alpha/2} (\sigma \|v\| + \tilde{\sigma} \|\tilde{v}\|)}{\sigma_{\dagger} (\|v\|^2 + \|\tilde{v}\|^2)^{1/2}}\bigg).
\]
\item If $Z = \tilde{Z}$, but we make no assumptions on the form of $m(Z)$ or $K(Z)$, then
\[ \label{eq:thm-overlap-linreg-2}
p(\mathrm{overlap}\mid Z,\tilde{Z}) = \Pr\bigg(|W| \leq \frac{z_{\alpha/2} \, (\sigma + \tilde{\sigma}) \|v\|}{\sqrt{2} (v^\top K(Z) v)^{1/2}}\bigg).
\]
\item If $K(Z) = \sigma_{\dagger}^2 I$, but we make no assumptions on the form of $m(Z)$, then
\[ \label{eq:thm-overlap-linreg-3}
p(\mathrm{overlap} \mid Z,\tilde{Z}) 
\leq \Pr\bigg(\Big\vert W + \frac{v^\top m(Z) - \tilde{v}^\top m(\tilde{Z})}{\sigma_{\dagger}\,(\|v\|^2 + \|\tilde{v}\|^2)^{1/2}} \Big\vert \leq \frac{z_{\alpha/2} \sqrt{\sigma^2 + \tilde{\sigma}^2}}{\sigma_{\dagger}}\bigg).
\]
\eenum
\enthm

\cref{eq:thm-overlap-linreg-1} shows that if the linear regression model is correctly specified, then the standard posterior satisfies the overlap criterion (\cref{eq:overlap-criterion}), since $\sigma = \tilde{\sigma} = \sigma_{\dagger}$ and therefore
$p(\mathrm{overlap} \mid Z,\tilde{Z}) \geq \Pr(|W| \leq z_{\alpha/2}) = 1-\alpha > (1-\alpha)^2$,
by the fact that $\|v\| + \|\tilde{v}\| \geq (\|v\|^2 + \|\tilde{v}\|^2)^{1/2}$.
If the model is correct but the variance is unknown, and consistent estimators of $\sigma_{\dagger}^2$ are plugged in for $\sigma^2$ and $\tilde{\sigma}^2$, then the overlap criterion is satisfied for all $\numobs$ sufficiently large.

However, when either the covariance $K(Z)$ or the mean function $m(Z)$ is misspecified, the standard posterior can violate the overlap criterion.
Consider the case where $Z = \tilde{Z}$, that is, the design matrix is the same across replicates; we refer to this as a fixed design setting.
\cref{eq:thm-overlap-linreg-2} shows that $p(\mathrm{overlap} \mid Z,\tilde{Z})$ does not depend on $m(Z)$,
so misspecification of the mean function has no effect on the overlap probability in this case.
Nonetheless, \cref{eq:thm-overlap-linreg-2} shows that $p(\mathrm{overlap}\mid Z,\tilde{Z})$ can be arbitrarily small 
when the covariance is misspecified, because the ratio $(\sigma + \tilde{\sigma}) \|v\| / (v^\top K(Z) v)^{1/2}$ can be arbitrarily small.
Clearly, this ratio will be small if $\sigma^2$ and $\tilde{\sigma}^2$ are blindly set too low, but it can also be small if these variances are estimated from the data.
For instance, if the true distribution exhibits heteroskedasticity (that is, $K(Z)$ has a nonconstant diagonal), then $p(\mathrm{overlap} \mid Z,\tilde{Z})$ can violate the overlap criterion even when $\sigma^2$ and $\tilde{\sigma}^2$ are estimated; see \cref{sec:simulations}.

Finally, consider the case where $Z$ and $\tilde{Z}$ are not necessarily equal and we make no assumptions on $m(Z)$.
To avoid trivial failure modes in which the choice of $Z$ and $\tilde{Z}$ leads to a nonnegligible differential bias $v^\top m(Z) - \tilde{v}^\top m(\tilde{Z})$ as $\numobs$ grows, assume a random design setting where the rows of $Z$ and $\tilde{Z}$ are independent and identically distributed.
Then even if $K(Z) = \sigma_{\dagger}^2 I$, so that there is no heteroskedasticity and no correlation among outcomes, the overlap criterion can still be violated.
As before, $p(\mathrm{overlap} \mid Z,\tilde{Z})$ can be arbitrarily small if 
$\sigma^2$ and $\tilde{\sigma}^2$ are blindly set too low, but it can also be small if these variances are estimated.
By \cref{eq:thm-overlap-linreg-3}, $p(\mathrm{overlap}\mid Z,\tilde{Z})$ will be small if the magnitude of
\[
\label{eq:overlap-linreg-offset}
\frac{v^\top m(Z) - \tilde{v}^\top m(\tilde{Z})}{(\|v\|^2 + \|\tilde{v}\|^2)^{1/2}}
= \frac{u^\top Z^{+} f(Z)\beta_{\dagger} - u^\top \tilde{Z}^{+} f(\tilde{Z})\beta_{\dagger}}{\big(u^\top (Z^\top Z)^{-1} u + u^\top (\tilde{Z}^\top \tilde{Z})^{-1} u\big)^{1/2}}
\]
is large relative to $\sqrt{\sigma^2 + \tilde{\sigma}^2}$, where $Z^{+} = (Z^\top Z)^{-1} Z^\top$ is the pseudoinverse.
A trivial way this can occur is if the entries of $\beta_{\dagger}$ are large.
More interestingly, however, \cref{eq:overlap-linreg-offset} can be large if the dimension $D$ grows with $\numobs$,
even if each entry of $\beta_{\dagger}$ has fixed magnitude.
Specifically, in \cref{sec:simulations} we present experiments demonstrating this when $\beta_{\dagger}$ consists of the first $D$ entries of a fixed sequence
$\beta_{\dagger,1},\beta_{\dagger,2},\ldots$ such that $\sum_{d=1}^D \beta_{\dagger,d}^2 \to\infty$ as $D\to\infty$.

\section{Asymptotic normality of the bagged posterior} \label{sec:theory-bvm}

In this section, we establish a Bernstein--Von Mises theorem for the bagged posterior 
under sufficiently regular finite-dimensional models (\cref{thm:bb-bvm}).
In particular, we show that while the standard posterior may be arbitrarily under- or over-confident when the model is misspecified,
the bagged posterior avoids overconfident uncertainty quantification by accounting for sampling variability.
More formally, consider a model $\{\likdist{\param} : \param \in \paramspace\}$ for independent and identically distributed (\iid)\ data 
$\obs{1},\dots,\obs{\numobs}$, where $\obs{n}\in\obsspace$ and $\paramspace\subset\reals^{D}$ is open.
Suppose $\likfun{\param}$ is the density of $\likdist{\param}$ with respect to some reference measure.
The standard Bayesian posterior distribution given $\dataarg{\numobs}$ is 
\[ 
\postdistfull{\dee\param}{\dataarg{\numobs}} 
\defined \frac{\prod_{n=1}^{\numobs}\lik{\obs{n}}{\param}}{\marginallikfull{\dataarg{\numobs}}}\priordist(\dee\param), \label{eq:iid-data-posterior}
\]
where $\priordist(\dee\param)$ is the prior distribution and
$\marginallikfull{\dataarg{\numobs}} \defined \int\{\prod_{n=1}^{\numobs}\lik{\obs{n}}{\param}\}\priordist(\dee\param)$ is the marginal likelihood. 
We often use the shorthand notation $\postdist{\numobs} \defined \postdistfull{\cdot}{\dataarg{\numobs}}$.

Assume the observed data $\obsrv{1},\dots,\obsrv{\numobs}$ is generated \iid\ from some unknown distribution $\obsdist$.
Suppose there is a unique parameter $\optparam$ that 
minimizes the Kullback--Leibler divergence from $\obsdist$ to the model, or equivalently, $\optparam = \argmax_{\param \in \paramspace}\EE\{\log \lik{\obsrv{1}}{\param}\}$.
Under regularity conditions, the maximum likelihood estimator $\mle{\numobs} \defined \argmax_{\param} \prod_{n=1}^{\numobs}\lik{\obsrv{n}}{\param}$ is asymptotically normal in the sense that 
\[ \label{eq:mle-asympotics}
\numobs^{1/2}(\mle{\numobs} - \optparam) \convD \distNorm(0,\Ehessloglik{\optparam}^{-1}\Vargradloglik{\optparam}\Ehessloglik{\optparam}^{-1}),
\]
where $\Ehessloglik{\param} \defined -\EE\{\grad_{\param}^2\log\lik{\obsrv{1}}{\param}\}$,
 $\Vargradloglik{\param} \defined \cov\{\grad_{\param}\log\lik{\obsrv{1}}{\param}\}$, and 
 $\Ehessloglik{\optparam}^{-1}\Vargradloglik{\optparam}\Ehessloglik{\optparam}^{-1}$ is known as the sandwich covariance~\citep{White:1982}.
Under mild conditions, the Bernstein--Von Mises theorem (\citealp[Ch.~10]{vanderVaart:1998} and \citealp{Kleijn:2012}) guarantees that for $\paramsample \dist \postdist{\numobs}$,
\[ \label{eq:bayes-asymptotics}
\numobs^{1/2}(\paramsample - \mle{\numobs}) \given \datarvarg{\numobs} \convD \distNorm(0, \Ehessloglik{\optparam}^{-1}).
\]
Hence, the standard posterior is correctly calibrated, asymptotically, if the covariance matrices of the Gaussian distributions in \cref{eq:mle-asympotics,eq:bayes-asymptotics}
coincide -- that is, if $\Ehessloglik{\optparam}^{-1}\Vargradloglik{\optparam}\Ehessloglik{\optparam}^{-1} =  \Ehessloglik{\optparam}^{-1}$,
which is implied by $\Vargradloglik{\optparam} = \Ehessloglik{\optparam}$.
In particular, if $\Vargradloglik{\optparam} = \Ehessloglik{\optparam}$, then Bayesian credible sets are (asymptotically) 
valid confidence sets in the frequentist sense: sets of posterior probability $1 - \alpha$ contain the true parameter with $\obsdist^{\infty}$-probability $1 - \alpha$, under mild conditions.

If the model is well-specified, that is, if $\obsdist = \likdist{\trueparam}$ for some parameter $\trueparam \in \paramspace$
(and thus $\optparam = \trueparam$ by the uniqueness assumption), then $\Vargradloglik{\optparam} = \Ehessloglik{\optparam}$ under very mild conditions.
On the other hand, if the model is misspecified -- that is, if $\obsdist \ne \likdist{\param}$ for all $\param \in \paramspace$ --
then although \cref{eq:bayes-asymptotics} still holds, typically $\Vargradloglik{\optparam} \ne \Ehessloglik{\optparam}$. 
If $\Vargradloglik{\optparam} \ne \Ehessloglik{\optparam}$, then the standard posterior is not correctly calibrated, and in fact, asymptotic Bayesian credible sets may be arbitrarily 
over- or under-confident. 

Our Bernstein--von Mises theorem shows that the bagged posterior does not suffer from the overconfidence of the standard posterior.
Let $\bsdatarvarg{\bsnumobs}$ denote a bootstrapped copy of $\datarvarg{\numobs}$ with $\bsnumobs$ observations;
that is, each observation $\obsrv{n}$ is replicated $\bscount{n}$ times in $\bsdatarvarg{\bsnumobs}$,
where $\bscounts \dist \distMulti(\bsnumobs, 1/\numobs)$ is a multinomial-distributed count vector of length $\numobs$.
We formally define the bagged posterior $\bbpostdistfull{\cdot}{\datarvarg{\numobs}}$ as
\[
\bbpostdistfull{A}{\datarvarg{\numobs}} \defined \EE\{\postdistfull{A}{\bsdatarvarg{\bsnumobs}} \given \datarvarg{\numobs} \}
\]
for all measurable $A \subseteq \paramspace$; this is equivalent to the informal definition in \cref{eq:bayesbag-definition}.
To avoid notational clutter, we suppress the dependence of $\bbpostdistfull{\cdot}{\datarvarg{\numobs}}$ on $\bsnumobs$.
We use the shorthand notation $\bbpostdist{\numobs} \defined \bbpostdistfull{\cdot}{\datarvarg{\numobs}}$ and
we let $\bbparamsample\given\datarvarg{\numobs} \dist \bbpostdist{\numobs}$ denote a random variable distributed according to the bagged posterior.
We assume $\postdist{\numobs}$ and $\bbpostdist{\numobs}$ have densities $\postdensity{\numobs}$ and $\bbpostdensity{\numobs}$, respectively, 
with respect to Lebesgue measure.  Note that $\bbpostdensity{\numobs}$ exists if $\postdensity{\numobs}$ exists. 

%

%
%

%
%

%
%
%
%
%
%
%
%
%
%

For a measure $\nu$ and function $f$, we use the shorthand $\nu(f) \defined \int f \dee \nu$.  
Let $\alldatarv$ denote the infinite sequence $(\obsrv{1},\obsrv{2},\dots)$,
and abbreviate $\loglikfun{\param} \defined \log \likfun{\param}$.

\bnthm \label{thm:bb-bvm}
Suppose $\obsrv{1},\obsrv{2},\dots\;\iid\dist\obsdist$ and assume that:
\begin{enumerate}[label=(\roman*)]
\item $(\param, \obs{}) \mapsto \loglik{\obs{}}{\param}$ is measurable
and $\param \mapsto \loglik{\obsrv{1}}{\param}$ is differentiable at $\optparam$ with probability 1;
\item there is an open neighborhood $U$ of $\optparam$
and a function $m_{\optparam} : \obsspace \to \reals$ such that for some $\delta>0$ $\obsdist(m_{\optparam}^{2+\delta}) < \infty$ and for all $\param,\param' \in U$,
$|\loglikfun{\param} - \loglikfun{\param'}| \le m_{\optparam}\twonorm{\param - \param'}$ a.s.$[\obsdist]$;
\item $-\obsdist(\loglikfun{\param} - \loglikfun{\optparam}) = \frac{1}{2}(\param - \optparam)^{\top}\Ehessloglik{\optparam}(\param - \optparam) + \littleo(\twonorm{\param - \optparam}^{2})$ as $\param \to \optparam$;
\item $\Ehessloglik{\optparam}$ is an invertible matrix;
\item conditionally on $\alldatarv$, for almost every $\alldatarv$, for every sequence of constants $\concconst\to\infty$,
\[
\EE\big[\postdistfull{\{ \param \in \paramspace \st \twonorm{\param - \optparam} > \concconst/\bsnumobs^{1/2}\}}{\bsdatarvarg{\bsnumobs}} \;\big\vert\; \datarvarg{\numobs}\big] \to 0;
\]
and
\item  $\bsscale \defined \lim_{\numobs \to \infty} \bsnumobs/\numobs \in (0,\infty)$.
\end{enumerate}
Then, letting $\bbparamsample \dist \bbpostdist{\numobs}$, we have that conditionally on $\alldatarv$, for almost every $\alldatarv$,
\[
\numobs^{1/2}(\bbparamsample - \optparam) -\Delta_{\numobs} \given \datarvarg{\numobs} \convD \distNorm(0, \Ehessloglik{\optparam}^{-1}/\bsscale + \Ehessloglik{\optparam}^{-1}\Vargradloglik{\optparam}\Ehessloglik{\optparam}^{-1}/\bsscale),
\]
where $\Delta_{\numobs} \defined \numobs^{1/2}\Ehessloglik{\optparam}^{-1}(\empdist - \obsdist)\grad_{\param}\loglikfun{\optparam}$
and $\empdist \defined \numobs^{-1}\sum_{n=1}^{\numobs}\delta_{\obsrv{n}}$.

The result also holds in the regression setting with random regressors where the data take the form $\obsrv{n} = (Y_{n}, Z_{n})$ 
and the models $\lik{y \given z}{\param}$ are conditional,
so $\loglikfun{\param}(x) \defined \log \lik{y \given z}{\param}$. 
\enthm

The proof of \cref{thm:bb-bvm} is in \cref{app:proofs}. 
\cref{prop:bb-bbvm-gaussian-location} is a simpler version of the same result for the univariate Gaussian location model,
for which the statement and our proof technique are more transparent. 
Our technical assumptions %
are essentially the same as those used by \citet{Kleijn:2012} to prove the Bernstein--Von Mises theorem under misspecification for the standard posterior. 
Of particular note, \citet{Kleijn:2012} require that (and give conditions under which) for every sequence of constants $\concconst\to\infty$,
\[
\EE\big[\postdistfull{\{ \param \in \paramspace \st \twonorm{\param - \optparam} > \concconst/\numobs^{1/2}\}}{\datarvarg{\numobs}}\big] \to 0. \label{eq:posterior-mass-concentration}
\]
We conjecture that under reasonable regularity assumptions, this expected posterior concentration condition
implies our condition (v). 

To interpret this result, it is helpful to compare it to the behavior of the standard posterior. 
Under the conditions of \cref{thm:bb-bvm}, if $\paramsample \dist \postdist{\numobs}$,
then $\numobs^{1/2}(\paramsample - \optparam) - \Delta_{\numobs} \given \datarvarg{\numobs} \convD \distNorm(0, \Ehessloglik{\optparam}^{-1})$ in probability 
by \citet[Theorem 2.1 and Lemma 2.1]{Kleijn:2012}.
Thus, the bagged posterior and the standard posterior for $\numobs^{1/2}(\param - \optparam)$  have the same asymptotic mean, $\Delta_{\numobs}$, 
but the bagged posterior has asymptotic covariance $\Ehessloglik{\optparam}^{-1}/\bsscale + \Ehessloglik{\optparam}^{-1}\Vargradloglik{\optparam}\Ehessloglik{\optparam}^{-1}/\bsscale$
instead of $\Ehessloglik{\optparam}^{-1}$.
Hence, asymptotically, the bagged posterior is never overconfident if $c = 1$ (for instance, if $\bsnumobs = \numobs$)
and by \cref{thm:overlap-bvm}, we expect $100(1 - \alpha)\%$ credible sets of the bagged posteriors to have overlap probability of at least $1 - \alpha$ when $0 < \bsscale \leq 2$.

\section{Simulations} \label{sec:simulations}

In this section, we validate our theoretical results through a simulation study with a linear regression model, 
which is ideal for investigating the properties of BayesBag since all computations of posterior quantities can be done in closed form.
The setup is similar to the linear regression model from \cref{sec:linear-regression} except we place proper priors on the regression coefficients and the outcome variance $\sigma^2$. 
The data consist of regressors $Z_{n} \in \reals^{D}$ and outcomes $Y_{n} \in \reals~(n=1,\dots,\numobs)$,
and the parameter is $\param = (\param_{0}, \dots, \param_{D}) = (\log \sigma^{2}, \beta_{1},\dots,\beta_{D}) \in \reals^{D+1}$.
Using conjugate priors, the assumed model is 
\[
\sigma^{2} &\dist \distInvGam(a_{0}, b_{0}) \\
\beta_{d} &\given \sigma^{2} \distiid\distNorm(0, \sigma^{2}/\lambda)  & d &=1,\dots,D, \\
Y_{n} &\given Z_{n}, \beta, \sigma^{2} \distind \distNorm(Z_{n}^{\top}\beta, \sigma^{2}) & n &=1,\dots,\numobs,
\]
where $a_{0} = 2, b_{0} = 1$, and $\lambda = 1$ are fixed hyperparameters. 

\paragraph{Data generating distribution.} 
We simulated data for a random design scenario by generating 
$Z_{n} \distiid G$, $\eps_{n} \distiid \distNorm(0,1)$, and 
\[
Y_{n} = f(Z_{n})^{\top}\beta_{\dagger} + \eps_{n}  \label{eq:simulated-linreg-data}
\]
for $n =1,\dots,N$,
where $\beta_{\dagger d} = 4/\sqrt{d}$ for $d = 1,\ldots,D$
and we used two settings for each of $f$ and $G$. %

\bitems
\item \textbf{Regression function $f$.} By default, we used a \textsf{linear} function $f(z) = z$ to simulate data for the well-specified setting.
Alternatively, we used the \textsf{nonlinear} function $f(z) = (z_{1}^{3},\ldots,z_{D}^3)^{\top}$ for a misspecified setting.

\item \textbf{Regressor distribution $G$.} By default, we used $G = \distNorm(0, I)$ to simulate data; we refer to this as the \textsf{uncorrelated} setting.
Alternatively, we used a \textsf{correlated-$\kappa$} setting, where, for $h = 10$, $Z \dist G$ was defined by generating
$\xi\dist\chi^{2}(h)$ and then 
$Z \given \xi \dist\distNorm(0,\Sigma)$ where $\Sigma_{dd'} = \exp\{-(d-d')^{2}/\kappa^{2}\} / (\xi_{d}\xi_{d'})$ and
$\xi_{d} = \sqrt{\xi / (h-2)}^{\indicatorfn(d \text{ is odd})}$.
The motivation for the \textsf{correlated-$\kappa$} sampling procedure is to generate correlated regressors that have different
tail behaviors while still having the same first two moments, since regressors are typically standardized to have mean 0 and variance 1. 
Note that, marginally, $Z_{1}, Z_{3}, \dots$ are each rescaled $t$-distributed random variables with $h$ degrees of freedom such that $\var(Z_{1}) = 1$,
and $Z_{2}, Z_{4},\dots$ are standard normal. 
\eitems

\begin{figure}[tbp]
\begin{center}
\begin{subfigure}[b]{.44\textwidth}
\centering
\includegraphics[height=1.65in,trim={0in 0in 1.45in 0in},clip]{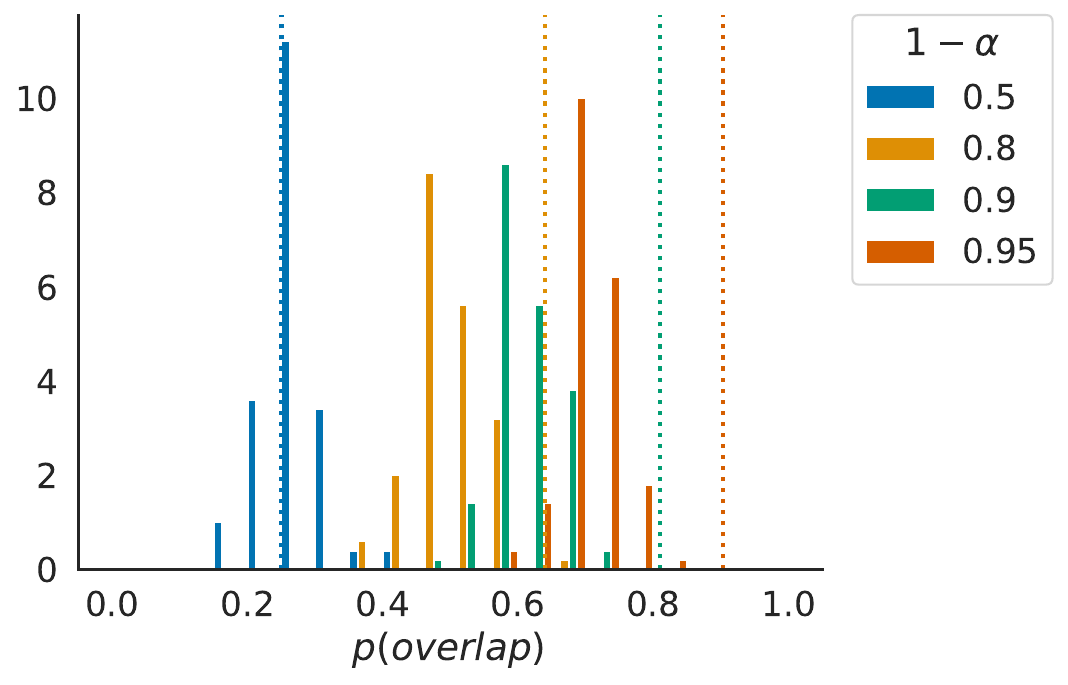}
\caption{$N = D = 250$, Bayes}
\end{subfigure}
\begin{subfigure}[b]{.55\textwidth}
\centering
\includegraphics[height=1.65in]{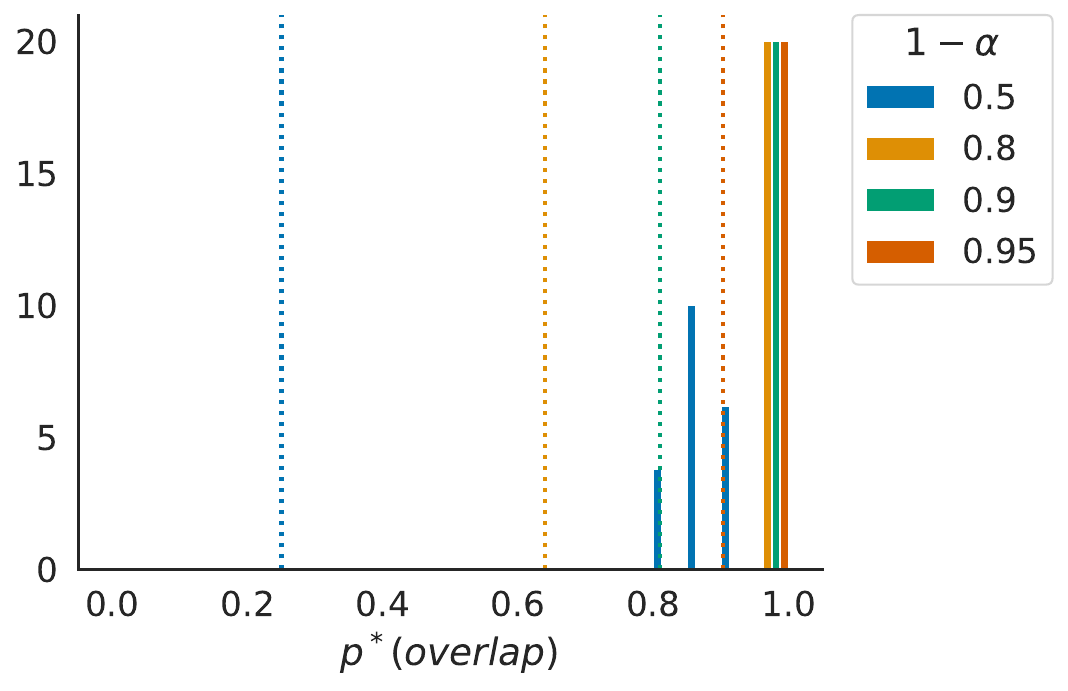}
\caption{$N = D = 250$, BayesBag}
\end{subfigure} \\
\begin{subfigure}[b]{.44\textwidth}
\centering
\includegraphics[height=1.65in,trim={0in 0in 1.45in 0in},clip]{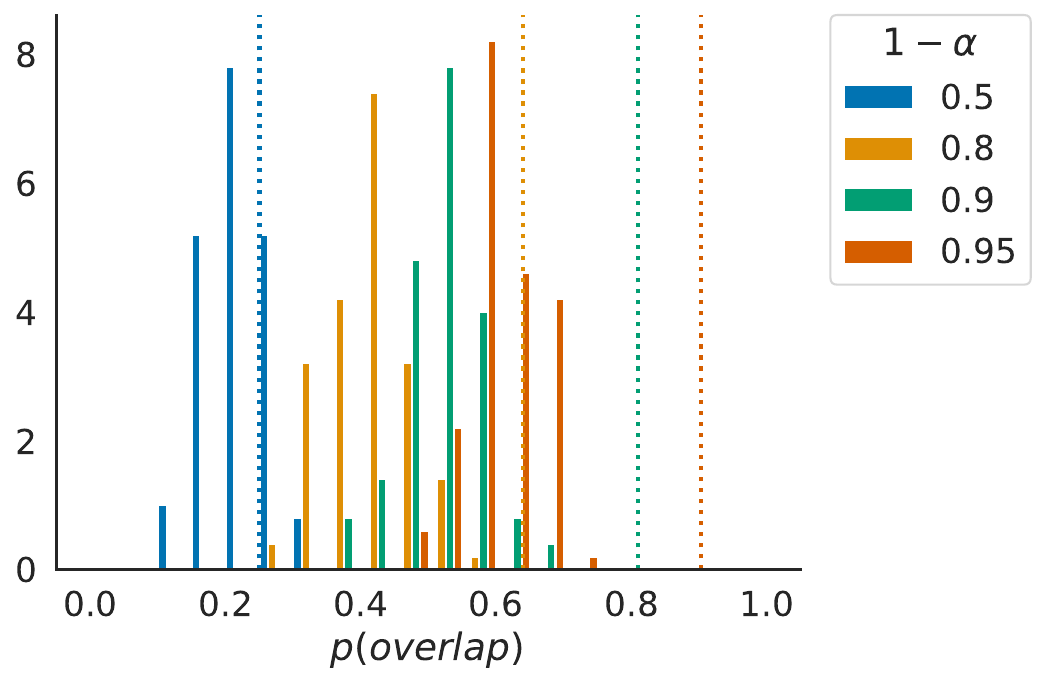}
\caption{$N = D = 500$, Bayes}
\end{subfigure}
\begin{subfigure}[b]{.55\textwidth}
\centering
\includegraphics[height=1.65in]{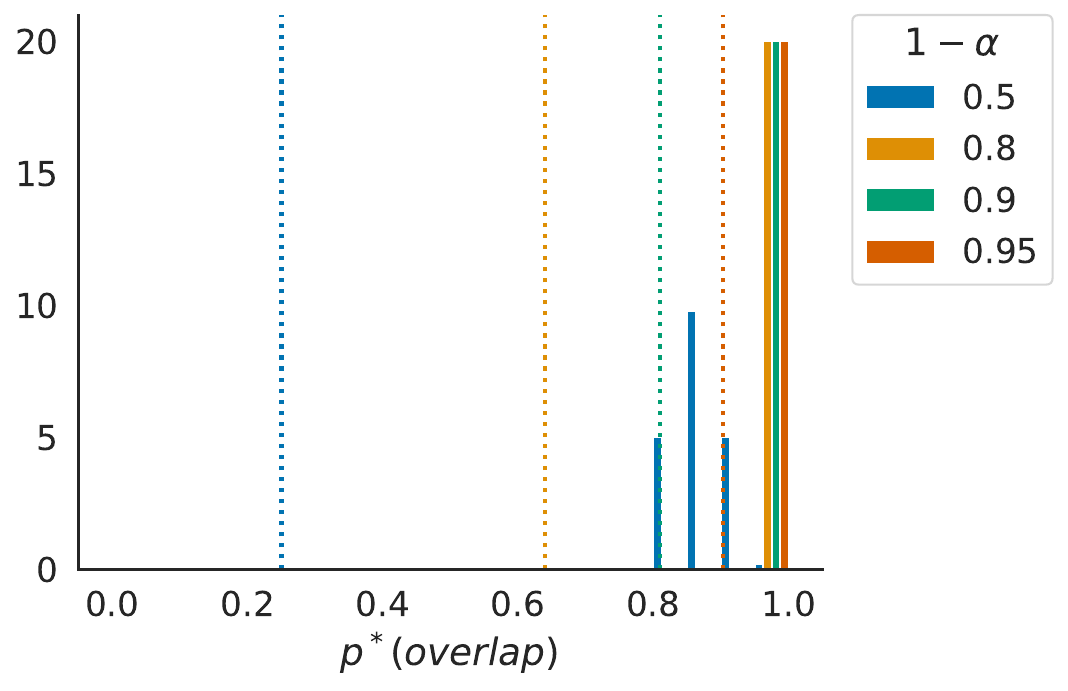}
\caption{$N = D = 500$, BayesBag}
\end{subfigure} 
\caption{Histograms of the probability of overlap of $1-\alpha$ credible sets for $(Z_{i}^{\mathrm{test}})^{\top} \beta$ ($i=1,\dots,100$) for the linear regression model with \textsf{nonlinear-uncorrelated} data. 
Vertical dotted lines indicate the overlap lower bounds $(1 - \alpha)^{2}$. 
}
\label{fig:linear-regression-uncorrelated-N-eq-D-histograms}
\end{center}
\end{figure}

\begin{figure}[tbp]
\begin{center}
\begin{subfigure}[b]{\textwidth}
\centering
\includegraphics[height=1.6in,trim={0in 0in 1.35in 0in},clip]{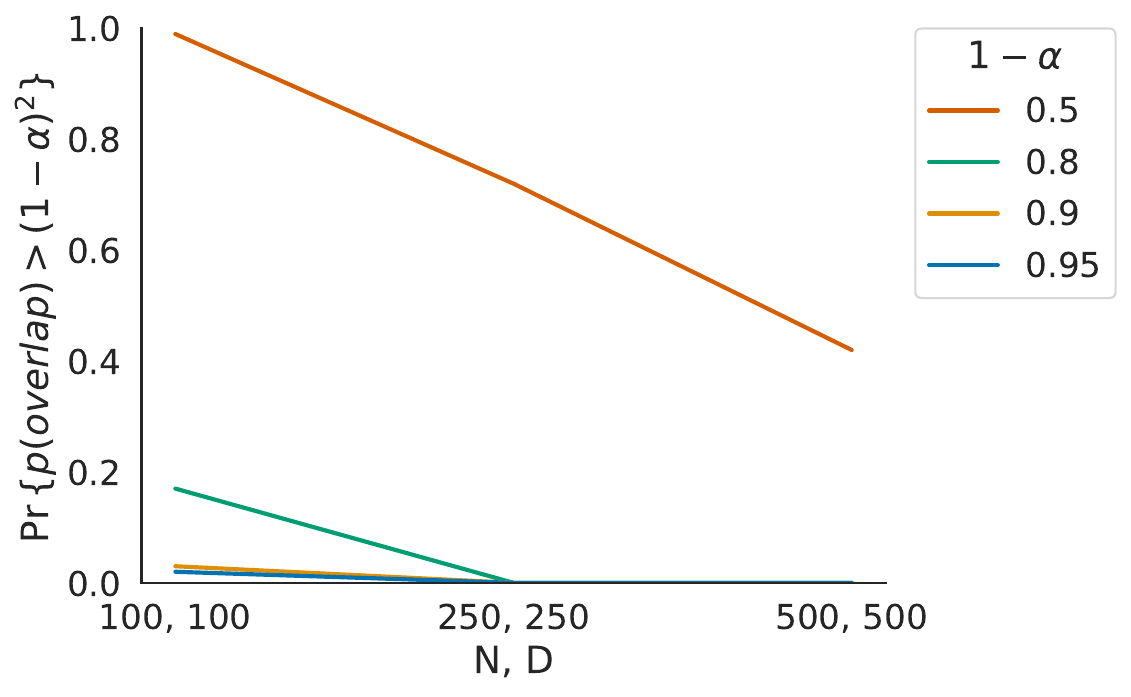}
\includegraphics[height=1.6in]{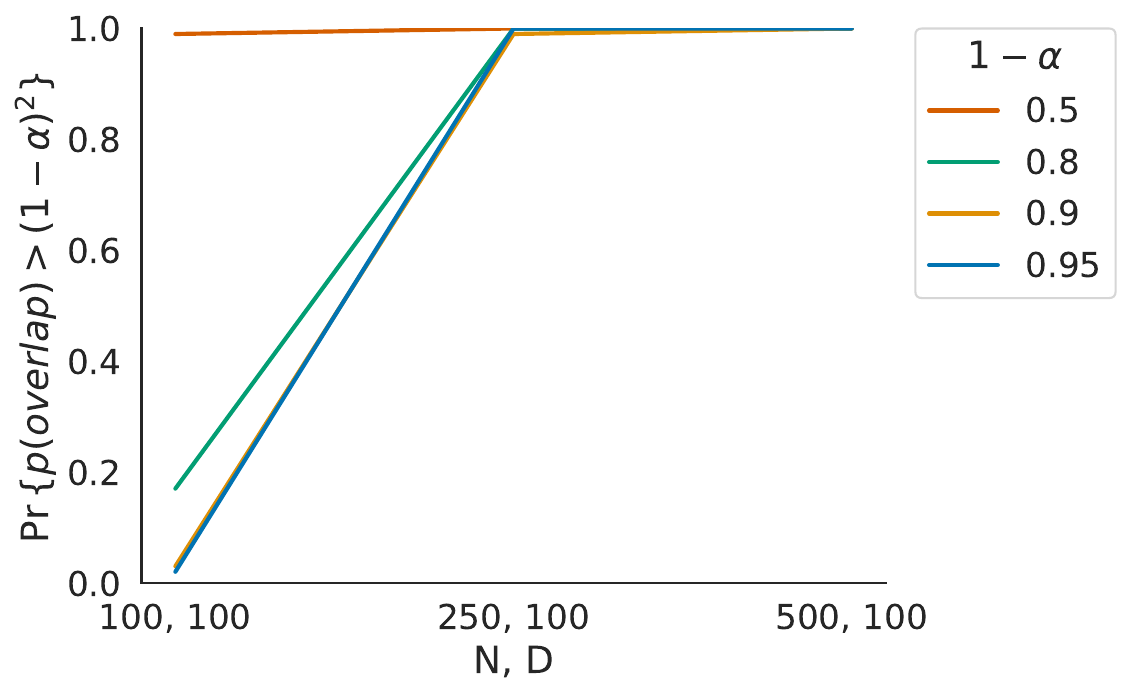}
\caption{\textsf{nonlinear-uncorrelated}}
\end{subfigure} \\
\begin{subfigure}[b]{\textwidth}
\centering
\includegraphics[height=1.6in,trim={0in 0in 1.35in 0in},clip]{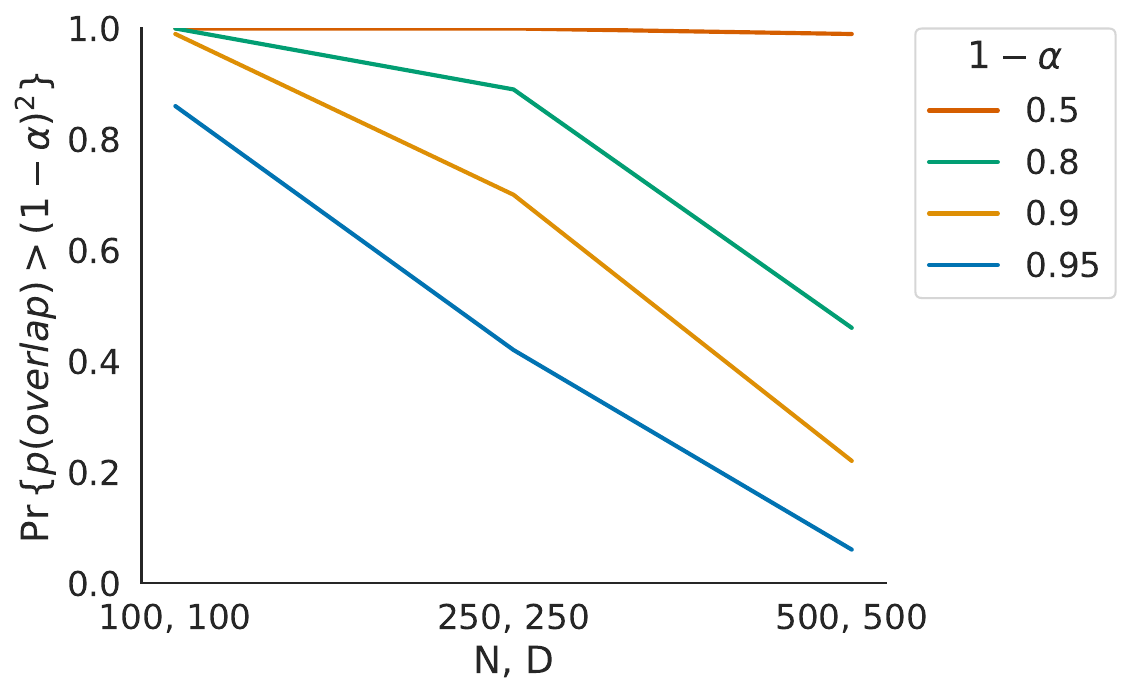}
\includegraphics[height=1.6in]{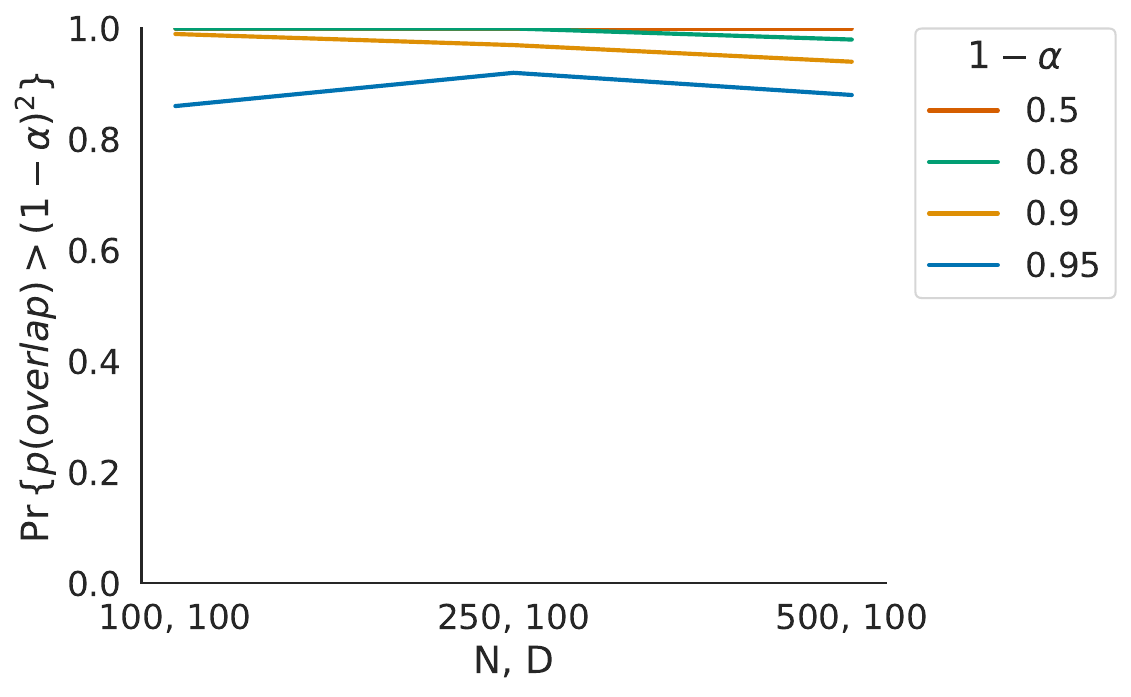}
\caption{\textsf{nonlinear-correlated-2}}
\end{subfigure}
\caption{Proportion of test points $Z_{i}^{\mathrm{test}}$ for which the Bayes overlap probability satisfies the lower bound. 
For BayesBag, the proportion is 1 in all cases.
}
\label{fig:linear-regression-N-eq-D}
\end{center}
\end{figure}

\paragraph{Overlap probabilities.} 
The primary objective in these experiments is to validate that the BayesBag posterior does not violate the probability of overlap lower bounds while the Bayesian posterior sometimes does.
Thus, for each data-generating distribution of interest, we estimate overlap probabilities by generating $R$ pairs of datasets $\{(Z^{(r,1)}_{1:\numobs}, Y^{(r,1)}_{1:\numobs}, Z^{(r,2)}_{1:\numobs}, Y^{(r,2)}_{1:\numobs})\}_{r=1}^{R}$ 
plus an additional 100 test points $Z_{1}^{\mathrm{test}},\dots, Z_{100}^{\mathrm{test}} \dist G$. 
If the $1 - \alpha$ posterior credible intervals for $(Z_{i}^{\mathrm{test}})^{\top}\beta$ conditioned on $(Z^{(r,1)}_{1:\numobs}, Y^{(r,1)}_{1:\numobs})$ and $(Z^{(r,2)}_{1:\numobs}, Y^{(r,2)}_{1:\numobs})$ overlap, set
the overlap indicator variable $O^{(r)}_{\alpha}(Z_{i}^{\mathrm{test}}) = 1$. 
Otherwise set $O^{(r)}_{\alpha}(Z_{i}^{\mathrm{test}}) = 0$. 
For each $i \in \{1,\dots,100\}$, we estimate the probability of overlap for $Z_{i}^{\mathrm{test}}$ as 
\[
\Pr\Big(\text{overlap of $(Z_{i}^{\mathrm{test}})^{\top}\beta$ at level $1-\alpha$}\Big) \approx R^{-1}\sum_{r=1}^{R} O^{(r)}_{\alpha}(Z_{i}^{\mathrm{test}}). 
\]
For all experiments we use $R = 100$.
\Cref{fig:linear-regression-uncorrelated-N-eq-D-histograms} shows that for \textsf{nonlinear-uncorrelated} data, BayesBag never violates the overlap lower bounds
while Bayes always or often violates the lower bounds, depending on the value of $1-\alpha$ (larger $1-\alpha$ leads to more violations). 
\Cref{fig:linear-regression-correlated2-N-eq-D-histograms,fig:linear-regression-correlated4-N-eq-D-histograms,fig:linear-regression-correlated-N-eq-D-histograms} in the Supplementary Material show similar results for \textsf{nonlinear-correlated-$\kappa$} data, although the problem with Bayes is less severe as the correlation increases. 
Moreover, as shown in \cref{fig:linear-regression-N-eq-D}, the problem becomes more severe as $N$ and $D$ jointly increase, but 
improves or stays the same if $D$ is fixed and $N$ increases.
These results emphasize how the misspecified high-dimensional regime is particularly problematic for the reproducibility of the standard posterior. 
We find similar results in the case of a fixed design matrix with heteroskedastic noise (see \cref{sec:fixed-simulations} in the Supplementary Material). 

\begin{figure}[tbp]
\begin{center}
\centering
\includegraphics[width=\textwidth]{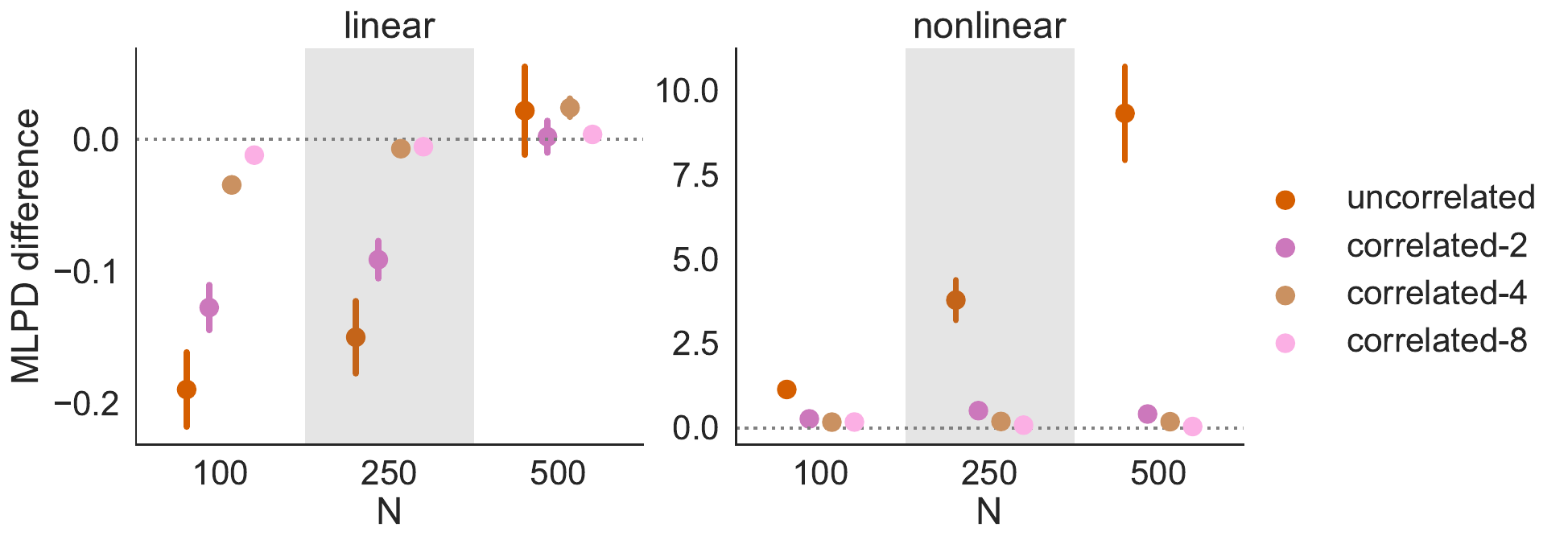}
\caption{99\% confidence intervals for difference in the mean log predictive densities of the standard and BayesBag posteriors (paired $t$ intervals), 
with values greater than zero indicating superior performance by BayesBag.
{\uline{Note the different scales for \textsf{linear} versus \textsf{nonlinear}.}}}
\label{fig:linear-regression-prediction}
\end{center}
\end{figure}

\paragraph{Predictive performance.} 
To complement our overlap probability analysis, we also computed the mean log predictive densities at the same test points. 
\Cref{fig:linear-regression-prediction} shows that while in well-specified \textsf{linear} settings the standard posterior can slightly outperform BayesBag (by roughly 0.2 nats or less), 
in the misspecified \textsf{nonlinear} settings BayesBag can be far superior (by 0.2 to nearly 10 nats).

\section{Application} \label{sec:application}

We next consider an application to community-level crime data from the United States using a Poisson regression model with log link function and
the spike-and-slab prior proposed by \citet{Piironen:2017}.
The data consist of $N = 1994$ observations containing 100 community-level covariates such as demographic summaries and local law enforcement statistics such as the number of police officers per capita. 
The goal is to predict the number of violent crimes per 100,000 persons in the population. 
We chose $\bsnumobs = \numobs$ and used $B=50$ bootstrap samples to approximate the bagged posterior. Nearly identical results were obtained with $B=25$, indicating that $B=50$ was sufficiently large.

To compute overlap probabilities, we held out 20\% of the observations as test points and randomly split the remaining observations into two equally sized data sets, from which we computed two posteriors to compare.
We generated $R = 50$ replicate experiments in this way, and followed the procedure in \cref{sec:simulations} to approximate the overlap probability for each replicate.

\begin{figure}[tbp]
\begin{center}
\begin{subfigure}[b]{.44\textwidth}
\centering
\includegraphics[height=1.6in,trim={0in 0in 1.45in 0in},clip]{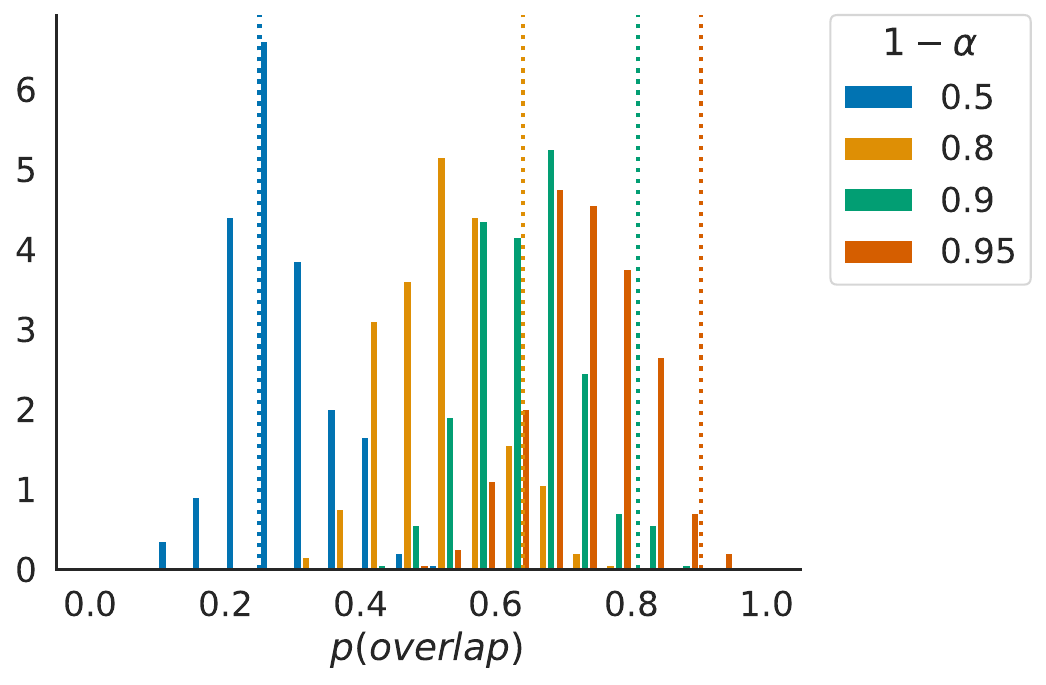}
\caption{Bayes}
\end{subfigure}
\begin{subfigure}[b]{.55\textwidth}
\centering
\includegraphics[height=1.6in]{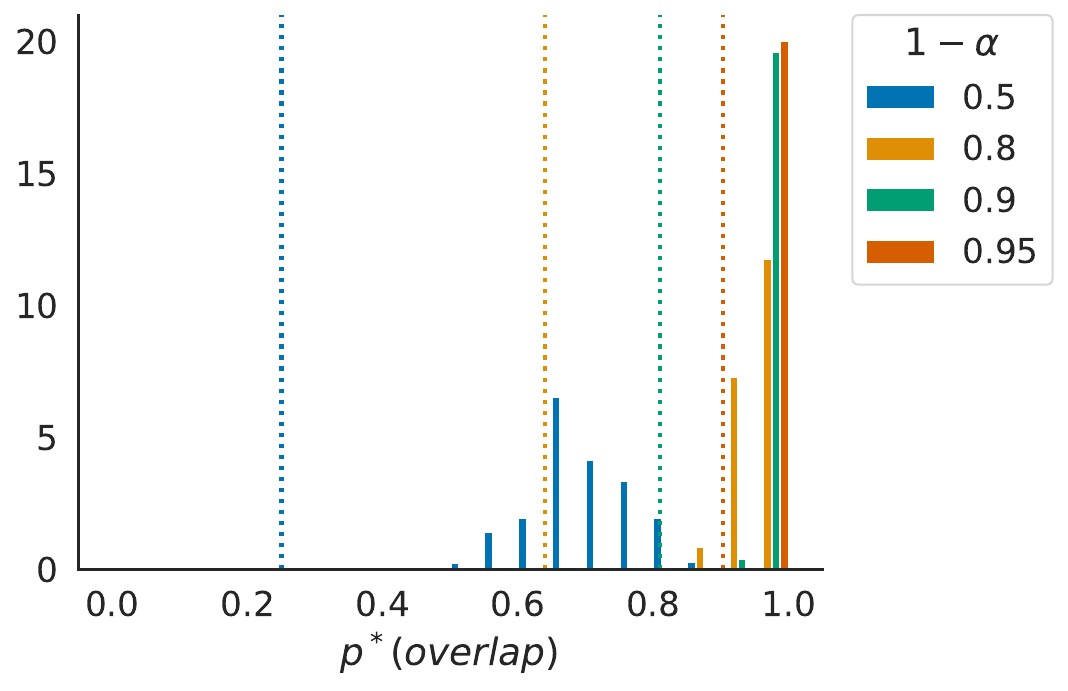}
\caption{BayesBag}
\end{subfigure} 
\caption{For crime data using a sparse Poisson regression model, shown are histograms of the overlap probability for $Z^{\top} \beta$ where $Z$ is drawn from a held-out test set. 
For most replicates, the overlap probabilities for the standard posteriors are below $(1 - \alpha)^{2}$ for $1-\alpha \in \{0.8, 0.9, 0.95\}$.
Meanwhile, for all replicates, the overlap probabilities for the bagged posteriors are greater than $(1 - \alpha)^{2}$. 
}
\label{fig:sparse-Poisson-regression}
\end{center}
\end{figure}

\Cref{fig:sparse-Poisson-regression} validates our theoretical results: 
the standard posterior is unstable across datasets, with overlap probabilities below  $(1 - \alpha)^{2}$ for $1-\alpha \in \{0.8, 0.9, 0.95\}$
in the vast majority of replicates. 
The bagged posteriors, on the other hand, have overlap greater than $(1 - \alpha)^{2}$ in all replicates. 
Moreover, BayesBag has superior predictive performance:
the mean log predictive densities for the standard and bagged posteriors are, respectively, $-5.4$ and $-4.3$ with a
99\% confidence interval for difference of $(1.043, 1.093)$ (paired $t$ interval).  

\begin{figure}[tbp]
\begin{center}
\begin{subfigure}[b]{.49\textwidth}
\centering
\includegraphics[height=1.6in]{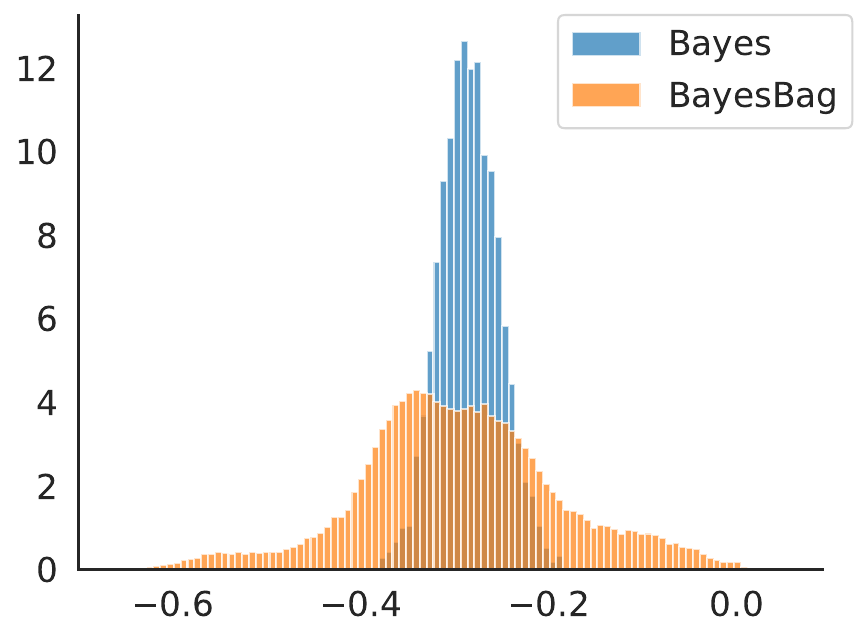}
\caption{\texttt{Lower Quartile Rent}}
\end{subfigure} 
\begin{subfigure}[b]{.49\textwidth}
\centering
\includegraphics[height=1.6in]{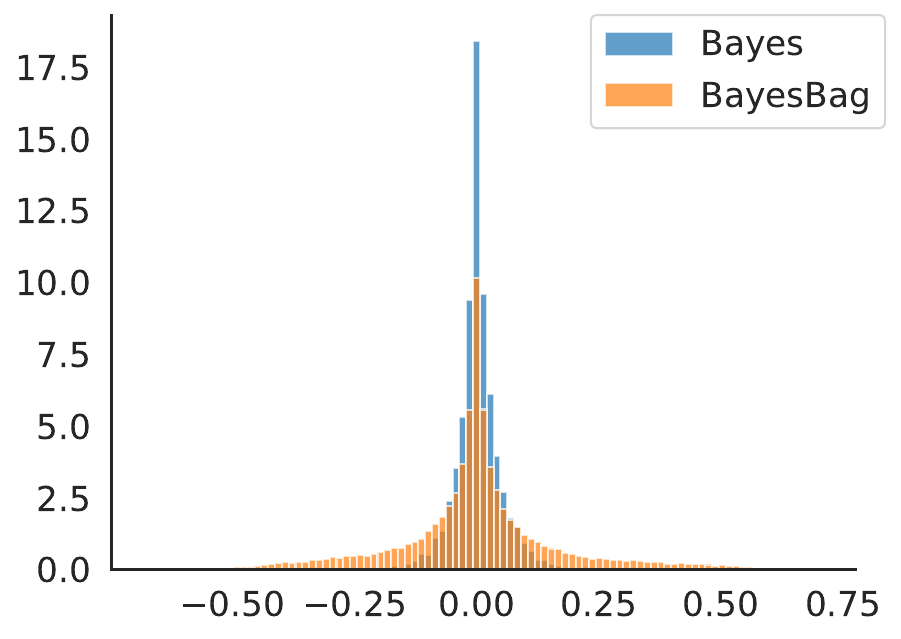}
\caption{\texttt{Median Rent}}
\end{subfigure} 
\begin{subfigure}[b]{.49\textwidth}
\centering
\includegraphics[height=1.6in]{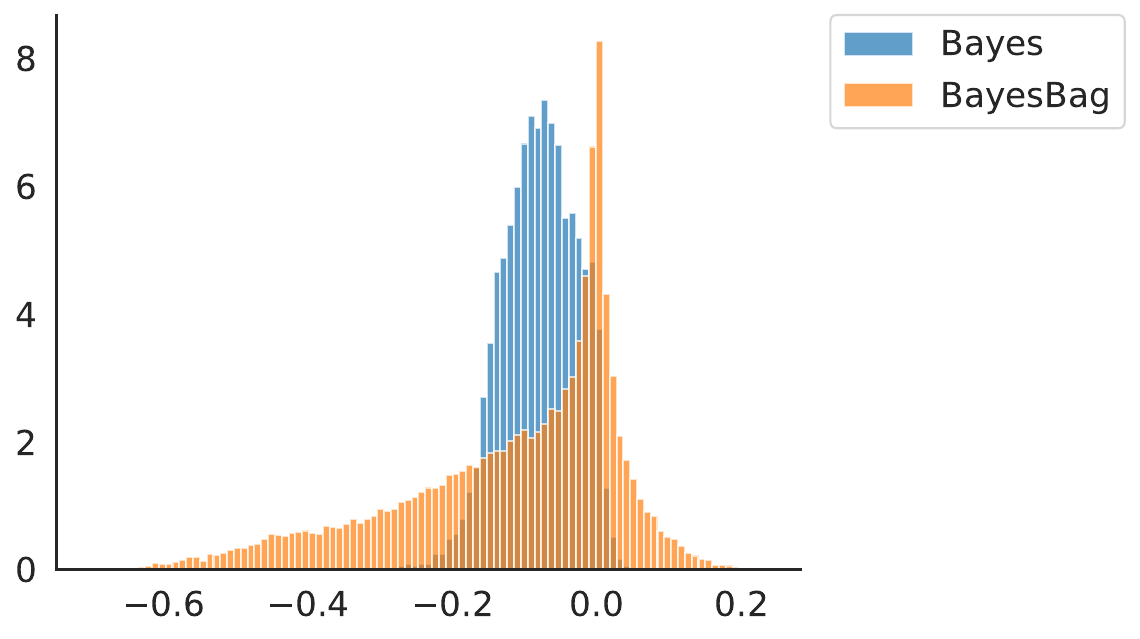}
\caption{\texttt{Upper Quartile Rent}}
\end{subfigure} 
\caption{The standard and bagged posterior marginals for three coefficients related to rental cost
for the data and model from \cref{sec:application}.
}
\label{fig:sparse-Poisson-regression-rent-marginals}
\end{center}
\end{figure}

\begin{figure}[tbp]
\begin{center}
\begin{subfigure}[b]{.49\textwidth}
\centering
\includegraphics[height=1.6in]{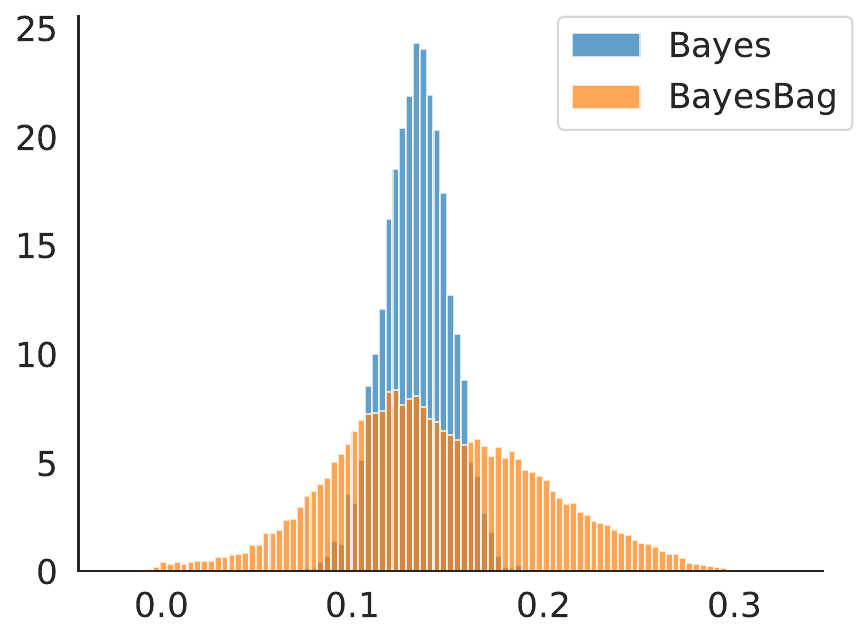}
\caption{\texttt{Percent Black}}
\end{subfigure} 
\begin{subfigure}[b]{.49\textwidth}
\centering
\includegraphics[height=1.6in]{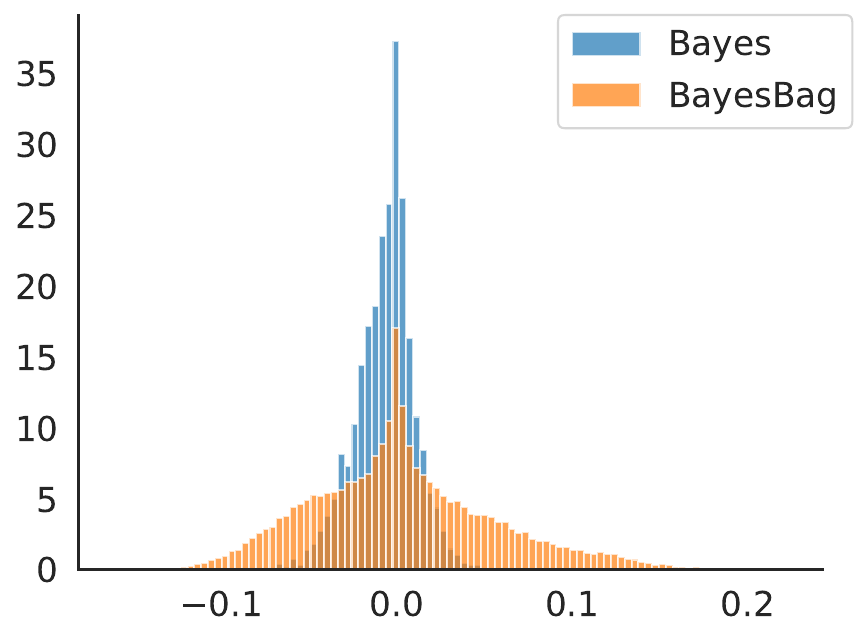}
\caption{\texttt{Percent White}}
\end{subfigure} 
\begin{subfigure}[b]{.49\textwidth}
\centering
\includegraphics[height=1.6in]{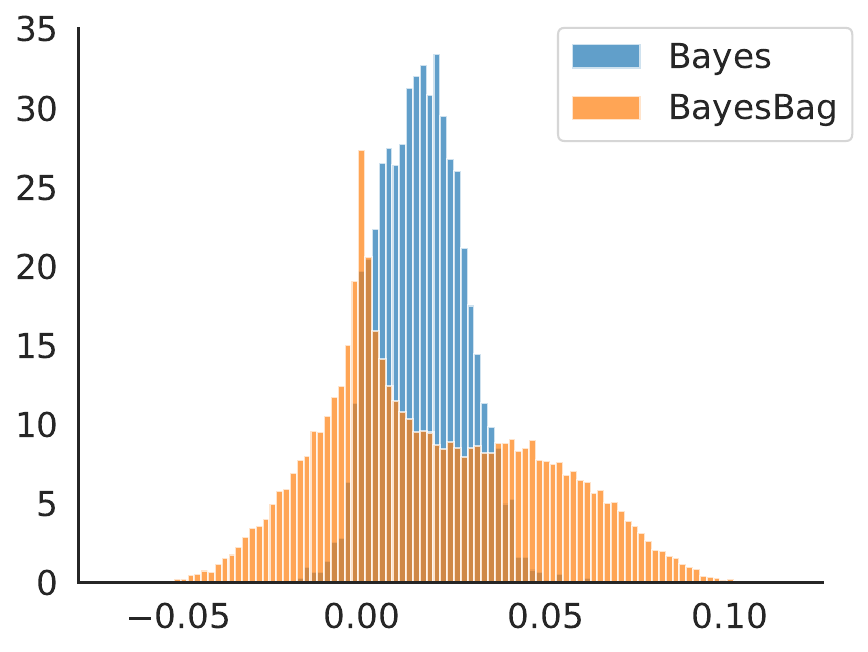}
\caption{\texttt{Percent Asian}}
\end{subfigure} 
\begin{subfigure}[b]{.49\textwidth}
\centering
\includegraphics[height=1.6in]{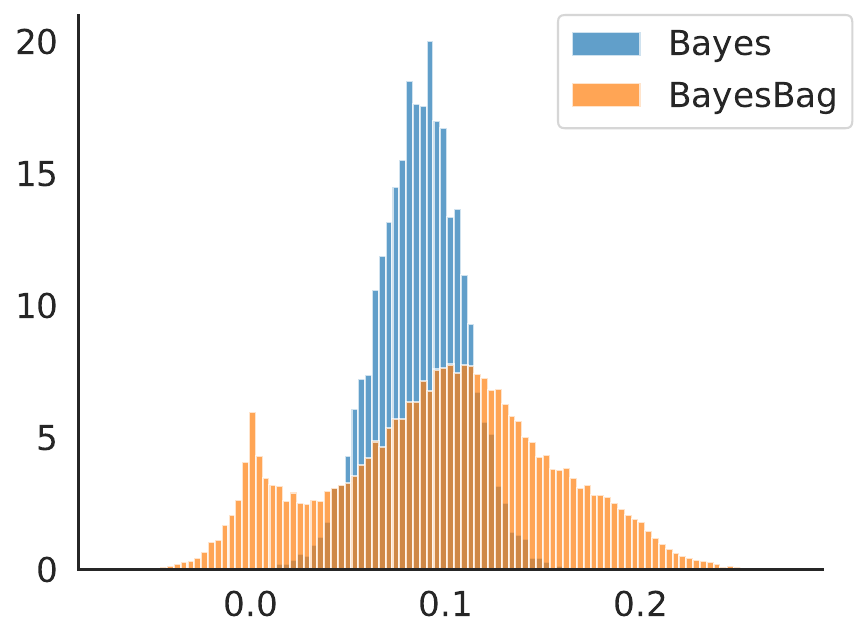}
\caption{\texttt{Percent Hispanic}}
\end{subfigure} 
\caption{The standard and bagged posterior marginals for three coefficients related to race
for the data and model from \cref{sec:application}.
}
\label{fig:sparse-Poisson-regression-race-marginals}
\end{center}
\end{figure}

To explore how using the bagged rather than the standard posterior might result in different conclusions, 
we compared the posterior marginals of the regression coefficients, with some representative results 
shown in \cref{fig:sparse-Poisson-regression-rent-marginals,fig:sparse-Poisson-regression-race-marginals}.
In all cases, the bagged posteriors were more diffuse, as would be expected.
In several cases, however, the BayesBag results are qualitatively different from the standard posterior results. 
The standard posterior for the coefficient of \texttt{Upper Quartile Rent} is symmetric and concentrated below zero while for the bagged posterior 
it has a sharp peak at zero and is skewed left (\cref{fig:sparse-Poisson-regression-rent-marginals}). 
Similarly, the standard posteriors are symmetric for the coefficients of covariates related to percent of different racial and ethnic groups (\cref{fig:sparse-Poisson-regression-race-marginals}).
Meanwhile, the bagged posterior for the coefficients of \texttt{Percent Asian} and \texttt{Percent Hispanic} are multimodal and have significantly more mass centered at zero.  
These examples illustrate how the bagged and standard posteriors may yield substantively different results in practice -- BayesBag is not merely inflating the posterior uncertainty.

\section{Discussion} \label{sec:discussion}

We conclude by first situating BayesBag in the wider literature on robust Bayesian inference,
and then, with that additional context in place, highlighting the strengths of our approach and suggest fruitful directions for future development. 

\subsection{Bayesian bagging}

Despite the similar sounding names, BayesBag is very different than Bayesian bagging \citep{Clyde:2001,Lee:2004}.
Bayesian bagging consists of applying the Bayesian bootstrap to a point estimator of a classification or regression model,
such as ordinary least squares.
In other words, it is a slight variant of traditional bagging where, instead of multinomial weights, one uses 
continuous weights drawn uniformly from the probability simplex.
In contrast, BayesBag uses traditional bagging on the posterior of an arbitrary Bayesian model.
In short, Bayesian bagging performs bagging using Bayes, whereas BayesBag performs Bayes using bagging.
Relatedly, in the same way that bagging expands the model space for a classification or regression method \citep{Domingos:1997},
BayesBag expands the posterior space for a Bayesian model.

\subsection{Bayesian uncertainty quantification with the bootstrap} 

The bootstrap has previously been employed to perform uncertainty quantification in Bayesian settings.
See \citet{Laird:1987} and references therein for uses of the bootstrap to 
adjust for underestimated uncertainties when using empirical Bayesian methods. 
Similar in spirit to the present work, \citet{Efron:2015} develops a variety of methods for obtaining frequentist
uncertainty quantification of Bayesian point estimates, including some that rely on bootstrapping. 

\subsection{Robust Bayesian inference}

Two common themes emerge when surveying existing methods for robust Bayesian inference. 
First, many methods require choosing a free parameter, %
and the proposals for choosing free parameters tend to be either (a) heuristic, (b) strongly dependent on being in the asymptotic regime, 
or (c) computationally prohibitive for most real-world problems. 
Second, those methods without a free parameter lose key parts of what makes the Bayesian approach attractive. 
For example, they strongly rely on asymptotic assumptions, make a Gaussian assumption, or do not incorporate a prior distribution. 

The power posterior is perhaps the most widely studied method for making the posterior robust to model misspecification \citep{Grunwald:2012,Holmes:2017,Grunwald:2017,Miller:2018:coarsening,Syring:2018,Lyddon:2019}. 
For a likelihood function $L(\param)$, prior distribution $\priordist$, and any $\zeta \ge 0$, the \emph{$\zeta$-power posterior} is defined as 
$\ppostdist{}{(\zeta)}(\dee\param) \propto L(\param)^{\zeta}\priordist(\dee\param)$.
Hence, $\ppostdist{}{(1)}$ is equal to the standard posterior and $\ppostdist{}{(0)}$ is equal to the prior. 
Typically, $\zeta$ is set to a value between these two extremes, as 
there is significant theoretical support for the use of power posteriors with 
$\zeta \in (0,1)$~\citep{Bhattacharya:2019,Walker:2001,Miller:2018:coarsening,Royall:2003,Grunwald:2012}.
However, there are two significant methodological challenges. 
First, computing the power posterior often requires new computational methods or additional approximations, particularly in latent variable models~\citep{AntonianoVillalobos:2013,Miller:2018:coarsening}.
Second, choosing an appropriate value of $\zeta$ can be difficult.
\citet{Grunwald:2012} proposes SafeBayes, a theoretically sound method which is evaluated empirically in \citet{Grunwald:2017} and \citet{deHeide:2019}. 
However, SafeBayes is computationally prohibitive except with simple models and very small datasets.
In addition, the underlying theory relies on strong assumptions on the model class. 
Many other methods for choosing $\zeta$ have been suggested, but they are either heuristic or rely on strong asymptotic
assumptions such as the accuracy of the plug-in estimator for the sandwich covariance~\citep{Royall:2003,Holmes:2017,Miller:2018:coarsening,Syring:2018,Lyddon:2019}.

More in the spirit of BayesBag are a number of bootstrapped point estimation approaches~\citep{Rubin:1981:BayesianBootstrap,Newton:1994,Chamberlain:2003,Lyddon:2018,Lyddon:2019}.
However, unlike BayesBag, these methods compute a collection of \emph{maximum a posteriori} (MAP) or \emph{maximum likelihood} (ML) estimates. 
The weighted likelihood bootstrap of \citet{Newton:1994} and a generalization proposed by \citet{Lyddon:2019} do not incorporate a prior, and therefore lose 
many of the benefits of Bayesian inference.
The related approach of \citet{Lyddon:2018}, which includes the weighted likelihood bootstrap and standard Bayesian inference as limiting cases, draws the bootstrap samples partially from
the posterior and partially from the empirical distribution. 
Unfortunately, there is no accompanying theory to guide how much the empirical distribution and posterior distribution should be weighted relative to each other -- nor rigorous 
robustness guarantees. 
Moreover, bootstrapped point estimation methods can behave poorly when the MAP and ML estimates are not well-behaved -- for example, 
due to the likelihood being peaked (or even tending to infinity) 
in a region of low posterior probability. 

\citet{Muller:2013} suggests replacing the standard posterior by a Gaussian distribution with covariance proportional to a plug-in estimate of the sandwich covariance.
A benefit of our approach is that it does not rely on a Gaussian approximation and does not require estimation of the sandwich covariance, making it suitable for small-sample settings.
While our theory does focus on Gaussian or asymptotically Gaussian posteriors, in practice BayesBag is applicable in non-asymptotic regimes where the posterior is highly non-Gaussian, as shown by the application in \cref{sec:application}.

\subsection{The benefits of BayesBag}

In view of previous work, the BayesBag approach has a number of attractive features that make it flexible, easy-to-use, and widely applicable. 
From a methodological perspective,  BayesBag is general-purpose.
It relies only on carrying out standard posterior inference, it is applicable to a wide range of models, and it can make full use of 
modern probabilistic programming tools -- the only added requirement is the design of a bootstrapping scheme. 
Although this paper focuses on using BayesBag with independent observations, future work can draw on the large literature devoted to adapting the bootstrap to more
complex models such as those involving time-series and spatial data.
BayesBag is also general-purpose in the sense that it is useful no matter whether the ultimate goal of Bayesian inference is parameter estimation, prediction, or model selection;
see \citet{Huggins:2019:BayesBagII} for how to use BayesBag for model selection. 

Another appeal of BayesBag as a methodology is that the only hyperparameter -- the bootstrap dataset size $\bsnumobs$ -- is straightforward to set.
Specifically, $\bsnumobs = \numobs$ is a natural, theoretically well-justified choice that, while slightly conservative, yields reproducible inferences.

In terms of computation, when using the approximation in \cref{eq:bayesbag-approximation}, there is an additional cost due to the need to compute 
the posterior for each bootstrapped dataset.
However, it is trivial to compute the bootstrapped posteriors in parallel.
As described in \cref{sec:choosing-B}, validating that the number of bootstrap datasets $B$ is sufficiently large only requires
computing simple Monte Carlo error bounds. 
Moreover, defaulting to $B=50$ or $100$ appears to be an empirically sound choice across a range of problems. 
Nonetheless, speeding up BayesBag with more specialized computational methods could be worthwhile in some applications.
For example, in \cref{app:computation}, we suggest one simple approach to speeding up Markov chain Monte Carlo (MCMC) runs when using BayesBag.
Pierre Jacob has proposed using more advanced unbiased MCMC techniques for potentially even greater 
computational efficiency.\footnote{\small\url{https://statisfaction.wordpress.com/2019/10/02/bayesbag-and-how-to-approximate-it/}}

Another benefit of BayesBag is that it incorporates robustness features of frequentist methods into Bayesian inference without sacrificing 
the core benefits of the Bayesian approach such as flexible modeling, straightforward integration over nuisance parameters, and the use of prior information. 
Further, our Jeffrey conditionalization interpretation establishes solid epistemological foundations for using BayesBag.
Thus, it provides an appealing and philosophically coherent synthesis of Bayesian and frequentist approaches
without introducing difficult-to-choose tuning parameters 
and without sacrificing the most useful parts of Bayesian inference.

\subsection*{Acknowledgments}

Thanks to Pierre Jacob for bringing P.~B\"uhlmann's BayesBag paper to our attention.
Thanks also to Ryan Giordano and Pierre Jacob for helpful feedback on an earlier version of this paper,
to Peter Gr\"unwald, Natalia Bochkina, Mathieu Gerber, and Anthony Lee for helpful discussions,
and to the Associate Editor and two referees whose comments led to substantial improvements to
scope and focus of the paper.

\bibliographystyle{imsart-nameyear}
\bibliography{library,../bayesbag}

\begin{thebibliography}{48}

\bibitem[\protect\citeauthoryear{Antoniano-Villalobos and
  Walker}{2013}]{AntonianoVillalobos:2013}
\begin{barticle}[author]
\bauthor{\bsnm{Antoniano-Villalobos},~\bfnm{Isadora}\binits{I.}} \AND
  \bauthor{\bsnm{Walker},~\bfnm{Stephen~G}\binits{S.~G.}}
(\byear{2013}).
\btitle{{Bayesian Nonparametric Inference for the Power Likelihood}}.
\bjournal{Journal of Computational and Graphical Statistics}
\bvolume{22}
\bpages{801--813}.
\end{barticle}
\endbibitem

\bibitem[\protect\citeauthoryear{Bhattacharya, Pati and
  Yang}{2019}]{Bhattacharya:2019}
\begin{barticle}[author]
\bauthor{\bsnm{Bhattacharya},~\bfnm{Anirban}\binits{A.}},
  \bauthor{\bsnm{Pati},~\bfnm{Debdeep}\binits{D.}} \AND
  \bauthor{\bsnm{Yang},~\bfnm{Yun}\binits{Y.}}
(\byear{2019}).
\btitle{{Bayesian fractional posteriors}}.
\bjournal{The Annals of Statistics}
\bvolume{47}
\bpages{39--66}.
\end{barticle}
\endbibitem

\bibitem[\protect\citeauthoryear{Bissiri, Holmes and
  Walker}{2016}]{Bissiri:2016}
\begin{barticle}[author]
\bauthor{\bsnm{Bissiri},~\bfnm{Pier~Giovanni}\binits{P.~G.}},
  \bauthor{\bsnm{Holmes},~\bfnm{Chris~C}\binits{C.~C.}} \AND
  \bauthor{\bsnm{Walker},~\bfnm{Stephen~G}\binits{S.~G.}}
(\byear{2016}).
\btitle{{A general framework for updating belief distributions}}.
\bjournal{Journal of the Royal Statistical Society: Series B (Statistical
  Methodology)}
\bvolume{78}
\bpages{1103--1130}.
\end{barticle}
\endbibitem

\bibitem[\protect\citeauthoryear{Box}{1979}]{Box:1979}
\begin{bincollection}[author]
\bauthor{\bsnm{Box},~\bfnm{G~E~P}\binits{G.~E.~P.}}
(\byear{1979}).
\btitle{{Robustness in the Strategy of Scientific Model Building}}.
In \bbooktitle{Robustness in Statistics}
\bpages{201--236}.
\bpublisher{Elsevier}.
\end{bincollection}
\endbibitem

\bibitem[\protect\citeauthoryear{Box}{1980}]{Box:1980}
\begin{barticle}[author]
\bauthor{\bsnm{Box},~\bfnm{George E~P}\binits{G.~E.~P.}}
(\byear{1980}).
\btitle{{Sampling and Bayes' Inference in Scientific Modelling and
  Robustness}}.
\bjournal{Journal of the Royal Statistical Society. Series A (General)}
\bvolume{143}
\bpages{383--430}.
\end{barticle}
\endbibitem

\bibitem[\protect\citeauthoryear{Breiman}{1996}]{Breiman:1996}
\begin{barticle}[author]
\bauthor{\bsnm{Breiman},~\bfnm{Leo}\binits{L.}}
(\byear{1996}).
\btitle{{Bagging Predictors}}.
\bjournal{Machine Learning}
\bvolume{24}
\bpages{123--140}.
\end{barticle}
\endbibitem

\bibitem[\protect\citeauthoryear{B{\"u}hlmann}{2014}]{Buhlmann:2014}
\begin{barticle}[author]
\bauthor{\bsnm{B{\"u}hlmann},~\bfnm{Peter}\binits{P.}}
(\byear{2014}).
\btitle{{Discussion of Big Bayes Stories and BayesBag}}.
\bjournal{Statistical Science}
\bvolume{29}
\bpages{91--94}.
\end{barticle}
\endbibitem

\bibitem[\protect\citeauthoryear{Chamberlain and
  Imbens}{2003}]{Chamberlain:2003}
\begin{barticle}[author]
\bauthor{\bsnm{Chamberlain},~\bfnm{Gary}\binits{G.}} \AND
  \bauthor{\bsnm{Imbens},~\bfnm{Guido}\binits{G.}}
(\byear{2003}).
\btitle{{Nonparametric applications of Bayesian inference}}.
\bjournal{Journal of Business Economic Statistics}
\bvolume{21}
\bpages{12--18}.
\end{barticle}
\endbibitem

\bibitem[\protect\citeauthoryear{Clyde and Lee}{2001}]{Clyde:2001}
\begin{binproceedings}[author]
\bauthor{\bsnm{Clyde},~\bfnm{Merlise}\binits{M.}} \AND
  \bauthor{\bsnm{Lee},~\bfnm{Herbert}\binits{H.}}
(\byear{2001}).
\btitle{Bagging and the Bayesian bootstrap}.
In \bbooktitle{International Workshop on Artificial Intelligence and
  Statistics}
\bpages{57--62}.
\bpublisher{PMLR}.
\end{binproceedings}
\endbibitem

\bibitem[\protect\citeauthoryear{Cox}{1990}]{Cox:1990}
\begin{barticle}[author]
\bauthor{\bsnm{Cox},~\bfnm{D~R}\binits{D.~R.}}
(\byear{1990}).
\btitle{{Role of Models in Statistical Analysis}}.
\bjournal{Statistical Science}
\bvolume{5}
\bpages{169--174}.
\end{barticle}
\endbibitem

\bibitem[\protect\citeauthoryear{De~Blasi and Walker}{2013}]{DeBlasi:2013}
\begin{barticle}[author]
\bauthor{\bsnm{De~Blasi},~\bfnm{Pierpaolo}\binits{P.}} \AND
  \bauthor{\bsnm{Walker},~\bfnm{Stephen~G}\binits{S.~G.}}
(\byear{2013}).
\btitle{{Bayesian asymptotics with misspecified models}}.
\bjournal{Statistica Sinica}
\bpages{1--19}.
\end{barticle}
\endbibitem

\bibitem[\protect\citeauthoryear{de~Heide et~al.}{2019}]{deHeide:2019}
\begin{barticle}[author]
\bauthor{\bparticle{de} \bsnm{Heide},~\bfnm{Rianne}\binits{R.}},
  \bauthor{\bsnm{Kirichenko},~\bfnm{Alisa}\binits{A.}},
  \bauthor{\bsnm{Mehta},~\bfnm{Nishant}\binits{N.}} \AND
  \bauthor{\bsnm{Gr{\"u}nwald},~\bfnm{Peter~D}\binits{P.~D.}}
(\byear{2019}).
\btitle{{Safe-Bayesian Generalized Linear Regression}}.
\bjournal{arXiv.org}
\bvolume{arXiv:1910.09227 [math.ST]}.
\end{barticle}
\endbibitem

\bibitem[\protect\citeauthoryear{Diaconis and Zabell}{1982}]{Diaconis:1982}
\begin{barticle}[author]
\bauthor{\bsnm{Diaconis},~\bfnm{P}\binits{P.}} \AND
  \bauthor{\bsnm{Zabell},~\bfnm{Sandy~L}\binits{S.~L.}}
(\byear{1982}).
\btitle{{Updating subjective probability}}.
\bjournal{Journal of the American Statistical Association}
\bvolume{77}
\bpages{822--830}.
\end{barticle}
\endbibitem

\bibitem[\protect\citeauthoryear{Domingos}{1997}]{Domingos:1997}
\begin{binproceedings}[author]
\bauthor{\bsnm{Domingos},~\bfnm{Pedro~M}\binits{P.~M.}}
(\byear{1997}).
\btitle{Why Does Bagging Work? A Bayesian Account and its Implications.}
In \bbooktitle{KDD}
\bpages{155--158}.
\end{binproceedings}
\endbibitem

\bibitem[\protect\citeauthoryear{Douady et~al.}{2003}]{Douady:2003}
\begin{barticle}[author]
\bauthor{\bsnm{Douady},~\bfnm{C~J}\binits{C.~J.}},
  \bauthor{\bsnm{Delsuc},~\bfnm{F}\binits{F.}},
  \bauthor{\bsnm{Boucher},~\bfnm{Y}\binits{Y.}},
  \bauthor{\bsnm{Doolittle},~\bfnm{W~F}\binits{W.~F.}} \AND
  \bauthor{\bsnm{Douzery},~\bfnm{E~J~P}\binits{E.~J.~P.}}
(\byear{2003}).
\btitle{{Comparison of Bayesian and Maximum Likelihood Bootstrap Measures of
  Phylogenetic Reliability}}.
\bjournal{Molecular Biology and Evolution}
\bvolume{20}
\bpages{248--254}.
\end{barticle}
\endbibitem

\bibitem[\protect\citeauthoryear{Durrett}{2019}]{Durrett:2019}
\begin{bbook}[author]
\bauthor{\bsnm{Durrett},~\bfnm{Richard}\binits{R.}}
(\byear{2019}).
\btitle{Probability: Theory and Examples}.
\bseries{Cambridge Series in Statistical and Probabilistic Mathematics}.
\bpublisher{Cambridge University Press}.
\end{bbook}
\endbibitem

\bibitem[\protect\citeauthoryear{Efron}{2015}]{Efron:2015}
\begin{barticle}[author]
\bauthor{\bsnm{Efron},~\bfnm{Bradley}\binits{B.}}
(\byear{2015}).
\btitle{{Frequentist accuracy of Bayesian estimates}}.
\bjournal{Journal of the Royal Statistical Society: Series B (Statistical
  Methodology)}
\bvolume{77}
\bpages{617--646}.
\end{barticle}
\endbibitem

\bibitem[\protect\citeauthoryear{Greco, Racugno and
  Ventura}{2008}]{Greco:2008:robust}
\begin{barticle}[author]
\bauthor{\bsnm{Greco},~\bfnm{Luca}\binits{L.}},
  \bauthor{\bsnm{Racugno},~\bfnm{Walter}\binits{W.}} \AND
  \bauthor{\bsnm{Ventura},~\bfnm{Laura}\binits{L.}}
(\byear{2008}).
\btitle{{Robust likelihood functions in Bayesian inference}}.
\bjournal{Journal of Statistical Planning and Inference}
\bvolume{138}
\bpages{1258 -- 1270}.
\bdoi{10.1016/j.jspi.2007.05.001}
\end{barticle}
\endbibitem

\bibitem[\protect\citeauthoryear{Gr{\"u}nwald}{2012}]{Grunwald:2012}
\begin{binproceedings}[author]
\bauthor{\bsnm{Gr{\"u}nwald},~\bfnm{Peter~D}\binits{P.~D.}}
(\byear{2012}).
\btitle{{The Safe Bayesian: Learning the Learning Rate via the Mixability
  Gap}}.
In \bbooktitle{Algorithmic Learning Theory}
\bpages{169--183}.
\end{binproceedings}
\endbibitem

\bibitem[\protect\citeauthoryear{Gr{\"u}nwald and van
  Ommen}{2017}]{Grunwald:2017}
\begin{barticle}[author]
\bauthor{\bsnm{Gr{\"u}nwald},~\bfnm{Peter~D}\binits{P.~D.}} \AND
  \bauthor{\bparticle{van} \bsnm{Ommen},~\bfnm{Thijs}\binits{T.}}
(\byear{2017}).
\btitle{{Inconsistency of Bayesian Inference for Misspecified Linear Models,
  and a Proposal for Repairing It}}.
\bjournal{Bayesian Analysis}
\bvolume{12}
\bpages{1069--1103}.
\end{barticle}
\endbibitem

\bibitem[\protect\citeauthoryear{Hoff and Wakefield}{2012}]{Hoff:2012}
\begin{barticle}[author]
\bauthor{\bsnm{Hoff},~\bfnm{Peter}\binits{P.}} \AND
  \bauthor{\bsnm{Wakefield},~\bfnm{Jon}\binits{J.}}
(\byear{2012}).
\btitle{Bayesian sandwich posteriors for pseudo-true parameters}.
\bjournal{arXiv preprint arXiv:1211.0087}.
\end{barticle}
\endbibitem

\bibitem[\protect\citeauthoryear{Holmes and Walker}{2017}]{Holmes:2017}
\begin{barticle}[author]
\bauthor{\bsnm{Holmes},~\bfnm{Christopher~C}\binits{C.~C.}} \AND
  \bauthor{\bsnm{Walker},~\bfnm{Stephen~G}\binits{S.~G.}}
(\byear{2017}).
\btitle{{Assigning a value to a power likelihood in a general Bayesian model}}.
\bjournal{Biometrika}
\bvolume{104}
\bpages{497--503}.
\end{barticle}
\endbibitem

\bibitem[\protect\citeauthoryear{Huggins and
  Miller}{2023}]{Huggins:2019:BayesBagII}
\begin{barticle}[author]
\bauthor{\bsnm{Huggins},~\bfnm{Jonathan~H}\binits{J.~H.}} \AND
  \bauthor{\bsnm{Miller},~\bfnm{Jeffrey~W}\binits{J.~W.}}
(\byear{2023}).
\btitle{{Reproducible Model Selection Using Bagged Posteriors}}.
\bjournal{Bayesian Analysis}
\bvolume{18}
\bpages{79--104}.
\end{barticle}
\endbibitem

\bibitem[\protect\citeauthoryear{Jeffrey}{1968}]{Jeffrey:1968}
\begin{bincollection}[author]
\bauthor{\bsnm{Jeffrey},~\bfnm{Richard~C}\binits{R.~C.}}
(\byear{1968}).
\btitle{{Probable Knowledge}}.
In \bbooktitle{The Problem of Inductive Logic}
(\beditor{\bfnm{Imre}\binits{I.}~\bsnm{Lakatos}}, ed.)
\bpages{166--180}.
\bpublisher{North-Holland}, \baddress{Amsterdam}.
\end{bincollection}
\endbibitem

\bibitem[\protect\citeauthoryear{Jeffrey}{1990}]{Jeffrey:1990}
\begin{bbook}[author]
\bauthor{\bsnm{Jeffrey},~\bfnm{Richard~C}\binits{R.~C.}}
(\byear{1990}).
\btitle{{The Logic of Decision}},
\bedition{2nd} ed.
\bpublisher{University of Chicago Press}.
\end{bbook}
\endbibitem

\bibitem[\protect\citeauthoryear{Jewson, Smith and
  Holmes}{2018}]{Jewson:2018:principles}
\begin{barticle}[author]
\bauthor{\bsnm{Jewson},~\bfnm{Jack}\binits{J.}},
  \bauthor{\bsnm{Smith},~\bfnm{Jim~Q.}\binits{J.~Q.}} \AND
  \bauthor{\bsnm{Holmes},~\bfnm{Chris}\binits{C.}}
(\byear{2018}).
\btitle{{Principles of Bayesian Inference Using General Divergence Criteria}}.
\bjournal{Entropy}
\bvolume{20}
\bpages{442}.
\end{barticle}
\endbibitem

\bibitem[\protect\citeauthoryear{Kallenberg}{2002}]{Kallenberg:2002}
\begin{bbook}[author]
\bauthor{\bsnm{Kallenberg},~\bfnm{Olav}\binits{O.}}
(\byear{2002}).
\btitle{{Foundations of Modern Probability}},
\bedition{2nd} ed.
\bpublisher{Springer}, \baddress{New York, NY}.
\end{bbook}
\endbibitem

\bibitem[\protect\citeauthoryear{Kleijn and van~der Vaart}{2012}]{Kleijn:2012}
\begin{barticle}[author]
\bauthor{\bsnm{Kleijn},~\bfnm{B~J~K}\binits{B.~J.~K.}} \AND
  \bauthor{\bparticle{van~der} \bsnm{Vaart},~\bfnm{A~W}\binits{A.~W.}}
(\byear{2012}).
\btitle{{The Bernstein-Von-Mises theorem under misspecification}}.
\bjournal{Electronic Journal of Statistics}
\bvolume{6}
\bpages{354--381}.
\end{barticle}
\endbibitem

\bibitem[\protect\citeauthoryear{Koehler, Brown and
  Haneuse}{2009}]{Koehler:2009}
\begin{barticle}[author]
\bauthor{\bsnm{Koehler},~\bfnm{Elizabeth}\binits{E.}},
  \bauthor{\bsnm{Brown},~\bfnm{Elizabeth}\binits{E.}} \AND
  \bauthor{\bsnm{Haneuse},~\bfnm{Sebastien J P~A}\binits{S.~J. P.~A.}}
(\byear{2009}).
\btitle{{On the Assessment of Monte Carlo Error in Simulation-Based Statistical
  Analyses}}.
\bjournal{The American Statistician}
\bvolume{63}
\bpages{155--162}.
\end{barticle}
\endbibitem

\bibitem[\protect\citeauthoryear{Laird and Louis}{1987}]{Laird:1987}
\begin{barticle}[author]
\bauthor{\bsnm{Laird},~\bfnm{Nan~M.}\binits{N.~M.}} \AND
  \bauthor{\bsnm{Louis},~\bfnm{Thomas~A.}\binits{T.~A.}}
(\byear{1987}).
\btitle{{Empirical Bayes Confidence Intervals Based on Bootstrap Samples}}.
\bjournal{Journal of the American Statistical Association}
\bvolume{82}
\bpages{739--750}.
\end{barticle}
\endbibitem

\bibitem[\protect\citeauthoryear{Lee and Clyde}{2004}]{Lee:2004}
\begin{barticle}[author]
\bauthor{\bsnm{Lee},~\bfnm{Herbert~KH}\binits{H.~K.}} \AND
  \bauthor{\bsnm{Clyde},~\bfnm{Merlise~A}\binits{M.~A.}}
(\byear{2004}).
\btitle{Lossless online Bayesian bagging}.
\bjournal{Journal of Machine Learning Research}
\bvolume{5}
\bpages{143--151}.
\end{barticle}
\endbibitem

\bibitem[\protect\citeauthoryear{Lehmann}{1990}]{Lehmann:1990}
\begin{barticle}[author]
\bauthor{\bsnm{Lehmann},~\bfnm{E~L}\binits{E.~L.}}
(\byear{1990}).
\btitle{{Model specification: the views of Fisher and Neyman, and later
  developments}}.
\bjournal{Statistical Science}
\bvolume{5}
\bpages{160--168}.
\end{barticle}
\endbibitem

\bibitem[\protect\citeauthoryear{Lyddon, Holmes and Walker}{2019}]{Lyddon:2019}
\begin{barticle}[author]
\bauthor{\bsnm{Lyddon},~\bfnm{S~P}\binits{S.~P.}},
  \bauthor{\bsnm{Holmes},~\bfnm{Chris~C}\binits{C.~C.}} \AND
  \bauthor{\bsnm{Walker},~\bfnm{S~G}\binits{S.~G.}}
(\byear{2019}).
\btitle{{General Bayesian updating and the loss-likelihood bootstrap}}.
\bjournal{Biometrika}
\bvolume{106}
\bpages{465--478}.
\end{barticle}
\endbibitem

\bibitem[\protect\citeauthoryear{Lyddon, Walker and Holmes}{2018}]{Lyddon:2018}
\begin{binproceedings}[author]
\bauthor{\bsnm{Lyddon},~\bfnm{S~P}\binits{S.~P.}},
  \bauthor{\bsnm{Walker},~\bfnm{S~G}\binits{S.~G.}} \AND
  \bauthor{\bsnm{Holmes},~\bfnm{Chris~C}\binits{C.~C.}}
(\byear{2018}).
\btitle{{Nonparametric learning from Bayesian models with randomized objective
  functions}}.
In \bbooktitle{Advances in Neural Information Processing Systems}.
\end{binproceedings}
\endbibitem

\bibitem[\protect\citeauthoryear{Miller and
  Dunson}{2018}]{Miller:2018:coarsening}
\begin{barticle}[author]
\bauthor{\bsnm{Miller},~\bfnm{Jeffrey~W}\binits{J.~W.}} \AND
  \bauthor{\bsnm{Dunson},~\bfnm{David~B}\binits{D.~B.}}
(\byear{2018}).
\btitle{{Robust Bayesian Inference via Coarsening}}.
\bjournal{Journal of the American Statistical Association}
\bvolume{114}
\bpages{1113--1125}.
\end{barticle}
\endbibitem

\bibitem[\protect\citeauthoryear{M{\"u}ller}{2013}]{Muller:2013}
\begin{barticle}[author]
\bauthor{\bsnm{M{\"u}ller},~\bfnm{Ulrich~K}\binits{U.~K.}}
(\byear{2013}).
\btitle{{Risk of Bayesian Inference in Misspecified Models, and the Sandwich
  Covariance Matrix}}.
\bjournal{Econometrica: Journal of the Econometric Society}
\bvolume{81}
\bpages{1805--1849}.
\end{barticle}
\endbibitem

\bibitem[\protect\citeauthoryear{Newton and Raftery}{1994}]{Newton:1994}
\begin{barticle}[author]
\bauthor{\bsnm{Newton},~\bfnm{Michael~A}\binits{M.~A.}} \AND
  \bauthor{\bsnm{Raftery},~\bfnm{Adrian~E}\binits{A.~E.}}
(\byear{1994}).
\btitle{{Approximate Bayesian Inference with the Weighted Likelihood
  Bootstrap}}.
\bjournal{Journal of the Royal Statistical Society. Series B (Methodological)}
\bvolume{56}
\bpages{3--46}.
\end{barticle}
\endbibitem

\bibitem[\protect\citeauthoryear{Piironen and Vehtari}{2017}]{Piironen:2017}
\begin{barticle}[author]
\bauthor{\bsnm{Piironen},~\bfnm{Juho}\binits{J.}} \AND
  \bauthor{\bsnm{Vehtari},~\bfnm{Aki}\binits{A.}}
(\byear{2017}).
\btitle{{Sparsity information and regularization in the horseshoe and other
  shrinkage priors}}.
\bjournal{Electronic Journal of Statistics}
\bvolume{11}
\bpages{5018--5051}.
\end{barticle}
\endbibitem

\bibitem[\protect\citeauthoryear{Royall and Tsou}{2003}]{Royall:2003}
\begin{barticle}[author]
\bauthor{\bsnm{Royall},~\bfnm{Richard}\binits{R.}} \AND
  \bauthor{\bsnm{Tsou},~\bfnm{Tsung-Shan}\binits{T.-S.}}
(\byear{2003}).
\btitle{{Interpreting statistical evidence by using imperfect models: robust
  adjusted likelihood functions}}.
\bjournal{Journal of the Royal Statistical Society: Series B (Statistical
  Methodology)}
\bvolume{65}
\bpages{391--404}.
\end{barticle}
\endbibitem

\bibitem[\protect\citeauthoryear{Rubin}{1981}]{Rubin:1981:BayesianBootstrap}
\begin{barticle}[author]
\bauthor{\bsnm{Rubin},~\bfnm{Donald~B}\binits{D.~B.}}
(\byear{1981}).
\btitle{{The Bayesian Bootstrap}}.
\bjournal{The Annals of Statistics}
\bvolume{9}
\bpages{130--134}.
\end{barticle}
\endbibitem

\bibitem[\protect\citeauthoryear{Syring and Martin}{2019}]{Syring:2018}
\begin{barticle}[author]
\bauthor{\bsnm{Syring},~\bfnm{Nicholas}\binits{N.}} \AND
  \bauthor{\bsnm{Martin},~\bfnm{Ryan}\binits{R.}}
(\byear{2019}).
\btitle{{Calibrating general posterior credible regions}}.
\bjournal{Biometrika}
\bvolume{106}
\bpages{479--486}.
\end{barticle}
\endbibitem

\bibitem[\protect\citeauthoryear{van~der Vaart}{1998}]{vanderVaart:1998}
\begin{bbook}[author]
\bauthor{\bparticle{van~der} \bsnm{Vaart},~\bfnm{A~W}\binits{A.~W.}}
(\byear{1998}).
\btitle{{Asymptotic Statistics}}.
\bpublisher{University of Cambridge}.
\end{bbook}
\endbibitem

\bibitem[\protect\citeauthoryear{van~der Vaart and
  Wellner}{1996}]{vanderVaart:1996}
\begin{bbook}[author]
\bauthor{\bparticle{van~der} \bsnm{Vaart},~\bfnm{A~W}\binits{A.~W.}} \AND
  \bauthor{\bsnm{Wellner},~\bfnm{Jon~A}\binits{J.~A.}}
(\byear{1996}).
\btitle{{Weak Convergence and Empirical Processes}}.
\bseries{With Applications to Statistics}.
\bpublisher{Springer}, \baddress{New York}.
\end{bbook}
\endbibitem

\bibitem[\protect\citeauthoryear{Waddell, Kishino and Ota}{2002}]{Waddell:2002}
\begin{barticle}[author]
\bauthor{\bsnm{Waddell},~\bfnm{Peter~J}\binits{P.~J.}},
  \bauthor{\bsnm{Kishino},~\bfnm{Hirohisa}\binits{H.}} \AND
  \bauthor{\bsnm{Ota},~\bfnm{Rissa}\binits{R.}}
(\byear{2002}).
\btitle{{Very fast algorithms for evaluating the stability of ML and Bayesian
  phylogenetic trees from sequence data.}}
\bjournal{Genome informatics. International Conference on Genome Informatics}
\bvolume{13}
\bpages{82--92}.
\end{barticle}
\endbibitem

\bibitem[\protect\citeauthoryear{Walker}{2013}]{Walker:2013}
\begin{barticle}[author]
\bauthor{\bsnm{Walker},~\bfnm{Stephen~G}\binits{S.~G.}}
(\byear{2013}).
\btitle{Bayesian inference with misspecified models}.
\bjournal{Journal of statistical planning and inference}
\bvolume{143}
\bpages{1621--1633}.
\end{barticle}
\endbibitem

\bibitem[\protect\citeauthoryear{Walker and Hjort}{2001}]{Walker:2001}
\begin{barticle}[author]
\bauthor{\bsnm{Walker},~\bfnm{Stephen~G}\binits{S.~G.}} \AND
  \bauthor{\bsnm{Hjort},~\bfnm{Nils~Lid}\binits{N.~L.}}
(\byear{2001}).
\btitle{{On Bayesian consistency}}.
\bjournal{Journal of the Royal Statistical Society: Series B (Statistical
  Methodology)}
\bvolume{63}
\bpages{811--821}.
\end{barticle}
\endbibitem

\bibitem[\protect\citeauthoryear{White}{1982}]{White:1982}
\begin{barticle}[author]
\bauthor{\bsnm{White},~\bfnm{Halbert}\binits{H.}}
(\byear{1982}).
\btitle{{Maximum Likelihood Estimation of Misspecified Models}}.
\bjournal{Econometrica: Journal of the Econometric Society}
\bvolume{50}
\bpages{1--25}.
\end{barticle}
\endbibitem

\bibitem[\protect\citeauthoryear{Yang and Zhu}{2018}]{Yang:2018}
\begin{barticle}[author]
\bauthor{\bsnm{Yang},~\bfnm{Ziheng}\binits{Z.}} \AND
  \bauthor{\bsnm{Zhu},~\bfnm{Tianqi}\binits{T.}}
(\byear{2018}).
\btitle{{Bayesian selection of misspecified models is overconfident and may
  cause spurious posterior probabilities for phylogenetic trees}}.
\bjournal{Proceedings of the National Academy of Sciences}
\bvolume{115}
\bpages{1854--1859}.
\end{barticle}
\endbibitem

\end{thebibliography}

\newpage

\appendix 

\counterwithin{figure}{section}
\renewcommand{\thefigure}{\Alph{section}.\arabic{figure}}
\counterwithin{table}{section}
\renewcommand{\thetable}{\Alph{section}.\arabic{table}}

\section{Computation}
\subsection{Choosing the number of bootstrap datasets for BayesBag}  \label{sec:choosing-B}

If we wish to use \cref{eq:bayesbag-approximation}  to approximate the bagged posterior,
then %
we must select the number of bootstrap datasets $B$. 
Assume that we can approximate $\postdensityfull{\param}{\bsdatasample{b}}$ to high accuracy.
Then evaluating the accuracy of the BayesBag approximation given by \cref{eq:bayesbag-approximation} reduces to the well-studied problem of estimating the accuracy of
a simple Monte Carlo approximation~\citep[e.g.,][]{Koehler:2009}. 
In practice, we have found it sufficient to take $B = 50$ or $100$ since the quantities we wish to estimate seem to be fairly low-variance.
Thus, we suggest starting with $B = 50$, estimating the Monte Carlo error of any quantities of interest such as parameter 
means and variances, and then increasing $B$ if the estimated error is unacceptably large. 
On the other hand, in some scenarios it may be desirable to reduce computational expense by balancing the number of bootstrap samples $B$
versus the accuracy of the approximation to 
$\postdensityfull{\param}{\bsdatasample{b}}$ (e.g., in terms of the length of Markov chain Monte Carlo runs). 
We discuss this computational trade-off next. 

\subsection{A BayesBag sampling algorithm} \label{app:computation}

When the posterior can be computed in closed form, using BayesBag is straightforward.
If, however, approximate sampling methods such as Markov chain Monte Carlo are necessary, the computational cost could become substantial. 
In such cases we propose the basic scheme described in \cref{alg:bb-sampler}, although more advanced approaches could also be developed.
In short, the idea is to run a single long chain (or set of chains) on the standard posterior, then use the sampler hyperparameters
and posterior samples to initialize shorter chains that sample from many different bootstrap datasets.

If the approximation of $\postdensityfull{\param}{\bsdatasample{b}}$ is not very accurate
(e.g., because it requires a time-consuming Markov chain Monte Carlo run), then we face a tradeoff between the error due to 
approximating each $\postdensityfull{\param}{\bsdatasample{b}}$ and the Monte Carlo error due to the BayesBag approximation given in \cref{eq:bayesbag-approximation}.
When using Markov chain Monte Carlo, we recommend assessing on how accurate different length Markov chains are likely to be 
by running long chains for the standard posterior, then using this information to decide on the best trade off between the length of the Markov chains and number of bootstrap datasets. 
Such an approach should not result in much wasted computation since it is usually desirable to obtain a high-quality approximation to the standard posterior anyway.

\begin{algorithm}
\caption{Basic BayesBag Sampler}\label{alg:bb-sampler}
\begin{algorithmic}[1]
\Require A Markov chain Monte Carlo procedure \Call{MCMC}{$\data$, $T$, $\param_\text{init}$, $\beta_\text{init}$} that returns adapted sampler hyperparameters 
and $T$ approximate samples from $\postdistfull{\cdot}{\data}$, with the sampler initialized at $\param_\text{init}$ with hyperparameters $\beta_\text{init}$
\Require Data $\data$, ``large'' sample number $T$, ``small'' sample number $T^{\bbsym}$, number of bootstrap samples $\bsnumobs$, number of bootstrap datasets $B$, 
initial hyperparameters $\beta_\text{init}$

\State $\beta, \param_{1:T} \gets$ \Call{MCMC}{$\data$, $T$, $\beta_\text{init}$} %
\For{$b = 1,\dots,B$}
	\State Generate a new bootstrap dataset $\bsdatasample{b}$ of size $\bsnumobs$ from $\data$
	\State Sample $\param^{\bbsym}_{(b)\text{init}}$ uniformly from $\param_{1:T}$
	\State $\beta_{(b)}, \param^{\bbsym}_{(b)1:T^{\bbsym}} \gets$ \Call{MCMC}{$\bsdatasample{b}$, $T^{\bbsym}$, $\param^{\bbsym}_{(b)\text{init}}$, $\beta$}
\EndFor
\State $\theta^{\bbsym}_{1:BT^{\bbsym}} \gets \text{concatenate}(\param^{\bbsym}_{(1)1:T^{\bbsym}},\dots, \param^{\bbsym}_{(B)1:T^{\bbsym}})$
\State \Return posterior samples $\param_{1:T}$ and BayesBag samples $\theta^{\bbsym}_{1:BT^{\bbsym}}$
\end{algorithmic}
\end{algorithm}

\section{Additional experimental results}

\subsection{Additional linear regression simulations} 

\begin{figure}[tbp]
\begin{center}
\begin{subfigure}[b]{.44\textwidth}
\centering
\includegraphics[height=1.65in,trim={0in 0in 1.45in 0in},clip]{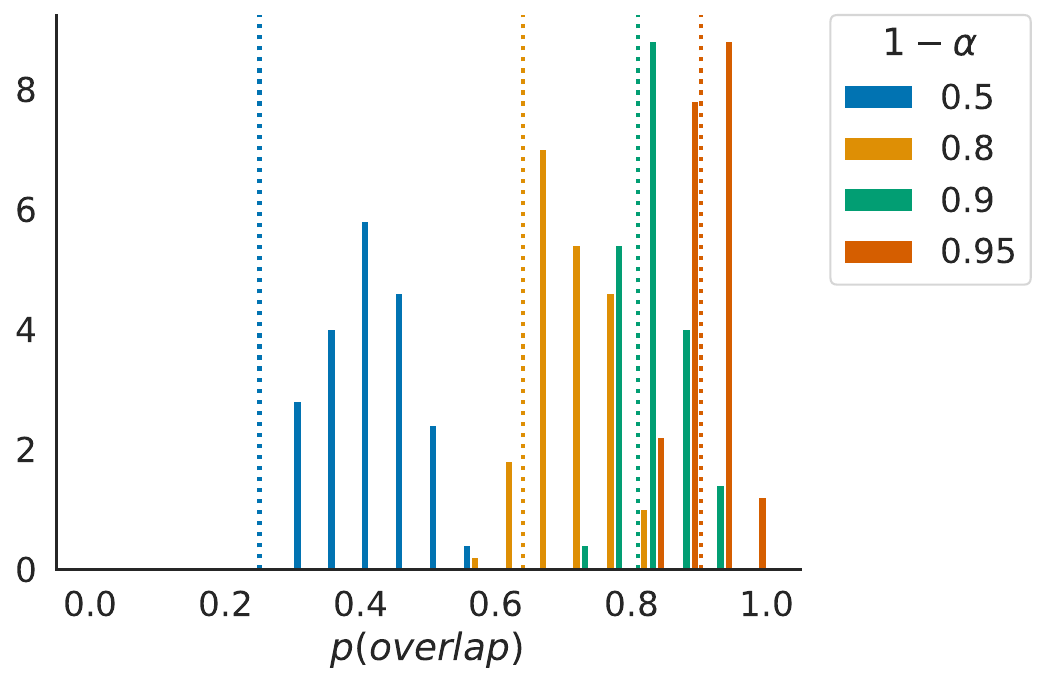}
\caption{$N = D = 250$, Bayes}
\end{subfigure}
\begin{subfigure}[b]{.55\textwidth}
\centering
\includegraphics[height=1.65in]{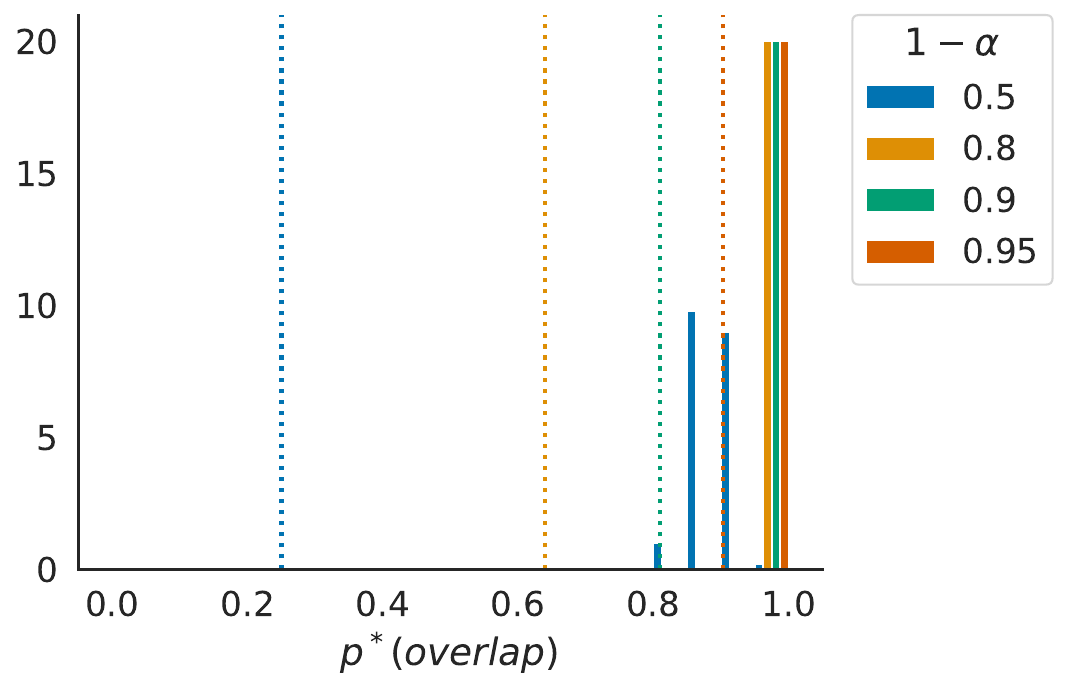}
\caption{$N = D = 250$, BayesBag}
\end{subfigure} \\
\begin{subfigure}[b]{.44\textwidth}
\centering
\includegraphics[height=1.65in,trim={0in 0in 1.45in 0in},clip]{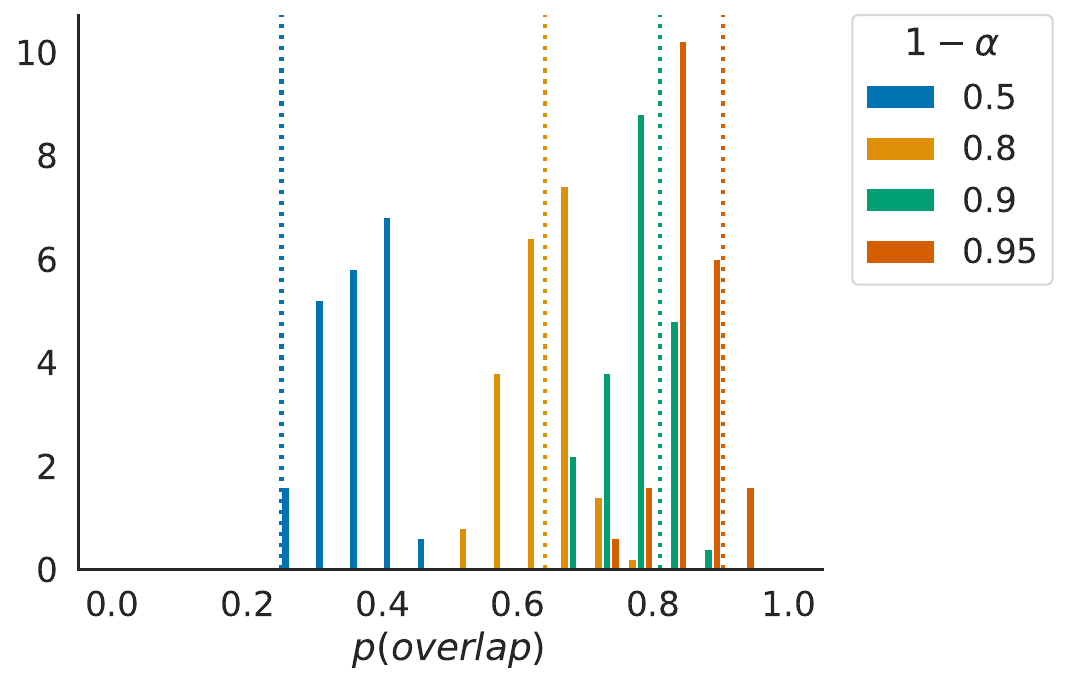}
\caption{$N = D = 500$, Bayes}
\end{subfigure}
\begin{subfigure}[b]{.55\textwidth}
\centering
\includegraphics[height=1.65in]{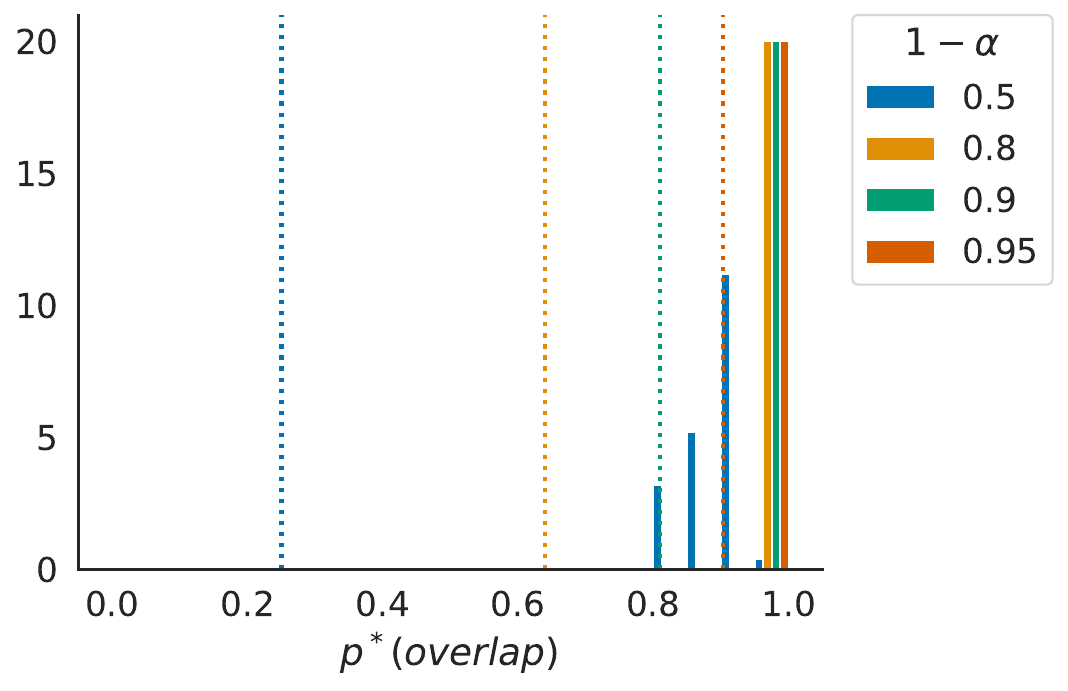}
\caption{$N = D = 500$, BayesBag}
\end{subfigure} 
\caption{Histograms of the probability of overlap of $1-\alpha$ credible sets for $(Z_{i}^{\mathrm{test}})^{\top} \beta$ ($i=1,\dots,100$) for linear regression with \textsf{nonlinear-correlated-2} data. 
}
\label{fig:linear-regression-correlated2-N-eq-D-histograms}
\end{center}
\end{figure}

\begin{figure}[tbp]
\begin{center}
\begin{subfigure}[b]{.44\textwidth}
\centering
\includegraphics[height=1.65in,trim={0in 0in 1.45in 0in},clip]{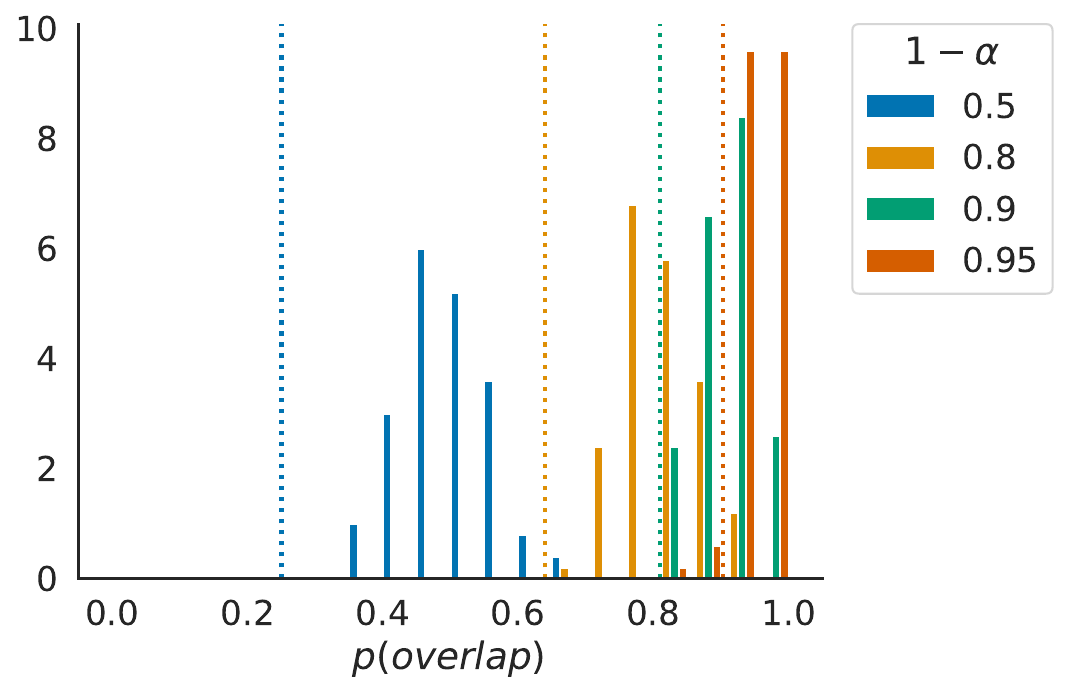}
\caption{$N = D = 250$, Bayes}
\end{subfigure}
\begin{subfigure}[b]{.55\textwidth}
\centering
\includegraphics[height=1.65in]{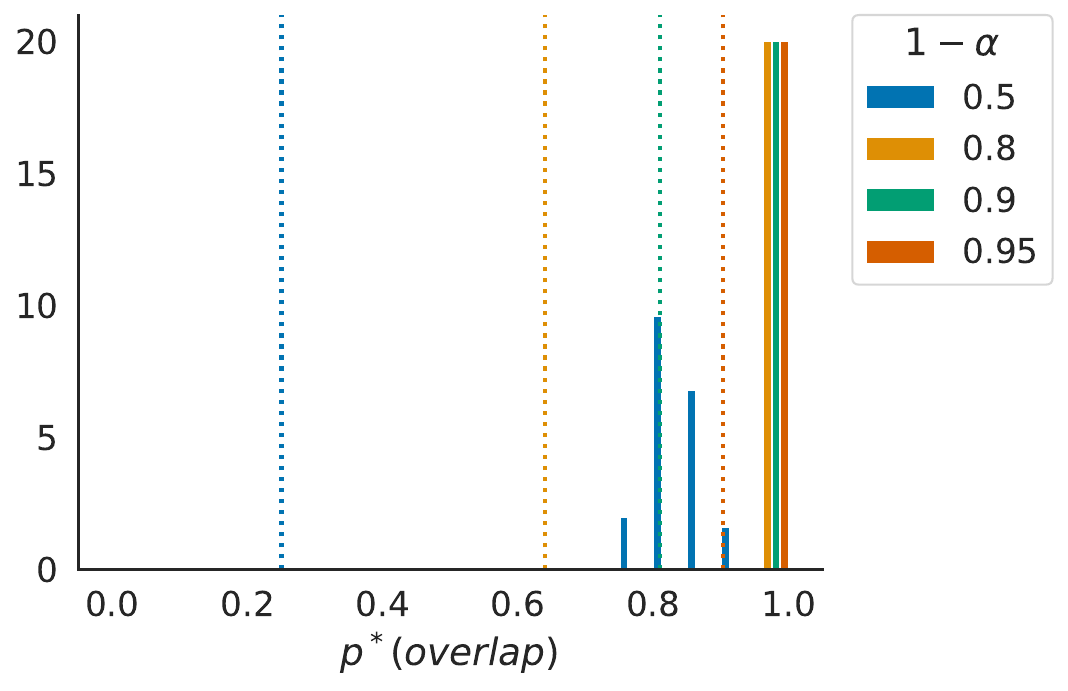}
\caption{$N = D = 250$, BayesBag}
\end{subfigure} \\
\begin{subfigure}[b]{.44\textwidth}
\centering
\includegraphics[height=1.65in,trim={0in 0in 1.45in 0in},clip]{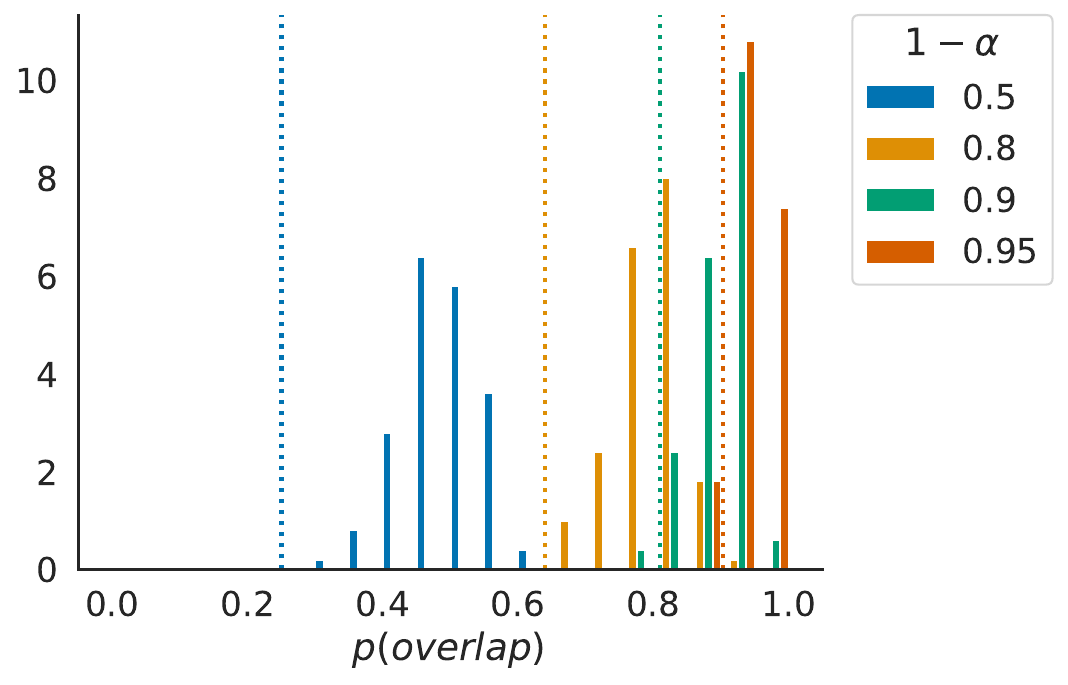}
\caption{$N = D = 500$, Bayes}
\end{subfigure}
\begin{subfigure}[b]{.55\textwidth}
\centering
\includegraphics[height=1.65in]{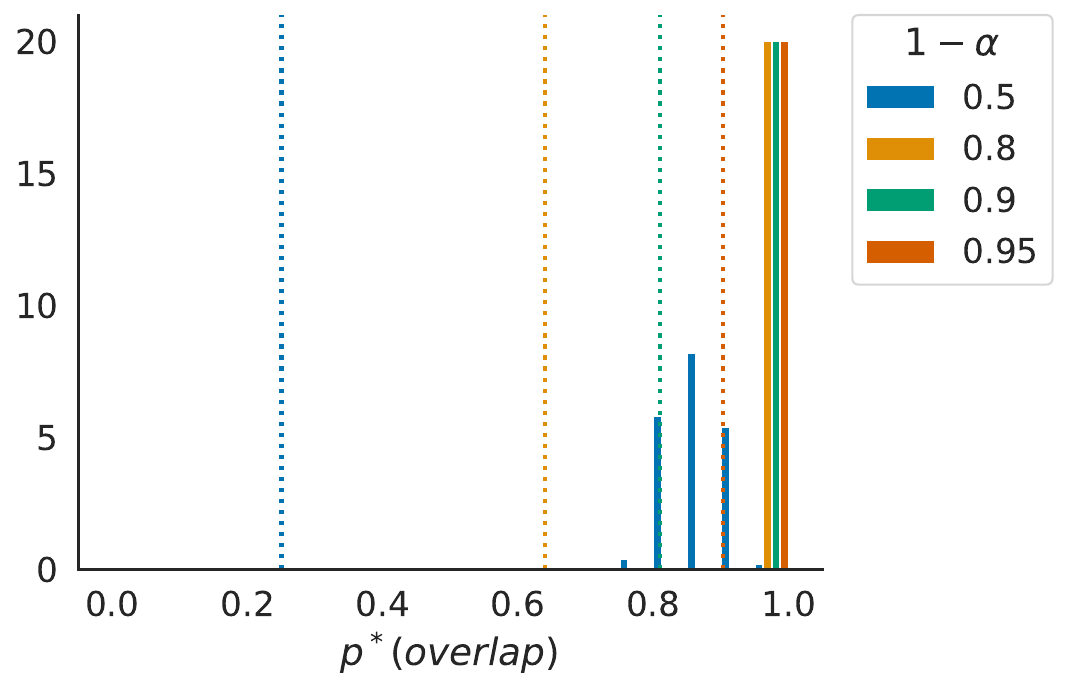}
\caption{$N = D = 500$, BayesBag}
\end{subfigure} 
\caption{Histograms of the probability of overlap of $1-\alpha$ credible sets for $(Z_{i}^{\mathrm{test}})^{\top} \beta$ ($i=1,\dots,100$) for linear regression with \textsf{nonlinear-correlated-4} data. 
}
\label{fig:linear-regression-correlated4-N-eq-D-histograms}
\end{center}
\end{figure}

\begin{figure}[tbp]
\begin{center}
\begin{subfigure}[b]{.44\textwidth}
\centering
\includegraphics[height=1.65in,trim={0in 0in 1.45in 0in},clip]{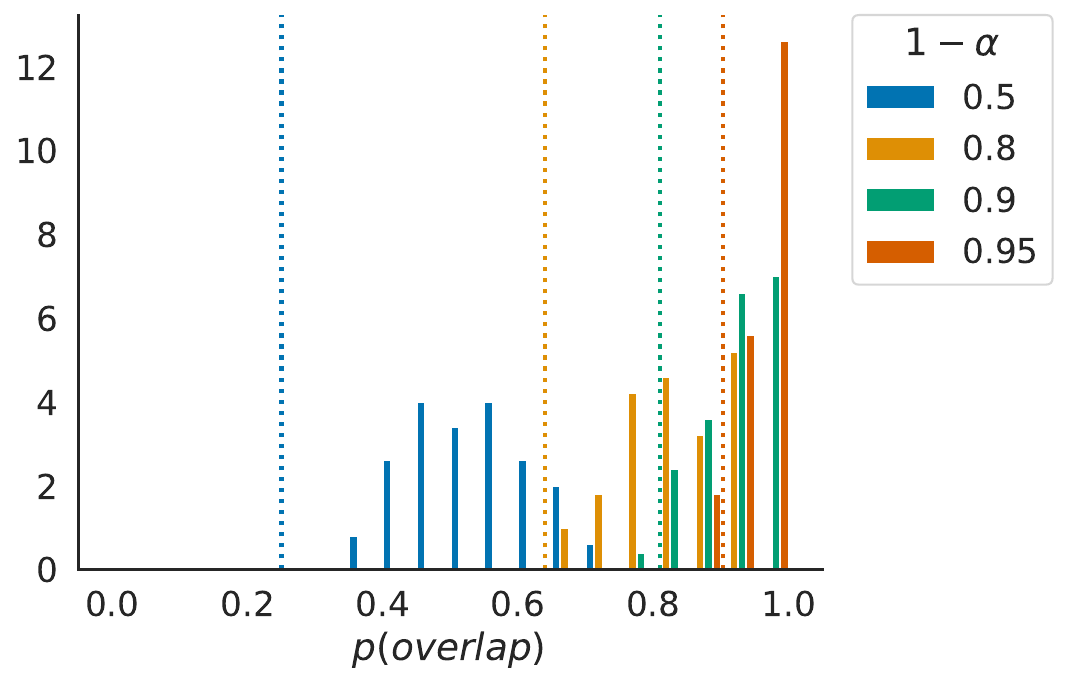}
\caption{$N = D = 250$, Bayes}
\end{subfigure}
\begin{subfigure}[b]{.55\textwidth}
\centering
\includegraphics[height=1.65in]{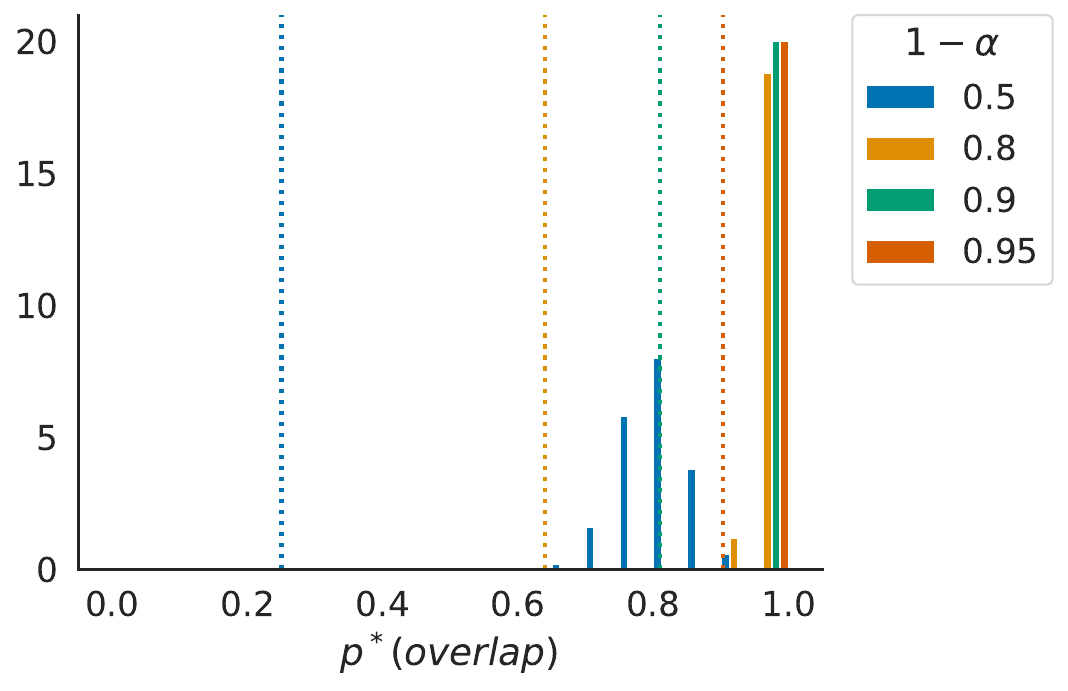}
\caption{$N = D = 250$, BayesBag}
\end{subfigure} \\
\begin{subfigure}[b]{.44\textwidth}
\centering
\includegraphics[height=1.65in,trim={0in 0in 1.45in 0in},clip]{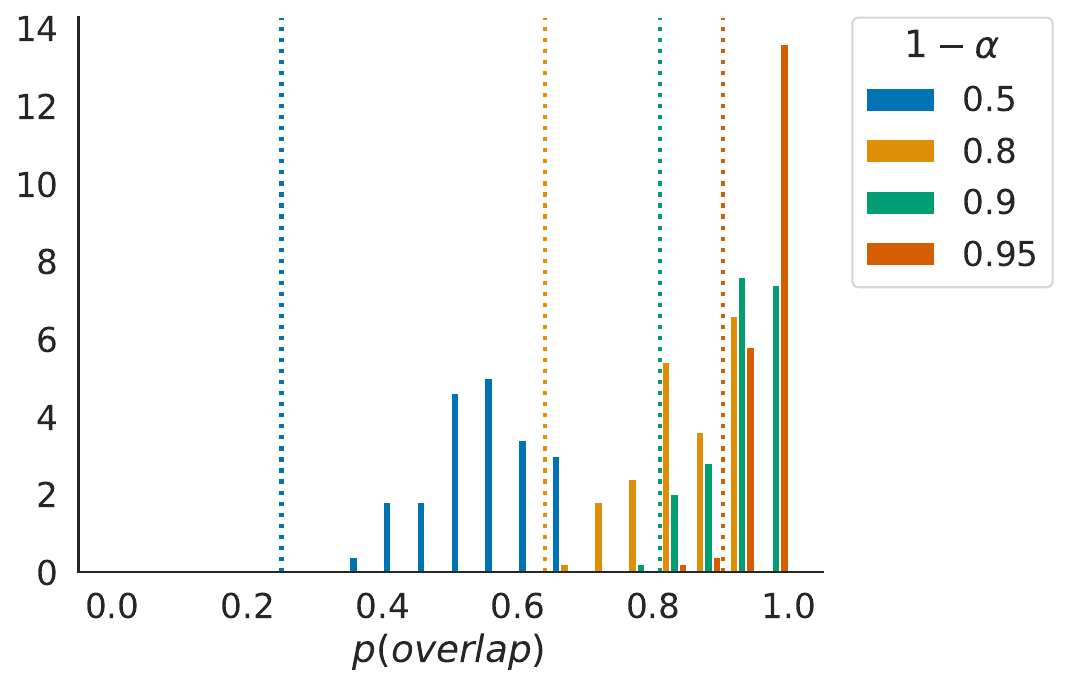}
\caption{$N = D = 500$, Bayes}
\end{subfigure}
\begin{subfigure}[b]{.55\textwidth}
\centering
\includegraphics[height=1.65in]{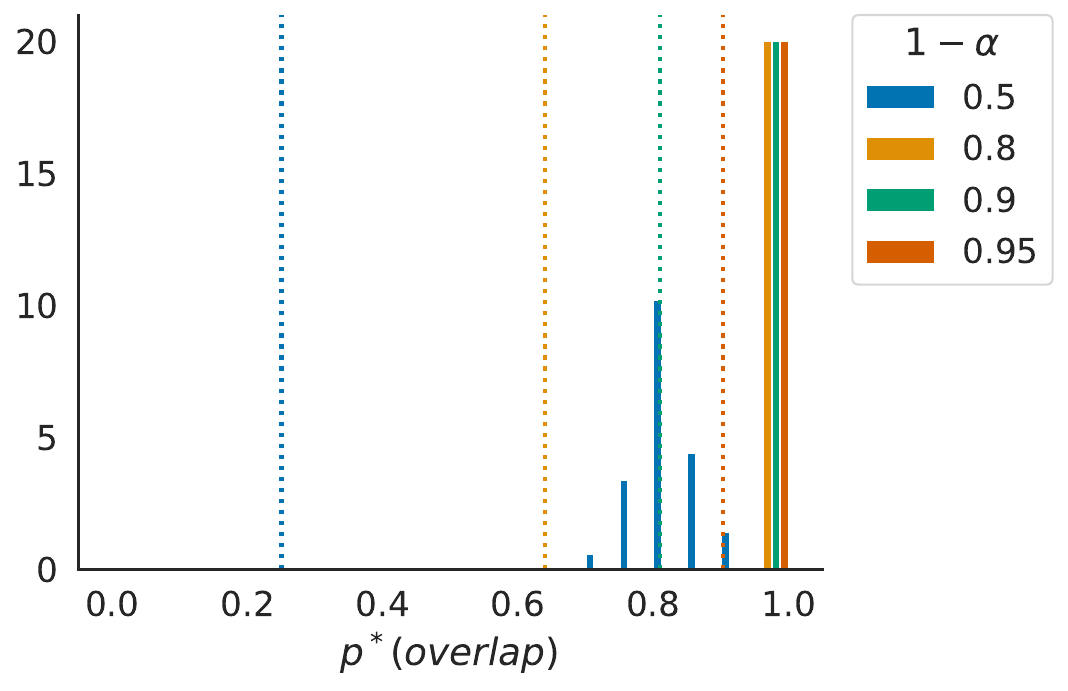}
\caption{$N = D = 500$, BayesBag}
\end{subfigure} 
\caption{Histograms of the probability of overlap of $1-\alpha$ credible sets for $(Z_{i}^{\mathrm{test}})^{\top} \beta$ ($i=1,\dots,100$) for linear regression with \textsf{nonlinear-correlated-8} data.
}
\label{fig:linear-regression-correlated-N-eq-D-histograms}
\end{center}
\end{figure}

\Cref{fig:linear-regression-correlated2-N-eq-D-histograms,fig:linear-regression-correlated4-N-eq-D-histograms,fig:linear-regression-correlated-N-eq-D-histograms} show similar results for \textsf{nonlinear-correlated-$\kappa$} data
to what \cref{fig:linear-regression-uncorrelated-N-eq-D-histograms} shows for \textsf{nonlinear-uncorrelated} data, although the problem with Bayes is less severe as the correlation increases. 

\clearpage

\subsection{Fixed design linear regression simulations} \label{sec:fixed-simulations}

\begin{figure}[tbp]
\begin{center}
\begin{subfigure}[b]{.44\textwidth}
\centering
\includegraphics[height=1.65in,trim={0in 0in 1.45in 0in},clip]{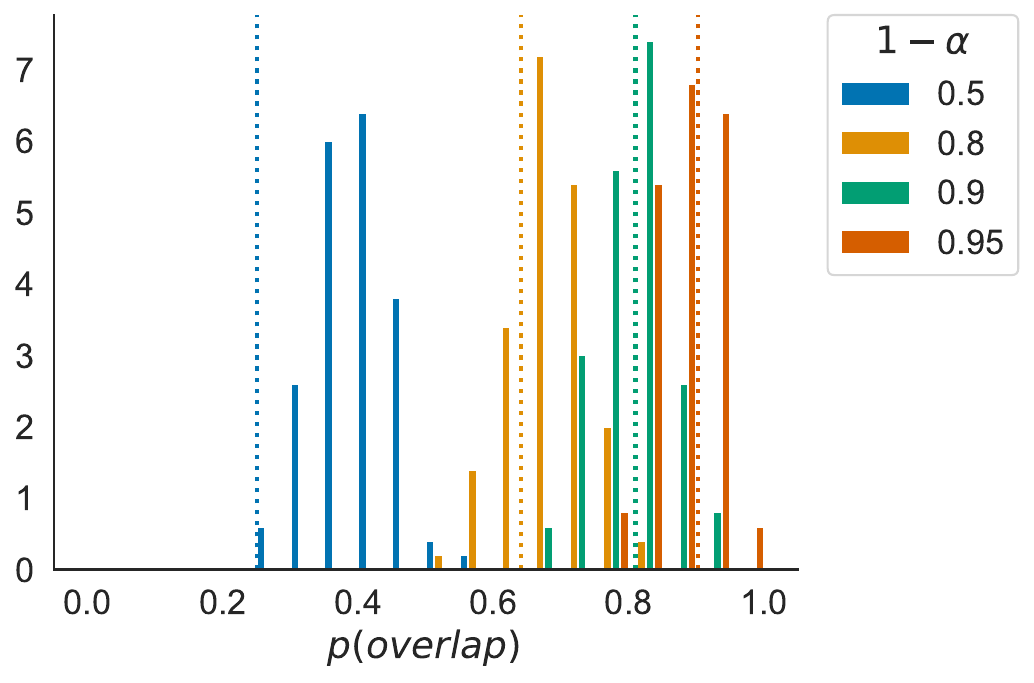}
\caption{$N = D = 256$, Bayes}
\end{subfigure}
\begin{subfigure}[b]{.55\textwidth}
\centering
\includegraphics[height=1.65in]{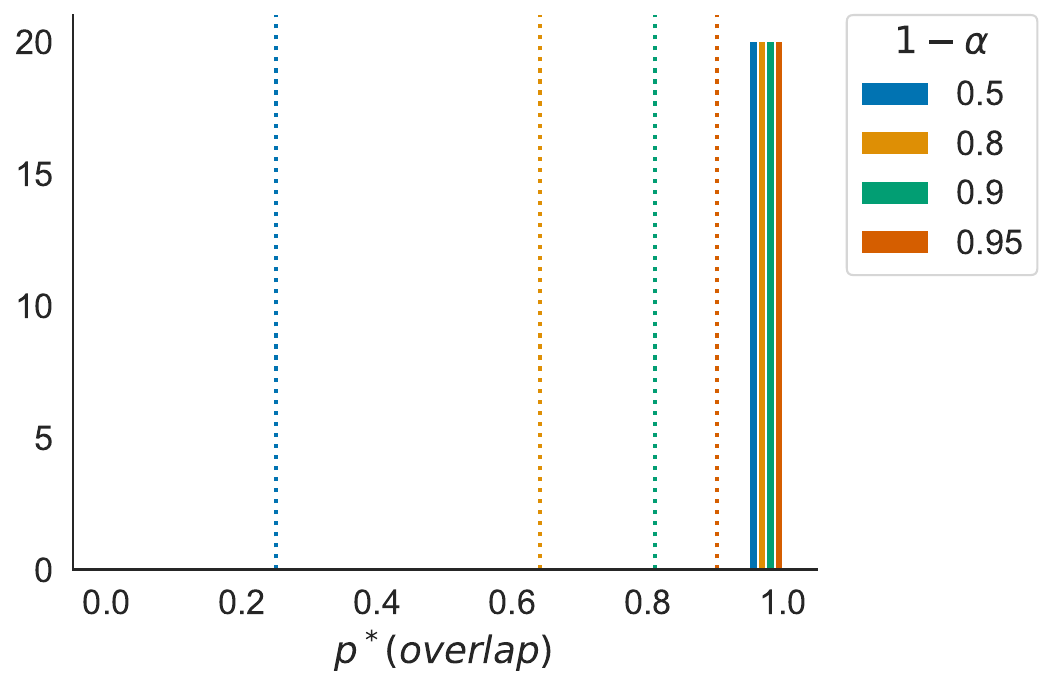}
\caption{$N = D = 256$, BayesBag}
\end{subfigure} \\
\begin{subfigure}[b]{.44\textwidth}
\centering
\includegraphics[height=1.65in,trim={0in 0in 1.45in 0in},clip]{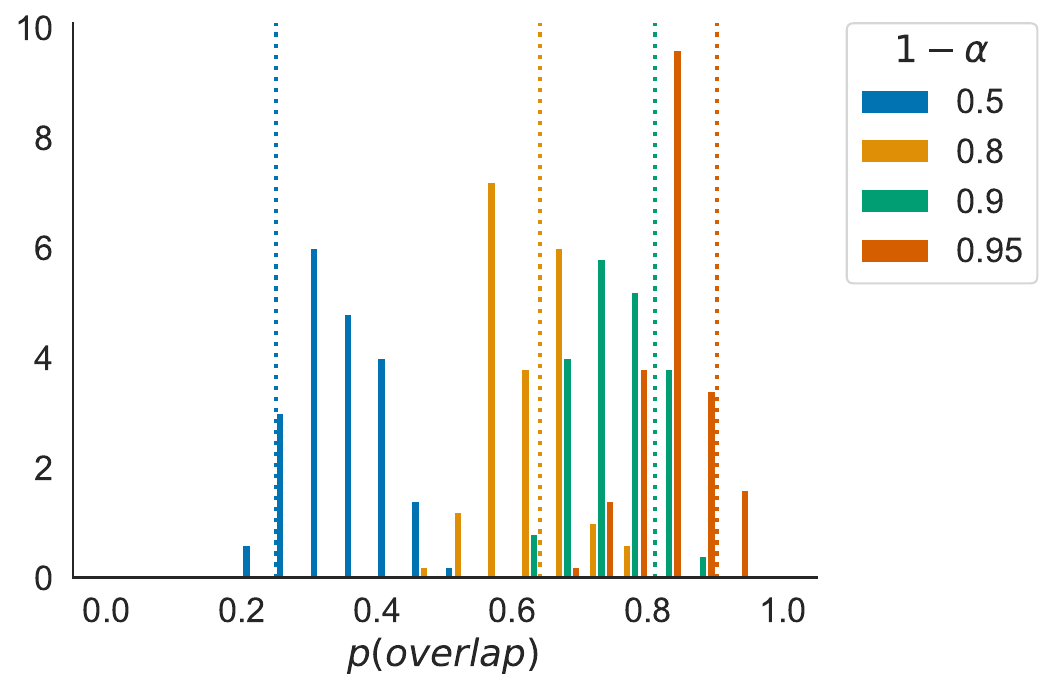}
\caption{$N = D = 400$, Bayes}
\end{subfigure}
\begin{subfigure}[b]{.55\textwidth}
\centering
\includegraphics[height=1.65in]{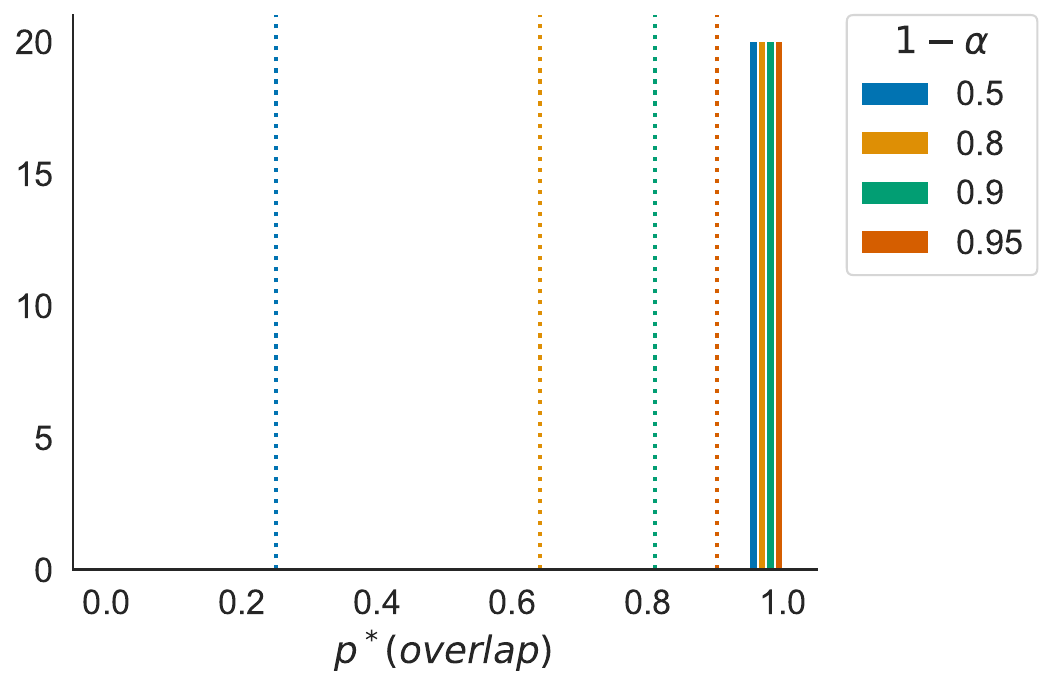}
\caption{$N = D = 400$, BayesBag}
\end{subfigure} 
\caption{Histograms of the probability of overlap of $1-\alpha$ credible sets for $(Z_{i}^{\mathrm{test}})^{\top} \beta$ ($i=1,\dots,100$) for linear regression with \textsf{linear} mean function $f$, fixed design, and heteroskedastic error. }
\label{fig:linear-regression-fixed-N-eq-D-histograms}
\end{center}
\end{figure}

To simulate data for a fixed design scenario, we set $z_{n 0} = 1$ to include an intercept, set covariates $z_{n 1}$ and $z_{n 2}$ to be a uniform grid on $[-2, 2] \times [-2, 2]$, and generate the remaining covariates as i.i.d.\ $\distNorm(0, 1)$.
We use the (well-specified) linear regression function $f(z) = z$ and
to introduce misspecification, we generate the outcomes as in \cref{eq:simulated-linreg-data} but with
heteroskedastic noise given by $\eps_{n} \given z_{n} \distind \distNorm(0, 1 + z_{n 1}^{2} + z_{n 2}^{2})$. 
\Cref{fig:linear-regression-fixed-N-eq-D-histograms} shows that standard Bayes exhibits poor overlap behavior, similar to the case of \textsf{nonlinear-correlated-2} 
data (\cref{fig:linear-regression-correlated2-N-eq-D-histograms}), whereas BayesBag has overlap probability very close to 1 at every test point. 
BayesBag also has superior predictive performance, with 99\% confidence intervals for the difference in
mean log predictive densities of (0.49, 0.69) and (1.00, 1.38) for, respectively, $\numobs = D = 256$ and $400$.

\clearpage

\section{BayesBag Bernstein--Von Mises Theorem for Gaussian Location Model}

\bnthm \label{prop:bb-bbvm-gaussian-location}
Let $\obsrv{1},\obsrv{2},\ldots\in\reals\;\iid$ such that for some $\delta > 0$, $\EE(|\obsrv{1}|^{2+\delta}) < \infty$.
Consider the Gaussian location model from \cref{sec:multivariate-gaussian-location}. 
Let $\bbparamsample\given\datarvarg{\numobs} \dist \bbpostdist{\numobs}$ and 
suppose $\bsscale \defined \lim_{\numobs \to \infty} \bsnumobs/\numobs \in (0,\infty)$ for $\bsnumobs = \bsnumobs(\numobs)$.
Then for almost every $(\obsrv{1},\obsrv{2},\ldots)$,
\[
\numobs^{1/2}\big\{\bbparamsample - \EE(\bbparamsample\given\datarvarg{\numobs})\big\}  \given \datarvarg{\numobs} \convD \distNorm(0, \;V/\bsscale + \var(\obsrv{1})/\bsscale).  \label{eq:bb-bbvm-gaussian-location}
\]
\enthm
In other words, with probability 1, the bagged posterior converges weakly to $\distNorm(0, \;V/\bsscale + \var(\obsrv{1})/\bsscale)$
after centering at its mean and scaling by $\numobs^{1/2}$.
\begin{proof}[\bf Proof of \cref{prop:bb-bbvm-gaussian-location}]
We use the classical characteristic function approach to proving central limit theorems. 
For $\mu \in \reals$ and $\sigma^{2} > 0$, the characteristic function of $\distNorm(\mu, \sigma^{2})$ is
\[
\cf{\distNorm(\mu, \sigma^{2})}(t) &= \exp(\ii \mu t - \sigma^{2}t^{2}/2), \qquad t \in \reals. \label{eq:gaussian-characteristic-alt}
\]
For $L \in \nats$ and $p_1,\ldots,p_K \ge 0$ with $\sum_{k=1}^{K}p_{k} = 1$, 
the characteristic function of the multinomial distribution $\distMulti(L, p)$ is
\[
\cf{\distMulti(L, p)}(t) &= \left(\sum_{k=1}^{K}p_{k}e^{\ii t_{k}}\right)^{L}, \qquad t \in \reals^{K}. \label{eq:multinomial-characteristic-alt}
\]
Let $\widetilde\Pi(\cdot \given \bsdatarvarg{\bsnumobs}) \defined \distNorm(\numobs^{1/2} R_{\bsnumobs}(\bsdatarvmeanarg{\bsnumobs} - \datarvmeanarg{\numobs}), \numobs V_{\bsnumobs})$,
noting that this is the distribution of $\numobs^{1/2}\{\bbparamsample - \EE(\bbparamsample\given\datarvarg{\numobs} )\} \given \bsdatarvarg{\bsnumobs}$.
Similarly, let $\widetilde\Pi^{\bbsym}(\cdot \given \datarvarg{\numobs})$ denote the distribution of $\numobs^{1/2}\{\bbparamsample - \EE(\bbparamsample\given\datarvarg{\numobs} )\} \given \datarvarg{\numobs}$.
Let $Y_{\numobs n} \defined \numobs^{1/2} R_{\bsnumobs}(\obsrv{n} - \datarvmeanarg{\numobs})$
and let $\bscounts \dist \distMulti(\bsnumobs, 1/\numobs)$.
Using \cref{eq:gaussian-characteristic-alt,eq:multinomial-characteristic-alt}, we have 
\[
\cf{\widetilde\Pi^{\bbsym}(\cdot \given \datarvarg{\numobs})}(t) 
&= \EE\{\cf{\widetilde\Pi(\cdot \given \bsdatarvarg{\bsnumobs})}(t)\given \datarvarg{\numobs}\} \\
&= \EE\big[\exp\!\big\{\ii t \bsnumobs^{-1} \textstyle{\sum_{n=1}^\numobs} \bscount{n} Y_{\numobs n} - \numobs V_{\bsnumobs}t^{2}/2\big\}\given \datarvarg{\numobs}\big] \\
&= \left\{\frac{1}{\numobs}\sum_{n=1}^{\numobs} \exp(\ii t \bsnumobs^{-1} Y_{\numobs n})\right\}^{\bsnumobs} \exp(-\numobs V_{\bsnumobs}t^{2}/2).
\label{eq:bb-gaussian-location-characteristic-part-1-alt}
\]
Let $\hat{V}_{\numobs} \defined \numobs^{-1}\sum_{n=1}^{\numobs} (\obsrv{n} - \datarvmeanarg{\numobs})^{2}$. 
By \citet[][Lemma 3.3.19]{Durrett:2019}, $e^{\ii s} = 1 + \ii s - s^{2}/2 + \mcR(s)$ where $\mcR(s) \le \min(|s|^{3}, 2|s|^{2})$. 
Since $\numobs^{-1} \sum_{n=1}^\numobs Y_{\numobs n} = 0$, 
the first factor of \cref{eq:bb-gaussian-location-characteristic-part-1-alt} can be expanded as 
\[
&\left\{\frac{1}{\numobs}\sum_{n=1}^{\numobs} \bigg(1 + \ii t \bsnumobs^{-1} Y_{\numobs n} - \tfrac{1}{2} t^2 \bsnumobs^{-2} Y_{\numobs n}^2 + \mcR(t \bsnumobs^{-1} Y_{\numobs n}) \bigg)\right\}^{\bsnumobs} \\
&= \left\{1 - \tfrac{1}{2} t^2 \frac{\numobs R_{\bsnumobs}^{2}}{\bsnumobs^{2}} \hat{V}_{\numobs} + \frac{1}{\numobs}\sum_{n=1}^{\numobs}\mcR(t \bsnumobs^{-1} Y_{\numobs n})\right\}^{\bsnumobs}.
\label{eq:bb-gaussian-location-characteristic-part-2-alt}
\]
The remainder term $\sum_{n=1}^{\numobs}\mcR(t \bsnumobs^{-1} Y_{\numobs n})$ is bounded by
\[
\frac{t^{2}R_{\bsnumobs}^{2} \numobs}{\bsnumobs^{2}}\max_{n \in \{1,\dots,\numobs\}} \min( \numobs^{1/2}|t| |\obsrv{n} - \datarvmeanarg{\numobs}|/\bsnumobs, 2) \sum_{n=1}^{\numobs} (\obsrv{n} - \datarvmeanarg{\numobs})^{2} 
\]
and by \cref{lem:almost-sure-maximum-to-zero}, $\limsup_{\numobs \to \infty}\max_{n = 1,\dots,\numobs} X_{n}/\numobs^{1/2} \convas 0$. 
By the strong law of large numbers 
\[
\limsup_{\numobs\to\infty} \frac{1}{\numobs}\sum_{n=1}^{\numobs} |\obsrv{n} - \datarvmeanarg{\numobs}|^{2} \overset{a.s.}{<} \infty.
\]
Combining these bounds with the that fact that $\bsnumobs/\numobs \to c$ and $R_{\bsnumobs} \to 1$, 
we conclude that for all $t \in \reals$, $\sum_{n=1}^{\numobs}\mcR(t \bsnumobs^{-1} Y_{\numobs n}) \to 0$ 
as $\numobs\to\infty$.

Further, note that 
$\hat{V}_{\numobs} \convas \var(\obsrv{1})$ as $\numobs\to\infty$.
Now, we use the fact that if $a_\numobs \to a$ and $c_\numobs \to c$, then $(1 + a_\numobs/\numobs)^{\numobs c_\numobs} \to \exp(a)^c$.
Thus, almost surely, for all $t$, \cref{eq:bb-gaussian-location-characteristic-part-2-alt} converges to $\exp(-\tfrac{1}{2} t^2 \var(\obsrv{1}) / c)$.
Combining this with \cref{eq:bb-gaussian-location-characteristic-part-1-alt},
and noting that $\numobs V_{\bsnumobs} \to V/c$, we have that almost surely, for all $t\in\reals$,
\[
\cf{\widetilde\Pi^{\bbsym}(\cdot \given \datarvarg{\numobs})}(t) 
\to \exp(-\tfrac{1}{2} t^2 (\var(\obsrv{1})/c + V/c)).
\]
The result follows by L\'evy's continuity theorem~\citep[Theorem 5.3]{Kallenberg:2002}. 
\end{proof}

\bnlem \label{lem:almost-sure-maximum-to-zero}
Suppose $\obsrv{1},\obsrv{2},\ldots\in\reals\;\iid$ such that  for some $\delta > 0$, $\EE(|\obsrv{1}|^{2+\delta}) < \infty$.
Then $\max_{n \in \{1,\dots,\numobs\}} |\obsrv{n} - \datarvmeanarg{\numobs}|/\numobs^{1/2} \convas 0$ as $\numobs\to\infty$. 
\enlem
\bprf
Define $Y_{n} \defined |\obsrv{n}|/n^{1/2}$.  By Markov's inequality, for all $\veps > 0$,
\[
\Pr(Y_{n} \ge \veps) \le \frac{\EE(|\obsrv{1}|^{2+\delta})}{n^{1+\delta/2}\veps^{2+\delta}}. 
\]
Hence, $\sum_{n=1}^{\infty}\Pr(Y_{n} \ge \veps) < \infty$, so by the Borel--Cantelli lemma, $\limsup_{n} Y_{n} \le \veps$ almost surely.
Since $\veps > 0$ is arbitrary, this implies that $\limsup_{n} Y_n = 0$ almost surely.
Now, let $\veps' > 0$ and define $\numobs_*$ (depending on $Y_1,Y_2,\ldots$) such that $\sup_{n > \numobs_*} Y_n \le \veps'$.
Letting $M = \max_{n \le \numobs_*} Y_n$, we have that almost surely,
\[
\max_{n \le \numobs} |\obsrv{n} - \datarvmeanarg{\numobs}|/\numobs^{1/2} 
&\le \max_{n \le \numobs} 2 |\obsrv{n}|/\numobs^{1/2} \\
&\le \max_{n \le \numobs_*} 2 |\obsrv{n}|/\numobs^{1/2} + \max_{\numobs_* < n \le \numobs} 2 |\obsrv{n}|/\numobs^{1/2} \\
&\le 2 M \numobs_*^{1/2}/\numobs^{1/2} + \max_{\numobs_* < n \le \numobs} 2 Y_n \\
&\le 3 \veps'
\]
for all $\numobs$ sufficiently large.
Therefore, since $\veps' > 0$ is arbitrary, 
\[
\limsup_{\numobs \to \infty}\max_{n \le \numobs} |\obsrv{n} - \datarvmeanarg{\numobs}|/\numobs^{1/2} = 0
\]
almost surely.
\eprf

\section{Proofs} \label{app:proofs}

\begin{proof}[\bf Proof of \cref{prop:overlap-bound}]
Since $X$ and $Y$ are independent given $\eta$, then $\indicatorfn(\eta\in A_X)$ and $\indicatorfn(\eta\in B_Y)$ are independent given $\eta$. Thus,
\begin{align}
\Pr(A_X\cap B_Y \neq \varnothing \mid \eta) &\geq \Pr(\eta \in A_X\cap B_Y \mid \eta)\notag\\
&= \Pr(\eta \in A_X,\, \eta\in B_Y \mid \eta)\notag\\
&= \Pr(\eta \in A_X \mid \eta)\Pr(\eta \in B_Y \mid \eta)\notag\\
&\geq (1-\alpha) (1-\alpha').\notag
\end{align}
\end{proof}

\begin{proof}[\bf Proof of \cref{thm:overlap-general}]
Since $X$ and $\tilde{X}$ are independent and identically distributed given $\paramsample$,
\begin{align}
\EE\big(\Pr(A_X\cap A_{\tilde{X}} \neq \varnothing \mid \paramsample)\big)
&\geq \EE\big(\Pr(\paramsample\in A_X, \, \paramsample\in A_{\tilde{X}} \mid \paramsample)\big) \\
&= \EE\big(\Pr(\paramsample\in A_X \mid \paramsample) \,\Pr(\paramsample\in A_{\tilde{X}} \mid \paramsample)\big) \\
&= \EE\big(\Pr(\paramsample\in A_X \mid \paramsample)^2 \big) \\
&\geq \EE\big(\Pr(\paramsample\in A_X \mid \paramsample) \big)^2 \\
&= \EE\big(\Pr(\paramsample \in A_X \mid X)\big)^2 \\
&\geq (1-\alpha)^2,
\end{align}
where in the last step we use that $\Pr(\paramsample\in A_x \mid x) \geq 1-\alpha$ for all $x$.
\end{proof}

\begin{proof}[\bf Proof of \cref{thm:overlap-gaussian}]
To handle both the standard Bayes and BayesBag cases simultaneously, consider a multivariate normal posterior on $\param$ with mean
$R_M \bar{X}_N \in \reals^{D}$ and covariance matrix $V_M + b M^{-1} R_M \hat{\Sigma}_N R_M$;
then standard Bayes is the case of $M=N$ and $b = 0$, while BayesBag is the case of $b = 1$.
The posterior of $u^\top \param$ is then 
\[
\distNorm(u^\top R_M \bar{X}_N, \, \sigma^2_{X_{1:N}})
\]
where $\sigma^2_{X_{1:N}} \defined u^\top V_M u + b M^{-1} u^\top R_M \hat{\Sigma}_N R_M u$.
Thus, a $100(1-\alpha)\%$ credible interval for $u^\top \param$ is given by 
\[
A_{X_{1:\numobs}}^b = u^\top R_M \bar{X}_N \pm z_{\alpha/2}\, \sigma_{X_{1:N}}.
\]
Letting $X_{1:N}$ and $Y_{1:N}$ be independent data sets drawn i.i.d.\ from $\obsdist$, we have 
\[
\label{eq:overlap-gaussian-proof}
\Pr(A_{X_{1:N}}^b \cap A_{Y_{1:N}}^b \neq \varnothing) &= \Pr\Big(\big\vert u^\top R_M \bar{X}_N - u^\top R_M \bar{Y}_N\big\vert \leq z_{\alpha/2} (\sigma_{X_{1:N}} + \sigma_{Y_{1:N}})\Big).
\]
By the central limit theorem, $N^{1/2}(\bar{X}_N - \bar{Y}_N) \convD \distNorm(0, 2\Sigma_{\optsym})$.
By assumption, $M/N \to c > 0$ as $N\to\infty$, which implies that $M\to\infty$.
Recalling that $R_M = (V_0^{-1} V/M + I)^{-1}$ and $V_M = (V_0^{-1} + M V^{-1})^{-1}$,
we have $R_M \to I$ and $N V_M \to V/c$ as $N\to\infty$.
Thus, by the strong law of large numbers,
$N^{1/2} \sigma_{X_{1:N}} \to (u^\top V u/c + b u^\top \Sigma_{\optsym} u / c)^{1/2}$ almost surely as $N\to\infty$,
and likewise for $N^{1/2}\sigma_{Y_{1:N}}$.
Therefore, by Slutsky's theorem, as $N\to\infty$,
\[
\Pr(A_{X_{1:N}}^b \cap A_{Y_{1:N}}^b \neq \varnothing) &\longrightarrow \Pr\Big(\big\vert \distNorm(0,2 u^\top\Sigma_{\optsym} u)\big\vert \leq z_{\alpha/2} 2 (u^\top V u/c + b u^\top \Sigma_{\optsym} u / c)^{1/2}\Big) \\
&= \Pr\bigg(|W| \leq z_{\alpha/2} \sqrt{2} \Big(\frac{u^\top ((V + b \Sigma_{\optsym})/c) u}{u^\top\Sigma_{\optsym} u}\Big)^{1/2}\bigg)
\]
where $W\sim\distNorm(0,1)$. This proves the theorem.
\end{proof}

\begin{proof}[\bf Proof of \cref{thm:overlap-gaussian-growing-dimension}]
Since $X_n,Y_n \sim \distNorm(0,\Sigma_{\optsym})$ i.i.d., we have $u^\top \bar{X}_N - u^\top \bar{Y}_N \sim \distNorm(0, 2 u^\top \Sigma_{\optsym} u / N)$.
Thus, for the standard posterior, by setting $V = I$, $V_0^{-1}=0$, $\|u\|=1$, $M = N$, and $b=0$ in \cref{eq:overlap-gaussian-proof}, we have
\[
\Pr(A_{X_{1:N}} \cap A_{Y_{1:N}} \neq \varnothing) &= \Pr\Big(\big\vert u^\top \bar{X}_N - u^\top \bar{Y}_N\big\vert \leq z_{\alpha/2} \, 2 / \sqrt{N}\Big) \\
&= \Pr\Big(\big\vert (2 u^\top \Sigma_{\optsym} u / N)^{1/2} W \big\vert \leq z_{\alpha/2} \, 2/\sqrt{N}\Big) \\
&= \Pr\big(|W| \leq z_{\alpha/2} \sqrt{2} / (u^\top \Sigma_{\optsym} u)^{1/2}\big)
\]
since $R_M = I$ and $\sigma_{X_{1:N}}^2 = 1/N$.  This proves the first part.

For the bagged posterior, define $X_n' = u^\top X_n$ and $Y_n' = u^\top Y_n$. Letting $s^2_{X'}$ denote the sample variance of $X'_n$, we have
\[
\sigma^2_{X_{1:N}} =  M^{-1} + M^{-1} u^\top \hat{\Sigma}_N u \geq  M^{-1} N^{-1} \sum_{n=1}^N (X'_n - \bar{X}'_N)^2 = M^{-1} s^2_{X'}
\]
and likewise for $\sigma^2_{Y_{1:N}}$, since $V = I$, $V_0^{-1}=0$, $\|u\|=1$, and $b=1$. Thus,
\[
\sigma_{X_{1:N}} + \sigma_{Y_{1:N}} \geq  (\sigma_{X_{1:N}}^2 + \sigma_{Y_{1:N}}^2)^{1/2} \geq M^{-1/2} (s^2_{X'} + s^2_{Y'})^{1/2}.
\]
Hence, by \cref{eq:overlap-gaussian-proof},
\[
\label{eq:overlap-gaussian-growing-dimension-proof}
\Pr(A^*_{X_{1:N}} \cap A^*_{Y_{1:N}} \neq \varnothing) 
&= \Pr\Big(\big\vert u^\top \bar{X}_N - u^\top \bar{Y}_N\big\vert \leq z_{\alpha/2} (\sigma_{X_{1:N}} + \sigma_{Y_{1:N}})\Big) \\
&\geq \Pr\Big(\big\vert\bar{X}'_N - \bar{Y}'_N\big\vert \leq z_{\alpha/2} \, M^{-1/2} (s^2_{X'} + s^2_{Y'})^{1/2}\Big).
\]
Letting $s^2 = u^\top \Sigma_{\optsym} u$, we have that $\sqrt{N}\bar{X}'_N / s = N^{-1/2} \sum_{n=1}^N X'_n/s \sim \distNorm(0,1)$ and $N s^2_{X'} / s^2 \sim \chi^2_{N-1}$ independently, by Cochran's theorem.  Since $X'_n$ and $Y'_n$ are independent, $\sqrt{N / (2 s^2)}(\bar{X}'_N - \bar{Y}'_N) \sim \distNorm(0,1)$
and $(N / s^2) (s^2_{X'} + s^2_{Y'}) \sim \chi^2_{2 N - 2}$ independently. Hence,
\[
\label{eq:overlap-gaussian-growing-dimension-proof-2}
\sqrt{N - 1} \frac{\bar{X}'_N - \bar{Y}'_N}{(s^2_{X'} + s^2_{Y'})^{1/2}} \eqD \frac{\distNorm(0,1)}{\sqrt{\chi^2_{2 N - 2} / (2 N - 2)}}
\eqD T_{2 N - 2}.
\]
Combining \cref{eq:overlap-gaussian-growing-dimension-proof,eq:overlap-gaussian-growing-dimension-proof-2} yields
\[
\Pr(A^*_{X_{1:N}} \cap A^*_{Y_{1:N}} \neq \varnothing) 
&\geq \Pr\bigg(\frac{|\bar{X}'_N - \bar{Y}'_N|}{(s^2_{X'} + s^2_{Y'})^{1/2}} \leq z_{\alpha/2} \, M^{-1/2} \bigg) \\
&= \Pr\big(|T_{2 N - 2}| \leq z_{\alpha/2} \sqrt{(N-1)/M} \big),
\]
as claimed.
\end{proof}

\begin{proof}[\bf Proof of \cref{thm:overlap-bvm}]
The proof is similar to the Gaussian location model, but simpler in some ways since we are already in the asymptotic regime.
First, consider the standard posterior.
Denote $V = \Ehessloglik{\optparam}^{-1}$ and $\Sigma = \Ehessloglik{\optparam}^{-1}\Vargradloglik{\optparam}\Ehessloglik{\optparam}^{-1}$.
Since $\numobs^{1/2}(\paramsample - \mle{\numobs}) \given \datarvarg{\numobs} \convD \distNorm(0, V)$ by assumption,
a $100(1-\alpha)\%$ credible interval for $u^\top \theta$ based on the asymptotic normal distribution is 
$A_{X_{1:\numobs}} = u^\top \mle{\numobs} \pm z_{\alpha/2} (u^\top V u / \numobs)^{1/2}$.
Likewise, $\numobs^{1/2}(\mle{\numobs} - \optparam) \convD \distNorm(0, \Sigma)$ by assumption.
Thus, letting $\mle{X}$ and $\mle{\tilde{X}}$ be the maximum likelihood estimators based on independent data sets $X_{1:\numobs}$ and $\tilde{X}_{1:\numobs}$,
we have $\numobs^{1/2} (u^\top \mle{X} - u^\top \mle{\tilde{X}}) \convD \distNorm(0, 2 u^\top \Sigma u)$
by the continuity theorem.
Thus, letting $W\sim \distNorm(0,1)$, the asymptotic overlap probability for the standard posterior is
\[
p_\infty(\mathrm{overlap}) &= \lim_{\numobs\to\infty} \Pr\Big(\big|u^\top \mle{X} - u^\top \mle{\tilde{X}}\big| \leq 2 z_{\alpha/2} (u^\top V u / \numobs)^{1/2}\Big) \\
&= \lim_{\numobs\to\infty} \Pr\Big(\big|\numobs^{1/2}(u^\top \mle{X} - u^\top \mle{\tilde{X}})\big| \leq 2 z_{\alpha/2} (u^\top V u)^{1/2}\Big) \\
&= \Pr\Big(\big|(2 u^\top \Sigma u)^{1/2} W\big| \leq 2 z_{\alpha/2} (u^\top V u)^{1/2}\Big) \\
&= \Pr\Big(|W| \leq \sqrt{2}z_{\alpha/2} \Big(\frac{u^\top V u}{u^\top \Sigma u}\Big)^{1/2}\Big),
\]
as claimed.
For the bagged posterior, we have $\numobs^{1/2}(\bbparamsample - \mle{\numobs}) \given \datarvarg{\numobs} \convD \distNorm(0, V/c + \Sigma/c)$ by assumption.
Hence, the proof is the same but with $V/c + \Sigma/c$ in place of $V$.
The claimed inequality for the bagged posterior holds because $u^\top V u \geq 0$, due to the fact that $V$ is positive semi-definite.
\end{proof}

\bprf[\bf Proof of \cref{thm:overlap-linreg}]
Let $\mu_{\dagger} = m(Z)$, $\tilde{\mu}_{\dagger} = m(\tilde{Z})$, $\Sigma_{\dagger} = K(Z)$, and $\tilde{\Sigma}_{\dagger} = K(\tilde{Z})$.
We have
\begin{align}
p(&\mathrm{overlap} \mid Z,\tilde{Z}) = \Pr(A\cap\tilde{A}\neq\varnothing \mid Z,\tilde{Z}) \notag \\
&= \Pr\Big(|v^\top Y - \tilde{v}^\top \tilde{Y}| \leq z_{\alpha/2} (\sigma\|v\| + \tilde{\sigma}\|\tilde{v}\|)  \;\Big\vert\; Z,\tilde{Z}\Big) \label{eq:overlap-linreg-standard}\\
&= \Pr\bigg(\Big\vert W + \frac{v^\top \mu_{\dagger} - \tilde{v}^\top \tilde{\mu}_{\dagger}}{(v^\top\Sigma_{\dagger} v + \tilde{v}^\top\tilde{\Sigma}_{\dagger} \tilde{v})^{1/2}} \Big\vert \leq \frac{z_{\alpha/2} (\sigma\|v\| + \tilde{\sigma}\|\tilde{v}\|)}{(v^\top\Sigma_{\dagger} v + \tilde{v}^\top\tilde{\Sigma}_{\dagger} \tilde{v})^{1/2}}\bigg). \notag
\end{align}
where $W\sim\distNorm(0,1)$.
If $m(Z) = Z \beta_{\dagger}$ and $K(Z) = \sigma_{\dagger}^2 I$, then 
\[ \label{eq:overlap-linreg-correct-spec1}
v^\top \mu_{\dagger} - \tilde{v}^\top \tilde{\mu}_{\dagger} 
= u^\top Z^{+} Z\beta_{\dagger} - u^\top \tilde{Z}^{+} \tilde{Z}\beta_{\dagger} = u^\top \beta_{\dagger} - u^\top \beta_{\dagger} = 0
\]
where $Z^{+} := (Z^\top Z)^{-1} Z^\top$ denotes the Moore--Penrose pseudoinverse of $Z$. Therefore, by \cref{eq:overlap-linreg-standard,eq:overlap-linreg-correct-spec1},
\[ \label{eq:overlap-linreg-correct-spec2}
p(\mathrm{overlap} \mid Z,\tilde{Z}) 
&= \Pr\bigg(|W| \leq \frac{z_{\alpha/2} (\sigma\|v\| + \tilde{\sigma}\|\tilde{v}\|)}{\sigma_{\dagger} (\|v\|^2 + \|\tilde{v}\|^2)^{1/2}}\bigg),
\]
which proves \cref{eq:thm-overlap-linreg-1}.

Suppose $Z = \tilde{Z}$, but we make no assumptions on the form of $m(Z)$ or $K(Z)$.
Then $v = \tilde{v}$, $\mu_{\dagger} = \tilde{\mu}_{\dagger}$, and $\Sigma_{\dagger} = \tilde{\Sigma}_{\dagger}$. Therefore, by \cref{eq:overlap-linreg-standard},
\[
p(\mathrm{overlap}\mid Z,\tilde{Z}) = \Pr\bigg(|W| \leq \frac{z_{\alpha/2} (\sigma + \tilde{\sigma}) \|v\|}{\sqrt{2} (v^\top \Sigma_{\dagger} v)^{1/2}}\bigg),
\]
proving \cref{eq:thm-overlap-linreg-2}.

Suppose $K(Z) = \sigma_{\dagger}^2 I$, but we make no assumptions on the form of $m(Z)$. Then by the Cauchy--Schwarz inequality,
\[ \label{eq:overlap-linreg-width}
\sigma\|v\| + \tilde{\sigma}\|\tilde{v}\| \leq \sqrt{\sigma^2 + \tilde{\sigma}^2} (\|v\|^2 + \|\tilde{v}\|^2)^{1/2} = \frac{\sqrt{\sigma^2 + \tilde{\sigma}^2}}{\sigma_{\dagger}} (v^\top\Sigma_{\dagger} v + \tilde{v}^\top\tilde{\Sigma}_{\dagger} \tilde{v})^{1/2},
\]
so by \cref{eq:overlap-linreg-standard},
\[
p(\mathrm{overlap} \mid Z,\tilde{Z}) 
\leq \Pr\bigg(\Big\vert W + \frac{v^\top \mu_{\dagger} - \tilde{v}^\top \tilde{\mu}_{\dagger}}{\sigma_{\dagger}\,(\|v\|^2 + \|\tilde{v}\|^2)^{1/2}} \Big\vert \leq \frac{z_{\alpha/2} \sqrt{\sigma^2 + \tilde{\sigma}^2}}{\sigma_{\dagger}}\bigg),
\]
which proves \cref{eq:thm-overlap-linreg-3}.
\eprf

The characteristic function of a distribution $\eta$ on $\reals^K$ is denoted $\cf{\eta}(t) \defined \int \exp(\ii t^\top x) \eta(\dee x)$ for $t\in\reals^K$.
We use $\convP$ to denote convergence in probability and $\convPouter$ to denote convergence in outer probability.

\begin{proof}[\bf Proof of \cref{thm:bb-bvm}]

We use the shorthand notation $\loglikfun{\param} \defined \log \likfun{\param}$,
and denote the gradient and Hessian by $\gradloglikfun{\param} \defined \grad_{\param}\loglikfun{\param}$
and $\hessloglikfun{\param} \defined \grad_{\param}^2\loglikfun{\param}$, respectively.
To de-clutter the notation, we abbreviate $\Ehessloglik{\optsym} \defined \Ehessloglik{\optparam}$, 
$\Vargradloglik{\optsym} \defined \Vargradloglik{\optparam}$,
and $\gradloglikfun{\optsym} \defined \gradloglikfun{\optparam}$.
Define
\[ 
\bbempdist &\defined \bsnumobs^{-1}\sum_{n=1}^{\numobs}\bscount{n}\delta_{\obsrv{n}}, \\
\Delta^{\bbsym}_{\numobs} &\defined \numobs^{1/2}\Ehessloglik{\optsym}^{-1}(\bbempdist - \empdist)\gradloglikfun{\optparam},
\]
the empirical process $\mathbb{G}_{\numobs} = \numobs^{1/2}(\empdist - \obsdist)$, and 
the bootstrap empirical process $\mathbb{G}^{\bbsym}_{\numobs} = \bsnumobs^{1/2}(\bbempdist - \empdist)$. 
The conditions of \citet[Lemma 19.31]{vanderVaart:1998} hold by assumption, so for any sequence $h_1,h_2,\ldots\in\reals^{D}$ bounded in probability, 
\[
\mathbb{G}_{\numobs}\{\numobs^{1/2}\lambda_{\numobs} - h_{\numobs}^{\top}\gradloglikfun{\optsym}\} \convP 0,
\]
where $\lambda_{\numobs} = \loglikfun{\optparam + h_{\numobs}/\numobs^{1/2}} - \loglikfun{\optparam}$.
By \citet[Theorem 3.6.3]{vanderVaart:1996}, for almost every $\alldatarv$, conditional on $\alldatarv$, $\mathbb{G}^{\bbsym}_{\numobs}$ and $\mathbb{G}_{\numobs}$ both 
converge weakly to the same limiting process.
For the remainder of the proof we condition on $\alldatarv$, so all statements will hold for almost every $\alldatarv$. 
It follows that
\[
\mathbb{G}^{\bbsym}_{\numobs}\{\numobs^{1/2}\lambda_{\numobs} - h_{\numobs}^{\top}\gradloglikfun{\optsym}\} \convPouter 0, \label{eq:bs-emp-proc-to-zero}
\]
where we recall that $\convPouter$ denotes convergence in outer probability. 
By the proof of \citet[Lemma 2.1]{Kleijn:2012}, 
\[
|N\empdist\lambda_{\numobs} - \mathbb{G}_{\numobs} h_{\numobs}^{\top}\gradloglikfun{\optsym} - \tfrac{1}{2}h_{\numobs}^{\top} \Ehessloglik{\optsym}h_{\numobs}| \convPouter 0
\]
and, following the same reasoning, we can expand the lefthand side of \cref{eq:bs-emp-proc-to-zero} and multiply though by $\bsscale^{1/2}$ to get 
\[
\bsscale^{1/2}(\numobs\bsnumobs)^{1/2} \bbempdist\lambda_{\numobs} - \bsscale^{1/2}\mathbb{G}^{\bbsym}_{\numobs} h_{\numobs}^{\top}\gradloglikfun{\optsym} 
 - \bsscale^{1/2}(\numobs\bsnumobs)^{1/2}\empdist\lambda_{\numobs} \convPouter 0
\]
and hence
\[
\bsnumobs \bbempdist\lambda_{\numobs} - (\bsscale^{1/2}\mathbb{G}^{\bbsym}_{\numobs} + \bsscale\mathbb{G}_{\numobs}) h_{\numobs}^{\top}\gradloglikfun{\optsym} - \tfrac{1}{2}h_{\numobs}^{\top}(\bsscale\Ehessloglik{\optsym})h_{\numobs} \convPouter 0. 
\]
Since $\bsscale\,\mathbb{G}_{\numobs} h_{\numobs}^{\top}\gradloglikfun{\optsym} = h_{\numobs}^{\top} (\bsscale \Ehessloglik{\optsym})\Delta_{\numobs}$ 
and $\bsscale^{1/2}\mathbb{G}^{\bbsym}_{\numobs} h_{\numobs}^{\top}\gradloglikfun{\optsym} (\bsscale \numobs / \bsnumobs)^{1/2} = h_{\numobs}^{\top} (\bsscale \Ehessloglik{\optsym})\Delta^{\bbsym}_{\numobs}$ by the definitions of $\Delta_{\numobs}$ and $\Delta^{\bbsym}_{\numobs}$,
it follows that for every compact $K \subset \paramspace$, 
\[
\sup_{h \in K} \Big\vert \bsnumobs \bbempdist(\loglikfun{\optparam + h/\numobs^{1/2}} - \loglikfun{\optparam}) - h^{\top} (\bsscale\Ehessloglik{\optsym}) (\Delta_{\numobs} + \Delta^{\bbsym}_{\numobs}) - \tfrac{1}{2}h^{\top}(\bsscale\Ehessloglik{\optsym})h\Big\vert \convPouter 0.
\]
We apply \citet[Theorem 2.1]{Kleijn:2012} to conclude that, letting $\bbparamsamplecopy|\bsdatarvarg{\bsnumobs} \dist \postdistfull{\cdot}{\bsdatarvarg{\bsnumobs}}$, 
the total variation distance between the distribution of $\numobs^{1/2}(\bbparamsamplecopy -  \optparam) \mid \bsdatarvarg{\bsnumobs}$
and $\distNorm(\Delta_{\numobs} + \Delta^{\bbsym}_{\numobs}, \Ehessloglik{\optsym}^{-1}/\bsscale)$ converges to zero in outer probability.
Compared to the notation of \citet[Theorem 2.1]{Kleijn:2012}, we have $\bsdatarvarg{\bsnumobs}$ in place of $X^{(n)}$, $\empdist^{\bsnumobs}$ in place of $P_{0}^{(n)}$,
$\bsscale\Ehessloglik{\optsym}$ in place of $V_{\theta^{*}}$, and $\Delta_{\numobs} + \Delta^{\bbsym}_{\numobs}$ in place of $\Delta_{n,\theta^{*}}$. 
Hence, uniformly in $t \in \reals^{D}$, the absolute difference in their characteristic functions must also converge to zero in outer probability.
Let $\eps_{\numobs}(t)$ (and similarly $\bar\eps_{\numobs}(t)$) denote a function that satisfies $\limsup_{\numobs \to \infty} \sup_{t \in \reals} \eps_{\numobs}(t) = 0$. 
We can therefore write the characteristic function of $\numobs^{1/2}(\bbparamsample - \optparam) - \Delta_{\numobs} \mid \datarvarg{\numobs}$ 
evaluated at $t \in \reals^{D}$ as
\[
\lefteqn{\EE\left[\exp\left\{\ii\Delta^{\bbsym \top}_{\numobs}t - t^{\top}\Ehessloglik{\optsym}^{-1}t/(2\bsscale)\right\}\given \datarvarg{\numobs}\right] + \eps_{\numobs}(t)} \\
\begin{split}
&= \EE\left[\exp\left\{\ii \numobs^{1/2} \bbempdist \gradloglikfun{\optsym}^{\top}\Ehessloglik{\optsym}^{-1}t \right\} \;\big|\; \datarvarg{\numobs}\right]  \exp\left\{- \ii \numobs^{1/2} \empdist\gradloglikfun{\optsym}^{\top}\Ehessloglik{\optsym}^{-1}t\right\} \\
&\phantom{=~}\times \exp\left\{ - t^{\top}\Ehessloglik{\optsym}^{-1}t/(2\bsscale)\right\} + \eps_{\numobs}(t).  \label{eq:bb-bvm-characteristic-function-initial}
\end{split}
\]
Letting $\delta\gradloglik{\obsrv{n}}{\optsym} \defined \gradloglik{\obsrv{n}}{\optsym} -\empdist\gradloglikfun{\optsym}$,
we can further expand the first line of \cref{eq:bb-bvm-characteristic-function-initial} to get 
\[
&\EE\left[\exp\left\{\ii \numobs^{1/2}\bsnumobs^{-1}\sum_{n=1}^{\numobs}\bscount{n}\gradloglik{\obsrv{n}}{\optsym}^{\top} \Ehessloglik{\optsym}^{-1}t \right\}\;\bigg|\; \datarvarg{\numobs}\right] \exp\left\{- \ii \numobs^{1/2} \empdist\gradloglikfun{\optsym}^{\top}\Ehessloglik{\optsym}^{-1}t\right\} \\
&= \left[ \frac{1}{\numobs} \sum_{n=1}^{\numobs} \exp\left\{\frac{\ii  \numobs^{1/2}\gradloglik{\obsrv{n}}{\optsym}^{\top}\Ehessloglik{\optsym}^{-1}t}{\bsnumobs}   \right\}\right]^{\bsnumobs} \exp\left\{- \ii \numobs^{1/2} \empdist\gradloglikfun{\optsym}^{\top}\Ehessloglik{\optsym}^{-1}t\right\} \\
&= \left[ \frac{1}{\numobs} \sum_{n=1}^{\numobs} \exp\left\{\frac{\ii  \numobs^{1/2}\delta\gradloglik{\obsrv{n}}{\optsym}^{\top}\Ehessloglik{\optsym}^{-1}t}{\bsnumobs}   \right\}\right]^{\bsnumobs}\\
&= \left[\frac{1}{\numobs}\sum_{n=1}^{\numobs}\left\{ 1 + \frac{\ii\numobs^{1/2}\delta\gradloglik{\obsrv{n}}{\optsym}^{\top}\Ehessloglik{\optsym}^{-1}t}{\bsnumobs} -  \frac{\numobs(\delta\gradloglik{\obsrv{n}}{\optsym}^{\top}\Ehessloglik{\optsym}^{-1}t)^{2}}{2\bsnumobs^{2}} + \mcR_{n} \right\}\right]^{\bsnumobs}  \\
&= \left\{1 -  \frac{\numobs t^{\top}\Ehessloglik{\optsym}^{-1}\empdist(\delta\gradloglikfun{\optsym}\delta\gradloglikfun{\optsym}^{\top})\Ehessloglik{\optsym}^{-1}t}{2\bsnumobs^{2}} + \mcR_{n} \right\}^{\bsnumobs},  \label{eq:bb-bvm-initial-taylor-expansion}
\]
where (recalling the notation from the proof of \cref{prop:bb-bbvm-gaussian-location})
\[
\mcR_{n} \defined \mcR\left(\frac{\ii\numobs^{1/2}\delta\gradloglik{\obsrv{n}}{\optsym}^{\top}\Ehessloglik{\optsym}^{-1}t}{\bsnumobs}\right).
\]
Arguing as in the proof of \cref{prop:bb-bbvm-gaussian-location} and using assumption (ii), we conclude that 
\[
\lim_{\numobs \to \infty} \sum_{n=1}^{\numobs}\mcR_{n} = 0.
\]

Note that $\bsnumobs/\numobs \to c$, and $\empdist(\delta\gradloglikfun{\optsym}\delta\gradloglikfun{\optsym}^{\top}) \convas \Vargradloglik{\optsym}$ as $\numobs\to\infty$.
Now, we use the fact that if $a_\numobs \to a$ and $c_\numobs \to c$, then $(1 + a_\numobs/\numobs)^{\numobs c_\numobs} \to \exp(a)^c$.
Combining all these observations with \cref{eq:bb-bvm-characteristic-function-initial,eq:bb-bvm-initial-taylor-expansion}, we have that, for all $t  \in \reals^{D}$, 
the characteristic function of $\numobs^{1/2}(\bbparamsample - \optparam) \mid \datarvarg{\numobs}$ evaluated at $t$ is
\[
\exp\left\{ \ii \Delta_{\numobs}^{\top}t - t^{\top}\Ehessloglik{\optsym}^{-1}t/(2\bsscale) -t^{\top}\Ehessloglik{\optsym}^{-1}\Vargradloglik{\optsym}\Ehessloglik{\optsym}^{-1}t/(2\bsscale) \right\} + \eps_{\numobs}(t) + \bar\eps_{\numobs}(t).
\]
The result follows from L\'evy's continuity theorem~\citep[Theorem 5.3]{Kallenberg:2002}.
\end{proof}

\end{document}